\documentclass[fleqn,usenatbib]{mnras}

\usepackage[T1]{fontenc}
\usepackage{ae,aecompl}

\DeclareRobustCommand{\VAN}[3]{#2}
\let\VANthebibliography\thebibliography
\def\thebibliography{\DeclareRobustCommand{\VAN}[3]{##3}\VANthebibliography}

\usepackage{graphicx}	
\usepackage{amsmath}	
\usepackage{amssymb}	
\usepackage{color}
\usepackage{algorithm}
\usepackage{booktabs}
\usepackage[noend]{algpseudocode}
\usepackage{pifont}
\usepackage{soul}
\usepackage{float}
\usepackage{multibib}
\newcites{Appendix}{Appendix References}
\usepackage{natbib}
\usepackage[flushleft]{threeparttablex}
\usepackage{mathrsfs}
\usepackage{widetext}
\usepackage{mathtools}
\usepackage{tikz}
\usepackage{esint}
\usepackage[draft]{todonotes}
\usepackage{lipsum}
\usepackage{longtable}
\usepackage{pifont}
\usepackage{bm}
\usepackage[makeroom]{cancel}

\usetikzlibrary{calc,shapes}

\graphicspath{{Figures/}}

\newcommand{\appropto}{\mathrel{\vcenter{
  \offinterlineskip\halign{\hfil$##$\cr
    \propto\cr\noalign{\kern2pt}\sim\cr\noalign{\kern-2pt}}}}}

\newcommand{\vecB}[1]{\mathrm{{\bmath{\mathit{#1}}}}}
\newcommand{\Exp}[2]{\left\langle{#1}\right\rangle_{#2}}

\renewcommand{\d}[1]{\ensuremath{\operatorname{d}\!{#1}}}

\DeclareMathOperator\V{\mathcal{V}}
\newcommand{\M}{\mathcal{M}}
\DeclareMathOperator\Ma{\mathcal{M}_{\text{A}}}

\newcommand{\MA}[1]{\mathcal{M}^{#1}_{\text{A}}}

\newcommand\ekin{E_{\text{kin}}}
\newcommand\ekinpar{E_{\text{kin},\parallel}}
\newcommand\ekinperp{E_{\text{kin},\perp}}
\newcommand\ekinb{\Exp{\ekin}{\V}}
\newcommand\ekinparb{\Exp{\ekinpar}{\V}}
\newcommand\ekinperpb{\Exp{\ekinperp}{\V}}
\newcommand\emag{E_{\rm mag}}
\newcommand\emagb{\Exp{\emag}{\V}}

\DeclareMathAlphabet\mathbfcal{OMS}{cmsy}{b}{n}

\newcommand{\Rm}{\text{Rm}}
\renewcommand{\Re}{\text{Re}}
\newcommand{\Pm}{\text{Pm}}
\newcommand{\Pms}{\text{Pm}_{\text{shear}}}
\newcommand{\Pmb}{\text{Pm}_{\text{bulk}}}
\renewcommand{\v}{\mathit{v}}
\renewcommand{\b}{\mathit{b}}

\newcommand{\lo}{\ell_0}

\newcommand{\nus}{\nu_{\text{shear}}}
\newcommand{\nub}{\nu_{\text{bulk}}}
\newcommand{\vh}{\vecB{\bar{v}}}
\newcommand{\bh}{\vecB{\bar{b}}}

\newcommand{\bnab}{\bm{\nabla}}

\newcommand{\St}{\mathbb{S}}
\newcommand{\Bt}{\mathbb{B}}
\newcommand{\It}{\mathbb{I}}
\newcommand{\At}{\mathbb{A}}

\newcommand{\vprp}{\vecB{v}_{\perp}}
\newcommand{\vpar}{\vecB{v}_{\parallel}}

\renewcommand{\th}{\bar{t}}
\newcommand{\rhoo}{\rho_0}
\newcommand{\rhoh}{\bar{\rho}}

\newcommand{\Res}{\text{Re}_{\text{shear}}}
\newcommand{\Reb}{\text{Re}_{\text{bulk}}}

\newcommand{\Pnu}{\text{P}\nu}

\newcommand{\first}{1^{\rm st}}
\newcommand{\second}{2^{\rm nd}}

\defcitealias{Brunt2010a}{BFP2010}
\defcitealias{Beattie2020}{BF2020}
\defcitealias{Kolmogorov1941}{K41}
\defcitealias{Menon2020}{MFK2020}
\defcitealias{Hopkins2013}{H2013}
\defcitealias{Skalidis2021}{S+2021}

\usepackage{scalerel,tikz}
\usetikzlibrary{svg.path}
\definecolor{orcidlogocol}{HTML}{A6CE39}
\tikzset{orcidlogo/.pic={\fill[orcidlogocol] svg{M256,128c0,70.7-57.3,128-128,128C57.3,256,0,198.7,0,128C0,57.3,57.3,0,128,0C198.7,0,256,57.3,256,128z}; \fill[white] svg{M86.3,186.2H70.9V79.1h15.4v48.4V186.2z} svg{M108.9,79.1h41.6c39.6,0,57,28.3,57,53.6c0,27.5-21.5,53.6-56.8,53.6h-41.8V79.1z M124.3,172.4h24.5c34.9,0,42.9-26.5,42.9-39.7c0-21.5-13.7-39.7-43.7-39.7h-23.7V172.4z} svg{M88.7,56.8c0,5.5-4.5,10.1-10.1,10.1c-5.6,0-10.1-4.6-10.1-10.1c0-5.6,4.5-10.1,10.1-10.1C84.2,46.7,88.7,51.3,88.7,56.8z};}}
\newcommand\orcidicon[1]{\href{https://orcid.org/#1}{\mbox{\scalerel*{
\begin{tikzpicture}[yscale=-1,transform shape]\pic{orcidlogo};
\end{tikzpicture}}{|}}}}

\newcommand{\aref}[1]{\hyperref[#1]{Appendix~\ref{#1}}}

\newcommand{\nquad}[1][1]{\hspace*{#1em}\ignorespaces}

\title[Bulk viscosity \& the dynamo]{Taking control of compressible modes: bulk viscosity and the turbulent dynamo}

\author[Beattie, et al., 2025]{
James R. Beattie$^{\orcidicon{0000-0001-9199-7771}\,2,3,1}$\thanks{E-mail: james.beattie@princeton.edu}, 
Christoph Federrath$^{\orcidicon{0000-0002-0706-2306}\,1,4}$,
Neco Kriel$^{\orcidicon{0000-0002-3558-3926}\,1}$,
Justin Kin Jun Hew$^{\orcidicon{0000-0002-5238-6115}\,1,5,6}$ 
\newauthor
\;\& Amitava Bhattacharjee$^{\orcidicon{0000-0001-6411-0178}\,2,7}$
\\
$^{1}$Research School of Astronomy and Astrophysics, Australian National University, Canberra, ACT 2611, Australia \\
$^{2}$Department of Astrophysical Sciences, Princeton University, Princeton, NJ 08544, USA \\
$^{3}$Canadian Institute for Theoretical Astrophysics, University of Toronto, 60 St. George Street, Toronto, ON M5S 3H8, Canada \\
$^{4}$Australian Research Council Centre of Excellence in All Sky Astrophysics (ASTRO3D), Canberra, ACT 2611, Australia \\
$^{5}$Space Plasma Power and Propulsion Laboratory, Department of Nuclear Physics and Accelerator Applications,\\ \; Research School of Physics, Australian National University, ACT 2601, Canberra, Australia \\
$^{6}$Mathematical Sciences Institute, Australian National University, Canberra, ACT, 2601, Australia \\
$^{7}$Princeton Plasma Physics Laboratory, Princeton University, Princeton, NJ 08544, USA
}

\date{Accepted XXX. Received YYY; in original form ZZZ}

\pubyear{2025}

\begin{document}
\label{firstpage}
\pagerange{\pageref{firstpage}--\pageref{lastpage}}
\maketitle

\begin{abstract}
    Many polyatomic astrophysical plasmas are compressible and out of chemical and thermal equilibrium, introducing a bulk viscosity into the plasma via the internal degrees of freedom of the molecular composition, directly impacting the decay of compressible modes, $\vecB{v}_{\parallel}(\bm{k})$. This is especially important for small-scale, turbulent dynamo processes in the interstellar medium, which are known to be sensitive to the effects of compression. To control the viscous properties of $\vecB{v}_{\parallel}(\bm{k})$, we perform trans-sonic, visco-resistive dynamo simulations with additional bulk viscosity $\nub$, deriving a new $\nub$ Reynolds number $\Reb$, and viscous Prandtl number $\Pnu \equiv \Reb / \Res$, where $\Res$ is the shear viscosity Reynolds number. We derive a framework for decomposing $\emag$ growth rates into incompressible and compressible terms via orthogonal tensor decompositions of $\bnab\otimes\vecB{v}$, where $\vecB{v}$ is the fluid velocity. We find that $\vecB{v}_{\parallel}(\bm{k})$ play a dual role, growing and decaying $\emag$, and that field-line stretching is the main driver of growth, even in compressible dynamos. In the absence of $\nub$ ($\Pnu \to \infty$), $\vecB{v}_{\parallel}(\bm{k})$ pile up on small-scales, creating a spectral bottleneck, which disappears for $\Pnu \approx 1$. As $\Pnu$ decreases, $\vecB{v}_{\parallel}(\bm{k})$ are dissipated at increasingly larger scales, in turn suppressing incompressible modes through a coupling between high-$k$ modes. We emphasise the importance of further understanding the role of $\nub$ in compressible astrophysical plasmas, which we estimate could be as strong as the shear viscosity in the cold ISM, and highlight that compressible direct numerical simulations without bulk viscosity have unresolved compressible mode dissipation scales.
\end{abstract}


\section{Introduction}\label{sec:intro}
    Compressibility changes the nature of turbulence, from the energy cascade \citep[e.g,][]{Kritsuk2007,Lithwick2001_compressibleMHD,Federrath2013_universality,Wang2018_inertial_range_shocks,Ferrand2020_compr_turb_flux,Federrath2021} to the magnetic field fluctuations embedded within the turbulent medium \citep[e.g.,][]{Mee2006_dynamo_expansion_waves,Federrath2016_dynamo,White2019_supersonic_dynamo_in_a_lab,Mandal2020,Beattie2020,Skalidis2021,Seta2021_saturation_supersonic_dynamo,Bott2021_inefficient_dynamo,Beattie2022_energy_balance,Beattie2022_ion_alfven_fluctuations}. Naturally, one way of parameterising the compressibility of a medium is through the root-mean-squared (rms) sonic Mach number $\M = \Exp{(v/c_s)^2}{\V}^{1/2}$ where $v$ is the fluid velocity, $\V$ is the system volume and $c_s$ is the sound speed within the volume. Directly through the Rankine-Hugoniot shock jump relations, the rms $\M$ tells us both the amplitude of a characteristic shock in a supersonic $\M > 1$ medium \citep{Lehmann2016,Park2019}, and by how much the shocks will compress the gas structure \citep{Landau1959,Padoan2011,Molina2012,Nolan2015,Federrath2015_polytropic,Beattie2020c,Hew2023_shock_dynamo,Davidovits2022_turbulence_generation_by_shocks}, which directly changes the statistics of all the fluid variables through flux freezing. Regardless of whether a plasma is compressible or not, in the presence of turbulent motions, magnetic fields will be grown and nourished through a small-scale, turbulent dynamo \citep[also called fluctuation dynamo, or just small-scale dynamo (SSD); for reviews,][]{Brandenburg2005_dynamo_review,Rincon2019_dynamo_theories,Tobias2021_review}.

    Broadly speaking,\footnote{Note we do not discuss the diffusion-free stage, which happens before the kinematic dynamo, establishing the \citet{Kazantsev1968} spectrum, where the magnetic field is successively folded to smaller-scales, until the peak of the magnetic energy spectrum reaches the resistive dissipation scale \citep{Schekochihin2002,Schekochihin2004_dynamo,StOnge2020S_weakly_collisional_dynamos}.} the turbulent dynamo converts turbulent kinetic energy in the plasma into magnetic energy through a three-stage process. First, while $|\vecB{b}| \ll |\vecB{v}|$ -- where $\vecB{b}$ is the magnetic field permeating through the fluid -- the induction equation becomes (approximately) linear, leading to exponential growth in the volume-averaged magnetic energy, $\emagb$,
    \begin{align} 
    \frac{\d{\emagb}}{\d{t}} &= 2\gamma_1 \emagb.
        && \text{(kinematic)}
        \label{eq:kinematic_growth}
    \intertext{Considering the standard incompressible phenomenology, once the Reynolds stress becomes comparable to the magnetic tension  $\bnab\cdot(\rho\vecB{v}\otimes\vecB{v}) \sim \bnab\cdot(\vecB{b}\otimes\vecB{b})$, the magnetic field exerts a back-reaction on the momentum equation (through the Lorentz force) in turn making the system of equations nonlinear, and suppressing the exponential growth until it becomes algebraic (specifically, linear) in time}
        \frac{\d{\emagb}}{\d{t}} &= 2\gamma_2.
        && \text{(nonlinear)}
        \label{eq:linear_growth}
    \intertext{Finally, with the field growth suppressed, the plasma resistivity balances with the growth terms and the velocity field maintains $\emagb$ in a statistically stationary state where}
        \frac{\d{\emagb}}{\d{t}} &= 0,
        && \text{(saturated)}
        \label{eq:saturated_no_growth}
    \end{align}
    where $\gamma_1$ and $\gamma_2$ are the growth rates of the respective stages \citep[e.g.,][]{Schober2015_saturation_of_turbulent_dynamo,Xu2016_dynamo,Seta2020_seed_magnetic_field,Galishnikova2022_saturation_and_tearing,Kriel2025_supersonic_scales}. Compared to incompressible, solenoidal turbulence, both the kinematic growth rate $\gamma_1$ and $t \rightarrow \infty$ ratio of the volume-averaged energy $\emagb/\ekinb$ are reduced if the medium is highly compressible \citep{Federrath2011_mach_dynamo,Schober2015_saturation_of_turbulent_dynamo,Seta2021_saturation_supersonic_dynamo,Sur2023_dynamo,Kriel2025_supersonic_scales} -- that is, it appears more difficult to grow and maintain a strong magnetic field in the presence of continuously driven shocks than in their absence. In \autoref{sec:dynamo_theory} we will discuss in more detail why this may be the case.

    Considering the effects of compression on the turbulent dynamo is important for many astrophysical phenomena. For example, the interstellar medium in our Galaxy is turbulent and trans-sonic in volume-average \citep{Gaesnsler_2011_trans_ISM}, and turbulent and supersonic by mass-average \citep[e.g.,][]{Klessen2011,Federrath2016_brick,Orkisz2017_Orion_Mach,Beattie2019b}. Hence, the coupling and flow of energy between the turbulence and magnetic fields that are embedded within the compressible interstellar medium are maintained with a super-to-trans-sonic dynamo \citep[e.g.,][]{Beck_2013_Bfield_in_gal,Xu2016_dynamo,McKee2020,Gent2023_dynamo_in_supernova_driven_turb,Beattie2023_GorD_P1}. This means that all phenomena that either depend upon the magnetic field structure or amplitude in the interstellar medium, e.g., cosmic rays trapped on magnetic field lines, transported through the medium \citep{Krumholz2020,Beattie2022_ion_alfven_fluctuations,Kempski2023_CR_transport,Lemoine2023_CR_transport,Sampson2023_streaming_CRs}, or stars forming in magnetised H$_2$ clouds \citep{Federrath2012,Krumholz2019,McKee2020,Mathew2021_IMF_bfields}, must have initial magnetic fields that are seeded and maintained by a supersonic turbulent dynamo (supersonic dynamo hereafter). Hence, all of these processes require an understanding of not just a dynamo, but a supersonic dynamo. 
    
    The supersonic dynamo is ubiquitous across many scales in an active galaxy. It has been shown that galaxy mergers excite these dynamos on $\text{kpc}$ scales in the interstellar medium, \citep{pakmor2014magnetic, pakmor2017magnetic, brzycki2019parameter, whittingham2021impact, whittingham2023impact, pfrommer2022simulating}, but they can also be found down at $\text{km}$ scales in the plasma environment between merging neutrons stars \citep{Miret2023_NS_mergers_winding,Chabanov2023_NSM_crustal_fields}. Of course, many of these phenomena contain more than a single dynamo operating in tandem. For example the magnetorotational instabilitiy dynamo, the $\alpha$ dynamo, shear current effect and $\omega$ dynamo \citep[see][for recent review]{Brandenburg2023_Galactic_Dynamos} have been found to all operate at the same time at least for some parameterisations of black hole accretion disks \citep{Jacquemin2023_BH_accretion_disk_dynamo}. The turbulent dynamo fits into these dynamo ecosystems by efficiently building and maintaining the energy reservoir through the electromotive force, $\Exp{\vecB{v}\times\vecB{b}}{\V}$, that in turn fuels the operation of the other dynamos \citep{Rincon2019_dynamo_theories,Brandenburg2023_Galactic_Dynamos}.

    Even though compressibility has been shown to have a strong effect on the statistics of the turbulent dynamo, previous supersonic dynamo experiments have solely focused on the shear viscosity, $\nus$, and the corresponding shear Reynolds number $\Res$ to control the viscous dissipative properties of the plasma. Changing $\Res$ directly controls the decay timescale for the solenoidal modes e.g., $t_{\nus} \sim (k_{\nus}^2\nus)^{-1}$ (where $k_{\nus}$ is the dissipation wavemode, with corresponding length scale $\ell_{\nus}$), and the growth rate in the kinematic stage, $\gamma_1 \sim v_{\nus}/\ell_{\nus} \sim (\Exp{v^2}{\V}^{1/2})\Res^{1/2}$ where the growth of the magnetic field is set by the dynamical timescale on $\ell_{\nus}$ \citep{Batchelor1950_dynamos,Yousef2007_exact_scaling_laws,Schekochihin2004_dynamo,Xu2016_dynamo,Galishnikova2022_saturation_and_tearing}. However, the decay timescale (and corresponding length scale) for compressible modes, $t_{\nub} \sim (k_{\nub}^2\nub)^{-1}$ (and $\ell_{\nub}$), have been left to decay on the same timescale as the incompressible modes, where $\nub$ is the bulk (volume or longitudinal) viscosity \citep[e.g.,][]{pan_johnsen_2017_bulk_viscosity_decaying_turb,Chen2019_bulk_visc_homogenous_turb,Sharma2023_bulk_viscosity}, which we discuss in much more detail in \autoref{sec:viscosity_theory}. Because the dynamical timescale of compressible turbulent modes are known to be fast, $t \sim \ell/v_{\ell} \sim \ell/\ell^{1/2} \sim \ell^{1/2}$ (assuming \citealt{Burgers1948}-type turbulence; but this may be even faster for low-$\M$ turbulence, where compressible modes follow a $k^{-3}$ spectrum, \citealt{Wang2017_psuedosound_spectrum}), compared to incompressible modes $t \sim \ell/\ell^{1/3} \sim \ell^{2/3}$ (assuming \citealt{Kolmogorov1941}-type turbulence), the diffusion timescale $t_{\nus}$ may be insufficient to decay the compressible modes within a simulation domain (e.g., where each $k$ modes has a dynamical/diffusion timescale associated with it). Therefore, even in direct numerical simulations that explicitly set $\Res$, the decay of the compressible modes may be governed completely by numerical dissipation, which is controlled by the numerical discretisation scheme for the fluxes in the simulation code, the solver used (e.g., how many waves are used in the Riemann solver), and the number of resolution elements in the simulation domain \citep{Grete2023_dynamical_range,Malvadi2023_numerical_dissipation,Grehan2025_numerical_resistivity}. 
    
    Theoretically, under the \citet{stokes_1850} hypothesis, $\nub = 0$ is justified, and exact for monoatomic gases \citep{Tiaza1942_stokes_relation}. However, if diatomic gases, such as H$_2$ and CO, which, for example, dominate the cold phase of the ISM \citep{Ferriere2001,Draine2011_physics_of_ISM}, or triatomic gases such as CO$_2$, HCN or HCO$^+$ present in many protostellar disks and stellar and planetary atmospheres, are out of chemical or thermal equilibrium, the magnitude of the bulk viscosity can vary between tens and thousands of times larger than the shear viscosity \citep{pan_johnsen_2017_bulk_viscosity_decaying_turb,Shama2019_bulk_viscosity_in_dilute_gases,Sharma2023_bulk_viscosity}. This is certainly the case for the plasma entrapped by merging neutron stars \citep{Giovanna2023_bulk_viscosity}, which support dynamos excited by Kelvin Helmholtz instabilities \citep{Chabanov2023_NSM_crustal_fields}, where strong bulk viscosities ($\nub/\nus \gtrsim 100$; \citealt{Most2022_microphysical_viscosity,Most2022_bulk_viscosity_NSM}) may play a role in the gravitational wave signatures of the merger events\citep{Chabanov2023_NSMerger_BV}. Beyond astrophysical plasmas, the exact role of bulk viscosity on the numerics of supersonic turbulence has remained largely unexplored, and for controlled, local box dynamo simulations, completely unexplored.

    In this study, we take the first steps toward describing the appropriate framework for bringing bulk viscosity into the context of turbulent dynamos. This includes defining new dimensionless plasma parameters, the bulk viscous Reynolds number, and the viscous Prandtl number, and making a number of important measurements that probe the role of this type of viscosity for the dynamo. Our results are not only broadly applicable to general polyatomic plasmas, but also highly important for doing controlled studies of supersonic, magnetised turbulence, where spatial and temporal properties of compressible modes are controlled through an explicit viscosity operator (a true compressible DNS). Our study is organised in the following manner. In \autoref{sec:viscosity_theory} we introduce the bulk viscosity through orthogonal decompositions of the velocity gradient tensor, following a somewhat pedagogical approach. In \autoref{sec:dynamo_theory} we go on to provide the theory for understanding compressibility in the context of key dynamo statistics -- the growth rate and the saturated energy ratio. In \autoref{sec:sims} we describe the visco(both shear and bulk)-resistive magnetohydrodynamic (MHD) simulation setup that we use to explore the role of bulk viscosity in turbulent dynamos. In \autoref{sec:integral_quants} we discuss the results of the saturation, dynamo growth rates, and in general, the volume integral quantities of the plasma, including decompositions into compressible and incompressible velocity modes. In \autoref{sec:kin_spectrum} we explore the total, compressible and incompressible kinetic energy spectra, highlighting the effects of bulk viscosity, not just on the compressible spectrum, but also on the coupling between the compressible and incompressible modes at high-$k$. In \autoref{sec:mag_spectra} we focus on the magnetic spectra, the evolution of the integral scales. In \autoref{sec:spectral_ratio} we combine the decomposed kinetic and magnetic energy spectra to explore how the dynamo saturates, on a mode-by-mode basis. Finally, in \autoref{sec:conclusion} we summarise and conclude.

    \section{Fluid Viscosity \& the velocity gradient tensor}\label{sec:viscosity_theory}
    
        \subsection{Bulk and shear viscosity}\label{sec:vel_grad_tensor}
            The viscous momentum equation for a fluid is,
            \begin{align}
                \frac{\partial \rho \vecB{v}}{\partial t} + \bnab\cdot\mathbb{F} = \bnab\cdot\bm{\sigma},
            \end{align}
            where $\mathbb{F}$ is a tensor of conserved momentum fluxes, and $\bm{\sigma}$ is viscous stress tensor. Before kinetic theory showed us the way to describe $\bm{\sigma}$ through perturbations away from thermal equilibrium in the form of the Maxwell-Boltzmann distribution function it was an empirical fact that for Newtonian fluids $\bm{\sigma}$ was proportional to the velocity gradient tensor \mbox{$\bnab\otimes\vecB{v}$}, as,
            \begin{align}
                \|\bm{\sigma}\| \sim \|\bnab\otimes\vecB{v}\|,
            \end{align}
            where $\bnab\otimes\vecB{v} = \partial_i v_j$ is the tensor product. By decomposing $\bnab\otimes\vecB{v}$ into the rate of expansion $\Bt$, the traceless, symmetric, rate of strain $\St$, and the traceless, antisymmetric, rate of rotation $\At$, tensor components, one can gain some insight into the structure of this type of viscosity, 
            \begin{align}
                \bnab\otimes\vecB{v} &= \Bt + \St + \mathbb{A}, \\
                \Bt &= \frac{1}{3} (\bnab\cdot\vecB{v}) \It \label{eq:B_v_grad_tensor},\\
                \St &= \frac{1}{2}\left(\bnab\otimes\vecB{v} + [\bnab\otimes\vecB{v}]^T\right) -\frac{1}{3} (\bnab\cdot\vecB{v}) \It, \label{eq:S_v_grad_tensor} \\
                \At &= \frac{1}{2}\left(\bnab\otimes\vecB{v} - [\bnab\otimes\vecB{v}]^T\right). \label{eq:A_v_grad_tensor} 
            \end{align}
            Because $\St$ is traceless, $\text{tr}(\St)=\St_{xx}+\St_{yy}+\St_{zz}=0$, it describes volume-preserving (incompressible and isochoric) transformations of fluid elements, and likewise for $\At$; where $\text{tr}\left\{\hdots\right\}$ is the trace operator. The difference is that all diagonal elements of $\At$ are identically zero, hence there are no components of $\At$ orthogonal to the fluid element, whereas $\text{tr}(\St)$ sums to zero, hence the orthogonal forces exist, but cancel out over the whole element. Furthermore, $\St$ is symmetric so $\St=\St^T$, and $\At$ is anti-symmetric, $\At = -\At^{T}$, which constrains the types of transformations that are allowable for each tensor.
            
            From this decomposition, it is clear that only $\Bt$ and $\St$, contribute to the fluid viscosity, since 
            \begin{align}\label{eq:A_vort_tensor}
                \At = \frac{1}{2}(\partial_jv_i - \partial_i v_j) =  -\frac{1}{2}\epsilon_{ijk}\omega_k= \frac{1}{2}\begin{pmatrix}
                 0 & -\omega_z & \omega_y   \\ 
                 \omega_z & 0 & -\omega_x \\ 
                 -\omega_y & \omega_x & 0
                \end{pmatrix},        
            \end{align}
            where $\bm{\omega} =\bnab \times \vecB{v}$ is the fluid vorticity, and $\epsilon_{ijk}$ is the Levi-Civita tensor. This is exactly the rate of rotation tensor, hence $\At$ cannot be a source of fluid viscosity (see our derivation in \aref{app:A_tensor}; note that $\At$ describes the rigid body rotation of a fluid element along its path -- formerly, antisymmetric matrices form a representation of the $\text{SO}(3)$ group, describing infinitesimal rotations of the fluid elements; \citealt{Choudhuri1998_plasma_text}). We may therefore write the viscous stress tensor as, 
            \begin{align}\label{eq:viscous+_stress_tensor}
                \bm{\sigma} =& 2\nus\left[\frac{1}{2}\left(\bnab\otimes\vecB{v} + [\bnab\otimes\vecB{v}]^T\right) -\frac{1}{3} (\bnab\cdot\vecB{v}) \It\right] \nonumber \\
                &\quad + \nub(\bnab\cdot\vecB{v}) \It,
            \end{align}
            where $\nus$ and $\nub$ are the coefficients of bulk viscosity and the shear viscosity, respectively\footnote{Note that based on this definition $\nub = \lambda + 2\nus/3$, where $\lambda$ is the second viscosity coefficient \citep[e.g.,][]{Emanuel1990_short_theory_review_bv,Shama2019_bulk_viscosity_in_dilute_gases}, but we prefer to parameterise our simulations in terms of $\nub$, which is equivalent to changing $\lambda$ for fixed $\nus$.}. Macroscopically, $\nub\Bt$ is the normal viscous stress associated with non-thermal-equilibrium changes to the volume of the fluid element. To get insight into the microscopic description of $\nub$, we turn to the standard procedure for calculating it \citep{Tiaza1942_stokes_relation,Cramer2012_estimates_for_bv,Shama2019_bulk_viscosity_in_dilute_gases}. Performing a Chapman–Enskog expansion of the Boltzmann equation about a local Maxwellian distribution function and only retaining terms that produce velocity divergence yields
            \begin{align}
                \nub = (\gamma - 1)^2\sum_i^N \frac{c_{V,i}\mu m_{\rm H}}{k_B}P_{\rm eq}\tau_i,
            \end{align}
            where $k_B$ is the Boltzmann constant, $m_{\rm H}$ the hydrogen mass, $\mu$ the mean molecular weight, $P_{\rm eq}$ is the equilibrium gas pressure, $\gamma$ is the ratio of specific heats at equilibrium, $c_{V,i}$ is the specific heat capacity of the $i^{\rm th}$ internal mode, $\tau_i$ is the relaxation time for each mode (where it assumed $\omega \ll \tau_i^{-1}$ and $\omega$ is the sound wave frequency), summed over $N$ rotational and vibrational degrees of freedom \citep{Tiaza1942_stokes_relation}. It is then easy to see that $\nub$ is associated with the summed time lags $\tau_i$ that each internal degree of freedom takes to relax to thermal equilibrium. Indeed, each degree of freedom may be decomposed into its own associated $\nub$, e.g.,
            \begin{align}
                \nub = \nub^r + \nub^v, 
            \end{align}
            for translational and vibrational degrees of freedom \citealt{Shang2020_bulk_viscosity_modes}. Following \citet{Cramer2012_estimates_for_bv}, assuming both degrees of freedom relax independently,
            \begin{align}
                \nub^r &= (\gamma - 1)^2\frac{f_r}{2}P_{\rm eq}\tau_r, \\
                \nub^v &= (\gamma - 1)^2\left(\frac{c_v \mu m_{\rm H}}{k_B} - \frac{1}{2}(f_r +3) \right)P_{\rm eq}\tau_v,
            \end{align}
            where $f_r$ are the rotational degrees of freedom, and $\tau_r$ and $\tau_v$ are the rotational and vibrational relaxation timescales, respectively. For monoatomic gases, there are no such degrees of freedom, so $\nub = 0$ (\citealt{stokes_1850} hypothesis), but for any other gas, this is no longer the case. Let us consider the cold ($T \approx 10-20\,\rm{K}$) interstellar molecular medium. It is composed primarily of H$_2$ $(f_r =2)$, followed by diatoms CO and O$_2$ \citep{Penteado2014_abundances,Krumholz2015}, among many other molecular species \citep{Herbst2009_organic_ISM}. Both for H$_2$ and CO, we estimate $\nub / \nus$ in \aref{app:bulk_viscosity_in_ISM}, utilising \citet{Parker1959_relaxation_of_diatomic_gases}'s theory for the relaxation for different degrees of freedom in diatomic gases. We estimate $\nub / \nus \approx 30$ for H$_2$ that has been shock-heated, due in part to the long rotational relaxation times, $\tau_v = \mathcal{O}(\rm{yrs})$, at molecular cloud densities, and $\nub / \nus \approx 1$ for CO in the ambient medium, due in part to the short rotational relaxation times ($\tau_v = \mathcal{O}(\rm{months})$. We note that both H$_2$ and CO do not have excited vibrational degrees of freedom until much larger $T \gtrsim 3\times 10^3\,\rm{K}$, so the bulk viscosity is almost completely controlled by the relaxation of the internal rotational degrees of freedom).
        
            As \citet{Sharma2023_bulk_viscosity} suggest in their review of bulk viscosity in fluid dynamics, the theory for bulk viscosity has largely been neglected in the literature. Here we summarise the few studies that are relevant to bulk viscosity in the context of fluid plasma turbulence. \citet{Chen2019_bulk_visc_homogenous_turb} studied forced homogeneous isotropic (and hydrodynamical) turbulence (HIT) for moderate sonic Mach numbers $\M \leq 0.6$ and Taylor miscroscale Reynolds numbers $\Res \approx 100$. Through the kinetic energy spectrum, they showed that for large $\nub/\nus$ ratios the turbulence almost returned to an incompressible state, i.e., the bulk viscosity acted primarily to decay the compressible modes. \citet{pan_johnsen_2017_bulk_viscosity_decaying_turb} studied the effect of bulk viscosity in decaying hydrodynamical turbulence, over a the same range of $\nub/\nus$ that we do in our study, finding that high bulk viscosities enhanced the total decay rate of the kinetic energy by both suppressing the compressible modes in magnitude, and through the compressible mode-pressure and the compressible-incompressible mode interactions. However, they found that the incompressible modes were reasonably insensitive to bulk viscosity. \citet{Touber2019_2d_turbulence} studied the eigenmodes of the Navier-Stokes equation with the addition of bulk viscosity, showing that in the Stokes (viscous) regime, significant bulk viscosity is able to enhance the small-scale enstrophy production in two-dimensional turbulence via the $\bm{\omega}(\bnab\cdot\vecB{v})$ coupling. However, no studies exist in the MHD turbulence regime, nor in the presence of the turbulent dynamo, but if there is a strong coupling of the bulk viscosity at $\ell_{\nus}$ (undoubtedly transitioning into the Stokes regime), then we should expect there to be an effect on the turbulent dynamo, since $\ell_{\nus}$ is responsible for the magnetic field growth. To situate the bulk viscosity into a regular SSD formalism, we now define the dimensionless numbers that we can use to parameterise the additional bulk viscous fluxes in the momentum equation, and thus, our simulations.

    \subsection{Dimensionless plasma numbers}\label{sec:derive_plasma_Reynolds}

        \subsubsection{Viscous plasma numbers}
        
        The bulk viscosity, as in \autoref{eq:viscous+_stress_tensor}, can be added simply as an additional viscous flux in the momentum equation of the MHD fluid model, which we write for an isothermal fluid,
        \begin{align}\label{eqn:momentum_eqn}
            \frac{\partial\rho\vecB{v}}{\partial t} &+ \bnab\cdot\Bigg[ \rho \vecB{v}\otimes\vecB{v} + \frac{1}{4\pi}\vecB{b}\otimes\vecB{b} + \left(c_s^2 \rho + \frac{b^2}{8\pi}\right)\It \nonumber\\
            &- 2\nus\rho\Big\{\frac{1}{2}\left(\bnab\otimes\vecB{v} + [\bnab\otimes\vecB{v}]^T\right) - \frac{1}{3}\left(\bnab\cdot\vecB{v}\right)\It \Big\} \nonumber\\
            &-\nub\rho\left(\bnab \cdot \vecB{v}\right)\It \Bigg] = \rho \vecB{S},
        \end{align}
        where $\rho$ is the mass density, and $\rho\vecB{S}$ is an arbitrary momentum source. One way of understanding the magnitude of each term in the momentum equation is to make it dimensionless to see how each term scales with dimensionless parameters. In light of this, we scale all length scales by $L$, all velocities by $V$, all times by $T = L/V$, all magnetic fields by $B$, and all densities by $\rho_0$, such that, e.g., $\vh = \vecB{v}/V$ (dimensionless quantities are denoted with bar notation). As a simple example, the time-derivative term of the momentum becomes,
        \begin{align}
            \frac{\partial \rho \vecB{v}}{\partial t} \rightarrow \frac{\rhoo V}{T}\frac{\partial \rhoh \vh}{\partial \th} = \frac{\rhoo V^2}{L} \frac{\partial \rhoh \vh}{\partial \th}.
        \end{align}
        Let us non-dimensionalise the momentum equation, rearranging the magnetic tension and pressure,
        \begin{align}
            \frac{\rhoo V^2}{L}\frac{\partial \rhoh\vh}{\partial \th} &+ \frac{1}{L}\bar{\bnab}\cdot\Bigg[ V^2\rhoo \rhoh \vh\otimes\vh - \frac{B^2}{4\pi}\left(\bh\otimes\bh + \frac{\bar{b}^2}{2}\It\right) \nonumber \\
            &+ c_s^2\rhoo \rhoh\It - \frac{2\nus\rhoo \rhoh V}{L}\left[\frac{1}{2}\left(\bar{\bnab}\otimes\vh + [\bar{\bnab}\otimes\vh]^T\right) \right. \nonumber \\
            &\left. -\frac{1}{3}\left(\bar{\bnab}\cdot\vh\right)\It \right] - \frac{\nub\rhoo\rhoh V}{L} \left(\bar{\bnab} \cdot \bar{\v}\right)\It \Bigg] = \frac{\rhoo V^2}{L} \rhoh \vecB{\bar{S}}. 
        \end{align}
        We can now divide the whole equation by $\rhoo V^2 / L$, appearing as the coefficient for the time derivative of the momentum, 
        \begin{align}
            \frac{\partial \rhoh\vh}{\partial \th} &+ \bar{\bnab}\cdot \Bigg[ \rhoh \vh\otimes\vh -\frac{B^2}{4\pi\rho_0 V^2}\left(\bh\otimes\bh - \frac{\bar{b}^2}{2}\It\right) + \frac{c_s^2}{V^2}\rhoh\It \nonumber \\
            &- 2\frac{\nus\rhoh}{VL}\left[\frac{1}{2}\left(\bar{\bnab}\otimes\vh + [\bar{\bnab}\otimes\vh]^T\right) - \frac{1}{3}\left(\bar{\bnab}\cdot\vh\right)\It \right] \nonumber \\
            &- \frac{\nub\rhoh}{VL}\left(\bar{\bnab} \cdot \bar{\v}\right)\It \Bigg] = \rhoh \vecB{\bar{S}}. 
        \end{align}
        Now we can make a number of observations. First, for the Lorentz force, $V_A^2 = B^2 / (4\pi\rhoo)$, and $V_A^2 / V^2 = 1/\MA{2}$ is the dimensionless Alfv\'en Mach number. Likewise, for the thermal pressure, $c_s^2 / V^2 = 1 / \M^2$ is the sonic Mach number. For the strain rate dissipation model $\Res \sim VL/\nus$ and for the bulk dissipation model $\Reb \sim VL/\nub$. Hence, we can write the dimensionless momentum equation as,
        \begin{align}
            \frac{\partial \rhoh\vh}{\partial \th} &+ \bar{\bnab}\cdot\Bigg[  \rhoh \vh\otimes\vh - \frac{1}{\MA{2}}\left(\bh\otimes\bh - \frac{\bar{b}^2}{2}\It\right) + \frac{\rhoh}{\M^2}\It \nonumber \\
            &- \frac{2\rhoh}{\Res}  \Bigg(\frac{1}{2}\Big[\bar{\bnab}\otimes\vh  + \left(\bar{\bnab}\otimes\vh\right)^T\Big] - \frac{1}{3}\left(\bar{\bnab}\cdot\vh\right)\It \Bigg) \nonumber \\
            & - \frac{\rhoh}{\Reb} \left(\bar{\bnab} \cdot \bar{\v}\right)\It \Bigg] = \rhoh \vecB{\bar{S}},
        \end{align}
        which demonstrates how each of the terms in the momentum equation are related to each of the dimensionless constants\footnote{Previous studies have adopted $\Res = 2\pi\Exp{v^2}{\V}^{1/2}/(k_0\nus)$, where $k_0 = \ell_0/(2\pi)$ is the turbulent driving scale in wavenumber space. This leads to a boost in Reynolds number by a factor of $2\pi$, which seems confounding, since all we have done is substituted $\ell_0$ for $k_0$ in the definition. However, note that real-space and wavenumber space statistics are linked by integrals, e.g.,
        \begin{equation}
            \Exp{v^2}{\V}^{1/2} = \int_{\forall k}\d{\vecB{k}}\,|\vecB{v}(\vecB{k})\vecB{v}^*(\vecB{k})|,
        \end{equation}
        Parsevel's theorem, where $|\vecB{v}(\vecB{k})\vecB{v}^*(\vecB{k})|$ is the power spectrum of $\vecB{v}$, and hence the velocity dispersion on $k_0$ is actually only a fraction of $\Exp{v^2}{\V}^{1/2}$, $\Exp{v^2}{k_0 + \d{k}}^{1/2} = \int_{k_0}^{k_0+\d{k}}\d{\vecB{k}}\,|\vecB{v}(\vecB{k})\vecB{v}^*(\vecB{k})|$. This means, to use a wavenumber definition of $\Res$, one has to also reduce $\Exp{v^2}{\V}^{1/2}$ such that one is associating the correct velocities with the correct wavemodes. This is unnecessarily complicated, so we always chose to use real-space Reynolds number definitions, which come directly from the momentum equation, as we have shown.} we use to parameterise our \textsc{flash} simulations, including the new definition of $\Reb$.
        
        More precisely than simple scaling arguments, the shear viscosity is controlled by changing the shear Reynolds number,
        \begin{align}\label{eq:Reshear_defn}
            \Res = \frac{|\bnab \cdot (\rho\vecB{v} \otimes \vecB{v})|}{|2\nus\bnab\cdot(\rho\mathbb{S})|} \sim \frac{\Exp{\v^2}{\V}^{1/2}\lo}{\nus},
        \end{align}
        and the bulk viscosity via the bulk Reynolds number,
        \begin{align}\label{eq:Rebulk_defn}
            \Reb = \frac{|\bnab \cdot (\rho\vecB{v} \otimes \vecB{v})|}{|\nub\bnab\cdot(\rho\mathbb{B})|} \sim \frac{\Exp{\v^2}{\V}^{1/2}\lo}{\nub},
        \end{align}
        which in ratio (keeping $\nus$ in the numerator, consistent with the definition of the magnetic Prandtl number) defines our viscous Prandtl number,
        \begin{align}\label{eq:Pnu_defn}
            \Pnu = \frac{\nus}{\nub} = \frac{\Reb}{\Res} \sim \frac{t_{\nub}}{t_{\nus}}.
        \end{align}
        With this definition, in the $\Pnu \leq 1$ regime, bulk viscosity dominates over the shear viscosity ($t_{\nub} \leq t_{\nus}$), and we therefore might expect that the compressible modes decay on larger scales than the incompressible modes, and vice versa for the $\Pnu > 1$ regime ($t_{\nub} > t_{\nus}$). We go through the natural implications of these two regimes on the energy spectra in the following \autoref{sec:dynamo_theory}. But before that, we turn our attention to the dimensionless parameters for the magnetic field.

    \begin{figure*}
        \centering
        \includegraphics{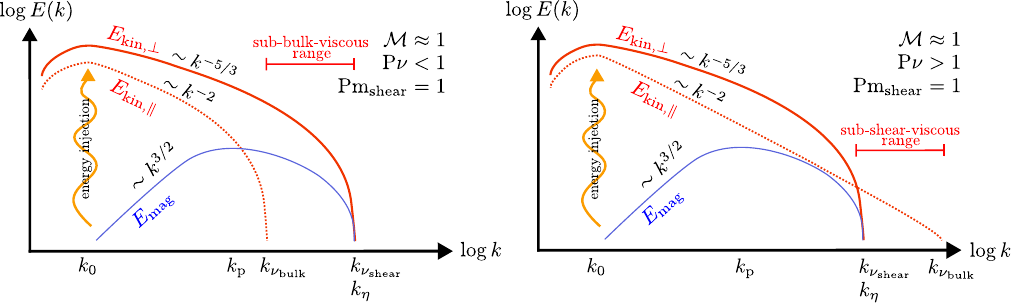}
        \caption{Schematics for the energy spectrum in the kinematic stage of the trans-sonic ($\M \approx 1$), turbulent dynamo with $\Pms = 1$ (\autoref{eq:Pm_shear_defn}), varying $\Pnu$ (\autoref{eq:Pnu_defn}), including the incompressible ($\ekinperp$; solid, red), compressible ($\ekinpar$; dotted, red) kinetic and magnetic spectra ($\emag$; blue). We annotate the energy injection scale $k_0$, peak magnetic energy scale $k_{\rm p}$, dissipation scale for the compressible modes $k_{\nub}$, dissipation scale for the incompressible modes $k_{\nus}$ and dissipation scale for the magnetic field $k_{\eta}$ on the $\log k$ axis. Note that for both the left and right panel, we have adopted a key result from \citet{Kriel2025_supersonic_scales}, who showed that $k_{\rm p} < k_{\nus}$ for the compressible turbulent dynamo, as opposed to the incompressible case where $k_{\rm p} \sim k_{\eta}$ \citep{Schekochihin2002_large_Pm_dynamos,Schekochihin2004_dynamo}. \textbf{Left:} The $\Pnu < 1$ case, where $k_{\nus} > k_{\nub}$, decaying the compressible modes faster than the incompressible modes and creating a sub-bulk-viscous range of scales $k_{\nub}< k < k_{\nus}$, where $\ekinperp(k) \gg \ekinpar(k)$. \textbf{Right:} The $\Pnu > 1$ case, where $k_{\nub} > k_{\nus}$, creating a sub-shear-viscous range in kinetic energies $k_{\nus}< k < k_{\nub}$, where $\ekinpar(k) \gg \ekinperp(k)$, and for $\Pms = 1$, $\ekinpar(k) \gg \emag(k)$. In \autoref{fig:decomp_vel_spect} we show how our simulations approximately follow these scales (at least over a small range of $k$ that may be considered as modes within the cascade). We note that \citet{Wang2017_psuedosound_spectrum} and \citet{Yoshiki2023_compressible_turbulence_spectrum} show that at low enough $\M \lesssim 0.4$, $\ekinpar(k)$ transitions into a $\propto k^{-3}$ spectrum, dominated by pseudo-acoustic wave turbulence, which we do not show in this schematic. Further, because $\M \approx 1$, we have not included the sonic scale and hence dichotomy between supersonic and subsonic $\ekinperp(k)$ cascades \citep{Federrath2021,Cernetic2024_sonic_scale_GPUs,Beattie2025_10k}.}
        \label{fig:spectra}
    \end{figure*}

        \subsubsection{Resistive plasma numbers}
        Performing the same analysis as above for fluid the induction equation with Ohmic resistivity gives,
        \begin{align}
            \frac{\partial \bar{\vecB{b}}}{\partial \bar{t}} - \bar{\bnab} \times (\bar{\vecB{v}}\times\bar{\vecB{b}})= \frac{1}{\Rm}\bar{\bnab}^2\bar{\vecB{b}},
        \end{align}
        where 
        \begin{align}\label{eq:Rm_defn}
            \Rm = \frac{|\bnab \times (\vecB{v}\times\vecB{b})|}{|\eta\bnab^2\vecB{b}|} \sim \frac{\Exp{v^2}{\V}^{1/2}\ell_0}{\eta},
        \end{align}
        is the magnetic Reynolds number for the Ohmic resistivity and $\eta$ is the Ohmic resistivity coefficient. Unlike $\Res$ and $\ell_{\nus}$, the dimensionless number $\Rm$ alone is not sufficient to define the resistive scale in the turbulence \citep{Schekochihin2004_dynamo,Galishnikova2022_saturation_and_tearing,Kriel2022_kinematic_dynamo_scales,Brandenburg2023_dissipative_structures,Kriel2025_supersonic_scales}. Instead, by balancing the stretching rate at the shear viscous-scale, $v_{\ell_{\nus}}/\ell_{\nus}$ with the resistive dissipation rate, $\eta/\ell_{\eta}^2$, $v_{\ell_{\nus}}/\ell_{\nus} \sim \eta/\ell_{\eta}^2$, we can rearrange to find,
        \begin{align}\label{eq:Pm_relation}
            \frac{\ell_{\nus}}{\ell_{\eta}} \sim \Pms^{1/2},
        \end{align}
        where 
        \begin{align}\label{eq:Pm_shear_defn}
            \Pms = \frac{\nus}{\eta} = \frac{\Rm}{\Res},
        \end{align}
        is the shear magnetic Prandtl number. \autoref{eq:Pm_relation} is independent from the effects of compressibility \citep{Kriel2025_supersonic_scales}. $\Pms$ parameterises the range of scales, $\ell_{\nus} > \ell > \ell_{\eta}$ (the classical sub-viscous range; \citealt{Schekochihin2002_saturation_evolution,Schekochihin2004_dynamo}). We can furthermore define a bulk magnetic Prandtl number,
        \begin{align}\label{eq:Pm_bulk_defn}
            \Pmb = \frac{\nub}{\eta} = \frac{\Rm}{\Reb},
        \end{align}
        and it follows that,
        \begin{align}
            \Pnu = \frac{\Pms}{\Pmb} = \frac{\Reb}{\Res},
        \end{align}
        which relates all three of the Prandtl numbers that we define in this study.  $\Pmb$, like $\Pms$ may be related to a scale separation $\ell_{\nub}/\ell_{\eta} \sim \Pmb^\alpha$, if the $v_{\ell_{\nub}} \sim \ell_{\nub}/t_{\ell_{\nub}}$ modes are responsible changing the resistive properties of the plasma, but because \citet{Kriel2025_supersonic_scales} found that \autoref{eq:Pm_relation} is universal for high and low $\M$, this is unlikely. We will constrain ourselves to a subset of the total parameter range in this study, keeping $\Res$ and $\Pms$ fixed, with $\Pms = 1$, but changing $\Pnu$, which means $\Pmb$ is solely determined by $\Pnu$, $\Pmb = \Pnu^{-1}$. This is because we are solely focusing on the role of bulk viscosity in the dynamo, and keeping $\Pms$ and $\Res$ fixed makes for a simple and informative experiment.
        
\section{Compressibility in the turbulent dynamo}\label{sec:dynamo_theory}
    In this section we: (1) construct a general picture for what a compressible small-scale dynamo with bulk viscosity looks like in $k$ space, using our parameterisations, $\Pnu$, $\Pms$ and $\M$; and (2), highlight the theoretical aspects of the effects from compressible modes on the growth rate and saturation of the turbulent dynamo. We adopt an approach that focuses on the magnetic energy equation, as done before in \citet{Schekochihin2004_dynamo}, \citet{Seta2021_saturation_supersonic_dynamo} and \citet{Sur2023_dynamo}, with some important differences\footnote{\label{fn:tension}\citet{Schekochihin2002} considered only the incompressible magnetic energy equation, and both \citet{Seta2021_saturation_supersonic_dynamo} and \citet{Sur2023_dynamo} did not decompose the velocity gradient tensor to completely separate the incompressible and compressible tensors, $\At$, $\St$ and $\Bt$, respectively, whilst trying to directly measure the effect of incompressible (stretching) and compressible effects in the dynamo. They only consider the effect of projecting the symmetric component of the velocity gradient tensor -- the rate of strain tensor -- the first term in \autoref{eq:growth_rate_decomp}, onto the magnetic field, critically neglecting $\Bt$. Furthermore, \citet{Sur2023_dynamo} do not make their stretching tensor traceless, just using $(\bnab\otimes\vecB{v} + [\bnab\otimes\vecB{v}]^T)/2$, which no longer makes $\St$ volume-preserving, and hence is contaminated by isotropic compression. Neglecting the additional compression within $\bnab\otimes\vecB{v}$ gives rise to overestimating the impact of compressions by a factor of $3$ in \citet{Seta2021_saturation_supersonic_dynamo} and $6$ in \citet{Sur2023_dynamo}, which may significantly modify the $\first$ moment interpretations and conclusions presented in \citet{Sur2023_dynamo}. \citet{Steinwandel2023_magnetised_ICM} also neglect the $\Bt$ contribution from within their $\hat{\vecB{b}}\otimes\hat{\vecB{b}}:\bnab\otimes\vecB{v}$ term in the magnetic field energy equation for their magnetised intracluster medium simulations, which impacts their decomposition in a similar, negligible way as \citet{Seta2021_saturation_supersonic_dynamo}.}. Let us begin with schematics of the dynamo energy spectrum.

    \subsection{Setting the scene for the compressible dynamo}

    We schematically represent the $\Pnu <1$ and $\Pnu > 1$ regimes in \autoref{fig:spectra}, with $\Pnu < 1$ in the left panel, and $\Pnu > 1$ in the right panel, respectively. Both panels correspond to $\Pms = 1 \implies \ell_{\nus} \sim \ell_{\eta}$, and $\M > 1$ in both panels. We further annotate the fundamental scales for the dynamo problem: the integral or energy injection scale $k_0$, the peak magnetic energy scale $k_{\rm p}$, following the results from supersonic dynamo experiments in \citet{Kriel2025_supersonic_scales}, the viscous dissipation scales, $k_{\nus}$ and $k_{\nub}$, and the magnetic energy dissipation scale $k_{\eta}$. We also annotate a \citet{Kolmogorov1941}-type energy spectrum $\sim k^{-5/3}$ for the incompressible mode spectrum $\ekinperp(k)$, giving rise to $\ell_0 \sim \Re^{3/4}\ell_{\nus}$, and a \citet{Burgers1948}-type spectrum $\sim k^{-2}$ for the compressible mode spectrum $\ekinpar(k)$, which gives $\ell_0 \sim \Re^{2/3}\ell_{\nub}$. We acknowledge that the $\ekinperp(k) \sim k^{-2}$ spectrum is not universally measured in compressible turbulence. For example, as demonstrated in \citet{Wang2017_psuedosound_spectrum} and \citet{Yoshiki2023_compressible_turbulence_spectrum}, at low-$\M$, $\ekinpar(k) \sim k^{-3}$ if the compressible modes are enslaved by incompressible modes (pseudo-acoustic wave turbulence), especially when background spatial inhomogeneities are weak, as demonstrated in the context of the solar wind by \citet{Bhattacharjee_1998}. \citet{Wang2017_psuedosound_spectrum} derived a critical $\M \approx 0.4$ for the $k^{-3}$ regime. \citet{Wang2013_cascade_of_kin_energy_2_long}, \citet{Touber2019_2d_turbulence}, and \citet{Beattie2025_10k} find a $\ekinpar(k) \sim k^{-2}$ spectrum at higher-$\M$, and even in trans-sonic supernova-driven turbulence \citep{Beattie2025_SNe_driven}, which they motivate through a \citet{Burgers1948}-type, shock-dominated spectrum. Many astrophysical plasmas that host dynamos are relatively supersonic, like the cold phase ISM ($\M \approx 10$;  \citealt{Krumholz2015,Ferriere2001}), the warm ionized ISM ($\M \approx 1-2$; \citealt{Gaesnsler_2011_trans_ISM}), and black-hole accretion disks with their surrounding plasma environment ($\M \approx 10 - 100$; \citealt{Hopkins2024_FIRE_zoom_in}), hence we adopt the $k^{-2}$ spectrum in our schematic.
    
    The simulations presented in this study roughly conform to the cascade exponents presented in the \autoref{fig:spectra} schematic, which we show later in \autoref{sec:decomposed_spectra}. However, we purposefully truncate the cascade to study the sub-viscous-scale dynamics, important for bulk viscous dynamics. See recent work by \citet{Beattie2025_10k} showing $\ekinperp(k)$ and $\ekinpar(k)$ spectra at much higher Reynolds numbers ($\Res \gtrsim 10^6$) and $\M \approx 4$ supersonic MHD simulations. For the magnetic field, we annotate a \citet{Kazantsev1968}-type spectrum $\sim k^{3/2}$. For $\Pnu < 1$ (left panel), where the compressible modes decay on shorter timescales than the incompressible modes, a new sub-bulk-viscous range of scales 
    \begin{align}\label{eq:sub-bulk-viscous_range}
     \ell_0 > \ell_{\nub} > \ell > \ell_{\nus} & & \text{(sub-bulk-viscous range)}      
    \end{align}
    emerges, where $\ekinperp(k) \gg \ekinpar(k)$. This regime is only achievable with explicit control over $k_{\nub}$, which we expect may lead to the small-scales becoming fundamentally incompressible. Conversely, for $\Pnu > 1$ (right panel), a sub-shear-viscous range of scales emerges in the kinetic energy, 
    \begin{align}\label{eq:sub-shear-bulk-viscous_range}
        \ell_0 >  \ell_{\nus} > \ell > \ell_{\nub} & & \text{(sub-shear-viscous range)} 
    \end{align}
    where compressible modes are dominant, $\ekinpar(k) \gg \ekinperp(k)$. For $\Pm = 1$, the compressible kinetic energy on these scales will also be beyond the magnetic energy, $\ekinpar(k) \gg \emag(k)$, making for an interesting range of scales that is naively fundamentally hydrodynamical, and dominated by compressions. If \autoref{fig:spectra} is representative of how bulk viscosity works, then the $\Pnu > 1$ regime ought to be the regime that all $\M > 1$ simulations are in without any explicit bulk viscosity. It is not obvious if this is desirable, but what is desirable will clearly depend strongly on the $\Pnu$ regime that exists in the turbulent plasma of interest. We aim to test the predictions outlined in this schematic in the remainder of the study.

    \begin{figure*}
            \centering
            \includegraphics[width=\linewidth]{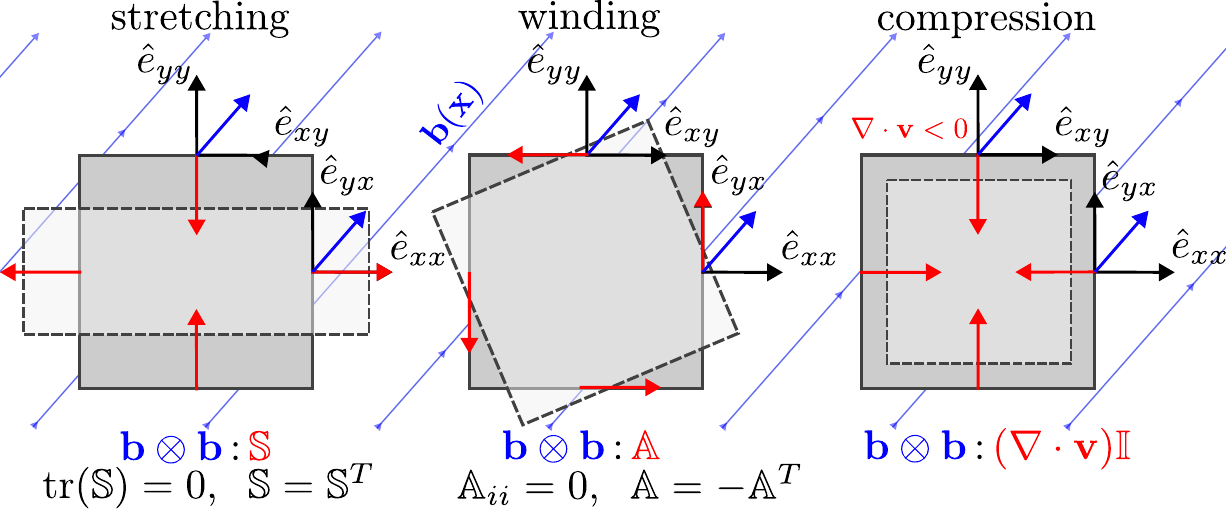}
            \caption{The three growth terms in \autoref{eq:growth_rate_decomp} visualised as orthogonal operations on the three 2D fluid elements (grey) that are evolving from the elements with solid lines to the elements with dashed lines. Key of the properties of the tensor operations are listed below the fluid elements. Each panel represents a geometrical projection between the blue (magnetic) vectors and the red vectors, which corresponds to the traceless rate of strain tensor $\St$ (\autoref{eq:S_v_grad_tensor}; leftmost panel), the traceless rate of rotation tensor $\At$ (\autoref{eq:A_v_grad_tensor}; central panel) and the rate of expansion tensor $\Bt$ (\autoref{eq:B_v_grad_tensor}; rightmost panel). The $\St$ and $\At$ tensors enforce volume-preserving, incompressible transformations of the fluid element, that is $\text{tr}(\St) = \text{tr}(\At)=0$ (orthogonal forces sum to zero). Note that we have represented only one possible stretching mode for $\St$, and $\St$ also contains all shearing modes. $\At$ only has components in the plane of the fluid element, with all orthogonal components being exactly zero, and all off-diagonal components being equal to the components of the fluid vorticity (\autoref{eq:A_vort_tensor}). Because $\At = - \At^T$, $\At$ corresponds to shear-free rotations of the fluid element. The final $(\bnab\cdot\vecB{v})\It$ tensor operation only has components in the orthogonal directions to the fluid element face. Each force must always point either away or into the fluid faces, because all components share the sign of $\bnab\cdot\vecB{v}$. Hence $(\bnab\cdot\vecB{v})\It$ is either a isotropic contraction, $\bnab\cdot\vecB{v}< 0$, as shown in the schematic, or dilation $\bnab\cdot\vecB{v} > 0$, changing the volume of the fluid element.}
            \label{fig:tensors}
    \end{figure*}

    Moreover, we further note that in the literature, previous studies have used the ratio $\nub/\nus$ to parameterise the different plasma regimes, \citep[e.g.,][]{Chen2019_bulk_visc_homogenous_turb}, but we prefer our definition\footnote{Note that because $\nub = (2/3)\nus + \lambda$, $\Pnu = 3\nus/(2\nus + 3\lambda)$, hence when $\Pnu = 1$, $\nus/\lambda = 3$, as expected from \autoref{eq:B_v_grad_tensor}. What this means is that at $\Pnu = 1/3$ the dissipation timescale of the bulk and shear viscosity is the same, which we will later find is consistent with our simulation results. \label{footnote:Pnudefn}}, which (1) explicitly defines the extra viscosity through a bulk viscosity Reynolds number, and (2) puts the bulk viscosity in a framework similar to the resistivity and magnetic Prandtl number.

    \subsection{Turbulent dynamo growth}\label{sec:growth}
    The $\first$ moment of the $\emag$ equation can be found by taking the scalar product of the magnetic field and induction equation and then averaging over space \citep[similiar to][but retaining the full velocity gradient tensor and the effects of compressibility; see \aref{app:first_moment_mag_eq} for our derivation]{Schekochihin2002,Seta2021_saturation_supersonic_dynamo}. Considering only Ohmic dissipation as a means of magnetic diffusion, it is,
    \begin{align}\label{eq:magnetic_energy_m1}
        \frac{\d{\Exp{\emag}{\V}}}{\d{t}} &= 2\gamma\emagb - 2\eta k_{\rm u}^2\emagb,\\
        \gamma &= \frac{\Exp{ (\vecB{b}\otimes\vecB{b}): [(\bnab\otimes\vecB{v}) - \frac{1}{2}(\bnab\cdot\vecB{v})\It]}{\V}}{\Exp{\b^2}{\V}}, \\
        k_{\rm u}^2 &= \frac{\Exp{ (\bnab\otimes\vecB{b}): (\bnab\otimes\vecB{b})}{\V}}{\Exp{\b^2}{\V}},
    \end{align}
    noting that for the incompressible, kinematic case, 
    \begin{align}
        \gamma t_0 \sim \Exp{\hat{(\vecB{b}}\otimes\hat{\vecB{b}}):(\bnab\otimes\vecB{v})}{\V}t_0 \sim t_0 / t_{\nus} \sim \Res^{1/2},
    \end{align}
    since $\bnab\otimes\vecB{v} \sim \St \sim t_{\nus}^{-1}$, as expected and discussed in \autoref{sec:intro} for \citet{Kolmogorov1941}-type turbulence. We know from \autoref{sec:vel_grad_tensor} that we can decompose $\bnab\otimes\vecB{v}$ into tensor components for the different modes, $\St$ and $\At$ for the incompressible modes, and $\Bt$ for the compressible modes. Hence we may write the $\gamma$ equation as
    \begin{align}
        \gamma &= \overbrace{\frac{\Exp{ (\vecB{b}\otimes\vecB{b}): (\St + \At) }{\V}}{\Exp{\b^2}{\V}}}^{\text{incompressible}} + \underbrace{\frac{\Exp{ (\vecB{b}\otimes\vecB{b}): [\Bt - \frac{1}{2}(\bnab\cdot\vecB{v})\It] }{\V}}{\Exp{\b^2}{\V}}}_{\text{compressible}}.
    \end{align}
    Splitting the incompressible term and using $\Bt$ we find that the growth rate can be understood as the combination of fundamental fluid element deformations,
    \begin{align}\label{eq:growth_rate_decomp}
        \gamma &= \overbrace{\frac{\Exp{ (\vecB{b}\otimes\vecB{b}): \St }{\V}}{\Exp{\b^2}{\V}}}^{\text{stretching}} +  \underbrace{\frac{\Exp{ (\vecB{b}\otimes\vecB{b}) : \At }{\V}}{\Exp{\b^2}{\V}}}_{\text{winding}} -\overbrace{\frac{\Exp{(\vecB{b}\otimes\vecB{b}) : (\bnab\cdot\vecB{v})\It }{\V}}{6\Exp{\b^2}{\V}}}^{\text{compression}}.
    \end{align}
    The first term in \autoref{eq:growth_rate_decomp} is the projection of the magnetic field onto the velocity gradient field that stretches it. One may expect that this term must dominate the growth of the magnetic field, since field-line-stretching is well-known to be the key growth mechanism in the turbulent dynamo \citep{Moffatt1969_fast_dynamo_growth,Zeldovich1984_fast_dynamo_growth,Schekochihin2004_dynamo,Seta2021_saturation_supersonic_dynamo,Kriel2022_kinematic_dynamo_scales,StOnge2020S_weakly_collisional_dynamos}. Next is the winding term, which is the projection of the magnetic field onto the rate of rotation tensor, \autoref{eq:A_vort_tensor}. Consider $\mathbb{M}=\vecB{b}\otimes\vecB{b}$, and note that $\mathbb{M} = \mathbb{M}^T$ is symmetric $(b_ib_j = b_jb_i)$. Then it follows from the definition of $\At$,
    \begin{align} \label{eq:no_winding_growth}
        \begin{split}
        \mathbb{M}:\At &= \text{tr}\left\{\mathbb{M}\At^T\right\} = -\text{tr}\left\{\mathbb{M}\At\right\}, \\
        &= -\frac{1}{2}\text{tr}\left\{\mathbb{M}(\bnab\otimes\vecB{v})\right\} + \frac{1}{2}\text{tr}\left\{\mathbb{M}(\bnab\otimes\vecB{v})\right\}\\
        &= 0,
            \end{split}
    \end{align}
    which means we expect $\Exp{(\vecB{b}\otimes\vecB{b}):\At}{\V}/\Exp{b^2}{\V} \equiv 0$. More broadly, this means turbulent magnetic fields are prevented from undergoing any kind of local winding to enhance or decay the field, quite unlike large-scale toroidal fields, which can undergo linear growth via winding \citep{Aguilera2023_winding}. The final term is the compression, which is the projection of the magnetic field onto the dilatational components of the fluid element. This term has a negative sign, with the total sign contribution of the term solely determined by the divergence, because 
    \begin{align}
        \Exp{(\vecB{b}\otimes\vecB{b}):(\bnab\cdot\vecB{v})\It }{\V} = \text{tr}\left\{(\vecB{b}\otimes\vecB{b})(\bnab\cdot\vecB{v})\right\} = b^2(\bnab\cdot\vecB{v}).
    \end{align}
    Hence, for $\bnab\cdot\vecB{v}< 0$, e.g., shocked fluid, the compressions enhance\footnote{It may look like this operator is isotropic, which it is. However, note that for compressions parallel to field lines, the growth rate from the compressions alone become $\gamma = -b^2 \partial_z v_z/\Exp{b^2}{\V}$, where we use $\hat{\vecB{z}}$ as our coordinate along the local field, which exactly cancels with the stretching contribution. Hence, it is only compressions perpendicular to field lines that contribute to the growth, as expected from standard flux-freezing considerations.} the growth rate by a local factor $(1/6)|\bnab\cdot\vecB{v}|$. However, for rarefactions this term reduces the growth rate by the same factor. Therefore, one can ask, on average, in a volume, is the fluid dilated $\Exp{\bnab\cdot\vecB{v}}{\V} > 0$ or compressed $\Exp{\bnab\cdot\vecB{v}}{\V} < 0$? It is well known that for supersonic compressible turbulence, most of the fluid in the volume is $\Exp{\bnab\cdot\vecB{v}}{\V} > 0$, with only a very small amount of the fluid contained with the $\bnab\cdot\vecB{v} < 0$ structures (\citealt{Passot1998,Schmidt2008,Hopkins2013,Squire2017,Mocz2019,Beattie2022_spdf} and also see the $\bnab\cdot\vecB{v}$ probability density function in \citealt{Chen2019_bulk_visc_homogenous_turb} and in Appendix A1 and A2 in \citealt{Sur2023_dynamo}). Hence as the compressions become stronger (for e.g., larger $\M$), this term, on average, ought to act to reduce the growth rate of the volume-averaged magnetic energy, which has been shown in numerical simulations \citep{Federrath2011_mach_dynamo}, however, not from evolving the bulk viscosity, but rather changing $\M$. 

    In the nonlinear induction stage \autoref{eq:linear_growth}, $\emagb \sim \ekin(\ell_{\rm eq})$, on energy equipartition scale $\ell_{\rm eq}$ \citep{StOnge2020S_weakly_collisional_dynamos,Galishnikova2022_saturation_and_tearing,Beattie2023_GorD_P1}. Hence the growth term in \autoref{eq:magnetic_energy_m1}, $2\gamma\emagb$, becomes 
    \begin{align}\label{eq:linear_growth_b_field_eq}
        2\gamma t_0\emagb \sim \frac{\emagb\ekin^{1/2}(\ell_{\rm eq})}{\ell_{\rm eq}}t_0 \sim \frac{\ekin^{3/2}(\ell_{\rm eq})}{\ell_{\rm eq}}t_0,
    \end{align}
    which is approximately constant, giving rise to linear growth as described in \autoref{sec:intro}. The constant $\ekin^{3/2}(\ell_{\rm eq})/\ell_{\rm eq} \sim v^3/\ell$ is the energy flux rate originating from the hydrodynamical cascade \citep{Schekochihin2002_saturation_evolution,Xu2016_dynamo,StOnge2020S_weakly_collisional_dynamos}. Hence the nonlinear dynamo allows the dynamo growth rate to become independent of the viscous- and sub-viscous-scale dynamics, and relates the magnetic flux generation directly to the turbulence itself. 

    \subsection{Dynamo saturation}\label{sec:saturation}
    Let us explore \autoref{eq:magnetic_energy_m1} in the saturated stage. Setting $\partial_t \emagb = 0$ for statistically stationary saturation,
    \begin{align}\label{eq:saturation_compressible}
        \Exp{ (\vecB{b}\otimes\vecB{b}) :\St}{\V} - &\frac{1}{6}\Exp{(\vecB{b}\otimes\vecB{b}):(\bnab\cdot\vecB{v})\It}{\V} = \nonumber\\ 
        &\quad\quad\quad\quad\quad\quad
        \eta\Exp{(\bnab\otimes\vecB{b}):(\bnab\otimes\vecB{b})}{\V},
    \end{align}
    previously derived in \citet{Yousef2007_exact_scaling_laws} for incompressible turbulence. Very simply put, when the growth terms balance with the dissipation, the dynamo saturates. However, these may balance in at least three different ways: (1) the growth terms may be suppressed such that they eventually find their way to $\eta\Exp{(\bnab\otimes\vecB{b}):(\bnab\otimes\vecB{b})}{\V}$, or (2) $\eta\Exp{(\bnab\otimes\vecB{b}):(\bnab\otimes\vecB{b})}{\V}$ may grow, eventually balancing with the growth terms, or (3) there is a mixture of both. For $\M \approx 0.2$ turbulence, \citet{Sur2023_dynamo} showed that $\eta\Exp{(\bnab\otimes\vecB{b}):(\bnab\otimes\vecB{b})}{\V}$ reduced by a small amount as the dynamo saturates\footnote{Note that for $\M \approx 10$ turbulence \citet{Seta2021_saturation_supersonic_dynamo} showed that the Ohmic dissipation probability density function (the current; not simply the $\first$ moment, that we are commenting upon here) reduced in the saturation. This is also shown for the supersonic runs in \citet{Sur2023_dynamo}.}, hinting that it is the $\first$ moment growth terms that need to shrink to eventually reach a saturation. Hence, the incompressible growth term, $\Exp{ (\vecB{b}\otimes\vecB{b}) :\St}{\V}$, is suppressed (presumably by the magnetic tension $\sim \bnab\cdot[\vecB{b}\otimes\vecB{b}]$). \citet{Schekochihin2004_dynamo_saturation_via_anisotropisation} showed that this could be achieved through anisotropising the velocity field along folded field structures in $\vecB{b}\otimes\vecB{b}$. Furthermore, \citet{Seta2021_saturation_supersonic_dynamo} showed that in the saturation, $|\vecB{u}\times\vecB{b}| \approx 0$ for both $\M = 0.1$ and $\M = 10$ dynamos, which suggests that all tensor projection growth terms weaken in the saturation. 
    
    Moreover, as shown in \citet{Sur2023_dynamo} for the $\M > 1$ dynamo, the compressible growth term, $(1/6)\Exp{(\vecB{b}\otimes\vecB{b}):(\bnab\cdot\vecB{v})\It}{\V}$, is suppressed by the magnetic field pressure gradients $\bnab b^2/(8\pi)$ that are preferentially parallel to $\bnab\rho$, opposing compression, since shocked gas can generate large $\bnab b^2$, organising the magnetic field into filamentary structures that are supported by magnetic pressure \citep{Molina2012,Beattie2022_spdf,Kriel2025_supersonic_scales}. This is relevant in regions of high $\bnab b^2$ (corresponding to high mass), but in the magnetic voids, $(1/6)\Exp{(\vecB{b}\otimes\vecB{b}):(\bnab\cdot\vecB{v})\It}{\V}$, which are the more volume-filling state of the gas, these terms act to reduce $\emagb / \ekinb$, which ought to decrease final saturation of the dynamo. We will revisit these volume-averaged terms, computing them through the temporal evolution of the simulations in both the kinematic and saturated stage in \autoref{sec:integral_quants}. 

    \begin{table*}
    \caption{Main simulation parameters and derived quantities.}
    \label{tab:sims}
    \begin{tabular}{l c c c c c c c c c c c c }
        \hline\hline
            \multicolumn{1}{c}{Sim. ID} & $\Res$ & $\Reb$ & $\Pnu$ & $\nus t_0/\ell_0^2$ & $\nub t_0/\ell_0^2$ & $\M$ & $2\gamma_1 t_0$ & $\left( \frac{\emag}{\ekin} \right)_{\text{sat}}$ & $(\Ma)_{\text{sat}}$ & $N_{\text{grid}}^3$  \\[0.4em]
            \multicolumn{1}{c}{(1)} & (2) & (3) & (4) & (5) & (6) & (7) & (8) & (9) & (10) & (11) \\
            \hline
                 \texttt{P$\nu\infty$}    & 1000    & $\infty$  & $\infty$  & $1\times 10^{-3}$  & -                  & 0.96 $\pm$ 0.03 & 1.48 $\pm$ 0.03 & 0.17 $\pm$ 0.03 & 2.49 $\pm$ 0.29  & $512^3$ \\
                 \texttt{P$\nu$1\_1024}   & 1000    & 1000       & 1        & $1\times 10^{-3}$  & $1\times 10^{-3}$  & 0.95 $\pm$ 0.03 & 1.42 $\pm$ 0.01  & 0.15 $\pm$ 0.04  & 2.62 $\pm$ 0.40 & $1024^3$ \\
                 \texttt{P$\nu$1}         & 1000    & 1000       & 1        & $1\times 10^{-3}$  & $1\times 10^{-3}$  & 0.95 $\pm$ 0.03 & 1.42 $\pm$ 0.02  & 0.14 $\pm$ 0.03  & 2.71 $\pm$ 0.30 & $512^3$ \\
                 \texttt{P$\nu$1\_256}    & 1000    & 1000       & 1        & $1\times 10^{-3}$  & $1\times 10^{-3}$  & 0.95 $\pm$ 0.03 & 1.36 $\pm$ 0.04  & 0.15 $\pm$ 0.02  & 2.61 $\pm$ 0.22 & $256^3$ \\
                 \texttt{P$\nu$1\_128}    & 1000    & 1000       & 1        & $1\times 10^{-3}$  & $1\times 10^{-3}$  & 0.94 $\pm$ 0.04 & 1.36 $\pm$ 0.03  & 0.14 $\pm$ 0.04  & 2.76 $\pm$ 0.38 & $128^3$ \\
                 \texttt{P$\nu$1e-1}      & 1000    & 100        & 0.1      & $1\times 10^{-3}$  & $1\times 10^{-2}$  & 0.94 $\pm$ 0.04 & 1.47 $\pm$ 0.05  & 0.20 $\pm$ 0.03  & 2.26 $\pm$ 0.30 & $512^3$ \\
                 \texttt{P$\nu$1e-1\_256} & 1000    & 100        & 0.1      & $1\times 10^{-3}$  & $1\times 10^{-2}$  & 0.92 $\pm$ 0.03 & 1.44 $\pm$ 0.04  & 0.18 $\pm$ 0.05 & 2.44 $\pm$ 0.33 & $256^3$ \\
                 \texttt{P$\nu$1e-1\_128} & 1000    & 100        & 0.1      & $1\times 10^{-3}$  & $1\times 10^{-2}$  & 0.93 $\pm$ 0.03 & 1.35 $\pm$ 0.05  & 0.12 $\pm$ 0.04  & 2.88 $\pm$ 0.36 & $128^3$ \\
                 \texttt{P$\nu$1e-2}      & 1000    & 10         & 0.01     & $1\times 10^{-3}$  & $1\times 10^{-1}$   & 0.90 $\pm$ 0.02 & 1.46 $\pm$ 0.03  & 0.19 $\pm$ 0.07  & 2.38 $\pm$ 0.39 & $256^3$ \\
                 \texttt{P$\nu$1e-3}      & 1000    & 1          & 0.001    & $1\times 10^{-3}$  & $1\times 10^{0\,\,\,\,}$   & 0.93 $\pm$ 0.04 & 1.28 $\pm$ 0.02 & - & - & $256^3$ \\ 
            \hline\hline
    \end{tabular}
    \begin{tablenotes}[para]
        \textit{\textbf{Notes.}} All simulations have magnetic Reynolds number $\Rm = 1000$, corresponding to Ohmic resistivity coefficient $\eta = 1\times 10^{-3} \ell_0^2 / t_0$, a $\Pms = 1$ (\autoref{eq:Pm_shear_defn}) and a $\Pmb = \Pnu^{-1}$ (\autoref{eq:Pm_bulk_defn}), hence we list only the viscous material properties in the table. \texttt{P$\nu$1e-3} was not run to saturation due to the computational cost (extremely viscous, very small diffusion integration time step over long integration times), so we do not report saturated statistics for that simulation. \textbf{Column (1):} the unique simulation ID. \textbf{Column (2):} the shear Reynolds number of the plasma. \textbf{Column (3):} the same as column (2), but for the bulk viscosity. \textbf{Column (4):} the viscous Prandtl number. \textbf{Column (5):} the coefficient of shear viscosity in units of turbulent correlation times on the driving scale of the turbulence. \textbf{Column (6):} the same as column (5) but for the coefficient of bulk viscosity. \textbf{Column (7):} the turbulent Mach number in the kinematic stage of the dynamo. \textbf{Column (8):} the growth rate of the kinematic stage in units of turbulent correlation times. \textbf{Column (9):} the ratio of the volume-averaged magnetic energy and volume-averaged kinetic energy in the saturated stage of the dynamo. \textbf{Column (10):} the same as column (9) but for the Alfv\'enic Mach number. \textbf{Column (11):} the number of grid points used in the numerical discretisation. We perform a convergence study in \aref{app:convergence} for both the \texttt{P$\nu$1e-1} and \texttt{P$\nu$1} simulations.
    \end{tablenotes}
\end{table*}

\section{Numerical simulations \& Methods}\label{sec:sims}
    \subsection{Visco-resistive magnetohydrodynamic fluid model}
        We use a modified version of the finite volume MHD code \textsc{flash} \citep{Fryxell2000,Dubey2008}, utilising a second-order conservative MUSCL-Hancock 5-wave approximate Riemann scheme \citep{Bouchut2010,Waagan2011}, and $\bnab \cdot \vecB{b}$ parabolic diffusion flux cleaning \citep{Marder1987_fluxcleaning} to solve the 3D, visco-resistive, isothermal, compressible MHD equations with a stochastic acceleration field acting to drive the turbulence,
        \begin{align}
            \frac{\partial \rho}{\partial t} + \bnab\cdot(\rho \vecB{v}) &= 0 \label{eq:continuity}, \\
                \frac{\partial \rho\vecB{v}}{\partial t}  + \bnab\cdot\left[ \rho \vecB{v}\otimes\vecB{v} - \frac{1}{4\pi}\vecB{b}\otimes\vecB{b} \right.\nquad[5]& \nonumber\\
                \left. + \left(c_s^2 \rho + \frac{b^2}{8\pi}\right)\It \right] &= \bnab\cdot\left(\rho\bm{\sigma}\right) + \rho \vecB{f},\label{eq:momentum} \\
                \frac{\partial \vecB{b}}{\partial t} + \bnab \cdot \left(\vecB{v}\otimes\vecB{b} - \vecB{b}\otimes\vecB{v} \right)&= -\eta\bnab\times\vecB{j},\label{eq:induction}\\
                \bnab \cdot \vecB{b} &= 0, \label{eq:div0}
        \end{align}
        where the viscous stress tensor is
        \begin{align}
            \bm{\sigma} &= 2\nu_{\rm shear}\St + \nu_{\rm bulk}\Bt, 
        \end{align}
        using the same definitions as in \autoref{sec:vel_grad_tensor}. We solve the equations on a triply periodic domain of dimension $\V \equiv L^3$, with uniform grids of $128^3-1024^3$~cells (we show a detailed convergence study in \aref{app:convergence}, and implementation of the bulk viscosity in \aref{app:implementation}), where $\Exp{\vecB{b}(t)}{\V}= \Exp{\rho\vecB{v}(t)}{\V} = 0$ (i.e., no net magnetic flux over $\V$, in the zero-momentum frame of $\V$), $c_s$ is the sound speed, $\vecB{j} = \bnab \times \vecB{b} / (4\pi)$ is the current density, and $\vecB{f}$ is the stochastic turbulent acceleration source term that drives the turbulence.

    \subsection{Turbulent driving}\label{sec:turb_driving}
        The turbulent source term $\vecB{f}$ follows an \citet{OU_process_1930} process with finite $e$-fold correlation time, $t_0 = \lo/\Exp{\v^2}{\V}^{1/2}$ \citep{Schmidt2009,Federrath2010_driving}. The field $\vecB{f}$ is constructed in Fourier space with energy isotropically injected on the peak scale $|\vecB{k}L/2\pi|=2$ (equivalently, $\lo = L/2$) and falls off to zero with a parabolic spectrum within $1 \leq |\vecB{k}L/2\pi| \leq 3$. On $\lo$ we use $t_0$ and the energy injection rate to control the rms velocity, which we set to $\Exp{\v^2}{\V}^{1/2}/c_s = \M = 1.0$ in the kinematic stage of the dynamo, ensuring that we have compressible modes for the bulk viscosity to control. Furthermore, we inject energy such that $\vecB{f}_{\perp}^2/\vecB{f}_{\parallel}^2=2$, corresponding to 2:1 energy in incompressible $\bnab \cdot \vecB{f}_{\perp} =0$ and compressible $\bnab \times \vecB{f}_{\parallel} =0$ mode components of $\vecB{f}$, ensuring that the compressible modes are being replenished on large-scales. See \citet{Federrath2010_driving} and \citet{Federrath2022_turbulence_driving_module} for more details about the \textsc{TurbGen} driving source. We integrate each of the experiments from $t/t_0=0$ to $t/t_0=100$, writing the 3D field variables to disk every $t/t_0=0.1$ for post-processing.
        
    \subsection{Setting Reynolds and Prandtl numbers}
        Following \autoref{sec:viscosity_theory}, we parameterise the visco-resistive properties of our simulations with five numbers. Throughout the study we fix $\Res = 1000$, corresponding to setting $\nus = 1\times 10^{-3} \ell_0^2 / t_0$ in \autoref{eq:Rebulk_defn}. This ensures that the incompressible modes are in a state of turbulence in our simulations \citep{Frisch1995}. We then vary $\Pnu$ between $\Pnu = \Reb = \infty$, where $\nub = 0$ (the dissipation is controlled purely by numerical means) to $\Pnu = 0.001$, where the $\nub = 10^3 \nus$ and $\Reb = 1$, which corresponds to $t_{\nub} \sim t_0$ (i.e., the diffusion is as fast as the dynamical time on $\ell_0$; see column~6 in \autoref{tab:sims} where we write the viscous coefficients in units of $t_0$). For $\Pnu > 1$, the simulations develop a sub-shear-viscous range, \autoref{eq:sub-shear-bulk-viscous_range}, where the compressible modes ought to dominate at high-$k$; and for $\Pnu < 1$, they develop a sub-bulk-viscous range, \autoref{eq:sub-bulk-viscous_range}, where compressible modes are suppressed whilst incompressible modes are left to dynamically evolve with the turbulence. The two ranges of scales are shown schematically in \autoref{fig:spectra}.

        We control the resistivity by setting the magnetic Reynolds number, $\Rm$ \autoref{eq:Rm_defn}. We choose $\eta = 1\times 10^{-3} \ell_0^2 / t_0$ corresponding to $\Rm = 1000$. The extent of the (classical) sub-viscous range is set through the shear magnetic Prandtl number, $\Pms$ (\autoref{eq:Rm_defn}; the regular $\Pm$ that has been studied in the small-scale dynamo; \citealt{Schekochihin2004_dynamo,Federrath2014_supersonic_dynamo,Galishnikova2022_saturation_and_tearing,Kriel2022_kinematic_dynamo_scales,Brandenburg2023_dissipative_structures,Kriel2025_supersonic_scales}). Since $\Res = \Rm = 1000$, $\Pms = 1$ for all of the simulations in our study, hence we decay the incompressible and magnetic modes similarly, without the development of a sub-viscous range, $\ell_{\eta} \leq \ell \leq \ell_{\nus}$. We leave the study of the bulk viscosity in the context of evolving $\Pms$ for future studies. Based on \autoref{eq:Pm_bulk_defn}, the corresponding bulk magnetic Prandtl number, $\Pmb$, is completely controlled by changing $\Pnu$, because $\Pmb = \Pnu^{-1}$ for $\Pms = 1$. Hence, for the remainder of the study, we focus only on the principle parameters that are evolved across the present simulations, $\Reb$ and $\Pnu$. All of our diffusion coefficients, $\nus$, $\nub$, $\eta$ are constant in time and space.

     \begin{figure*}
         \centering
         \includegraphics[width=\linewidth]{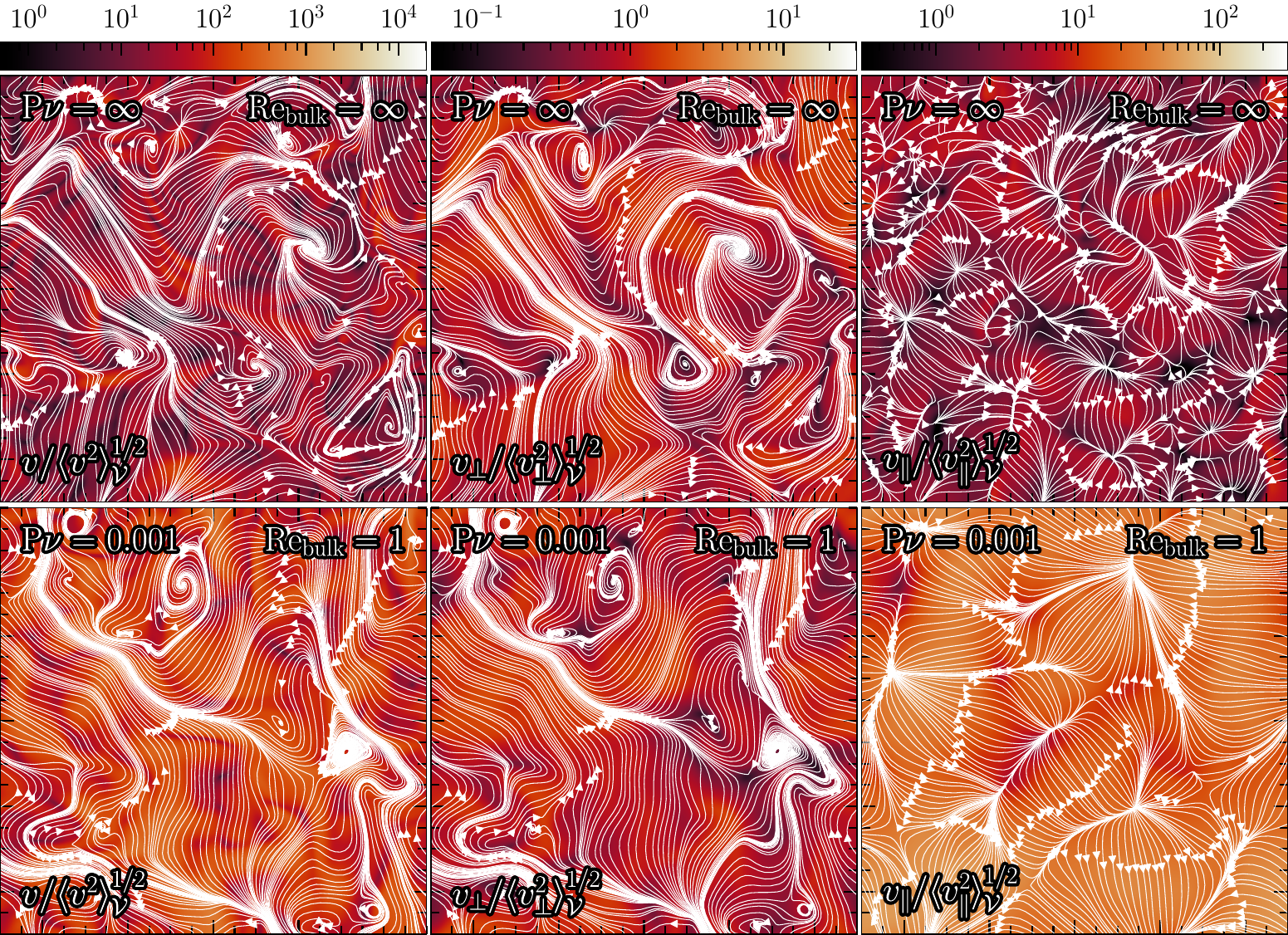}
         \caption{Two-dimensional slices of the total ($v$; left), incompressible ($v_{\perp}$; middle) and compressible ($v_{\parallel}$; right) velocity modes in the kinematic stage ($t/t_0 = 30$) of the turbulent dynamo. All velocities are scaled by the rms, which determines the colour of the plots, with the corresponding vector field slice of the modes overlaid. The top row corresponds to the $\Pnu = \infty$ simulation, where there is no explicit bulk viscosity, and the bottom row is for $\Pnu = 0.001$, where the the decay timescale of the bulk viscosity $t_{\nub}$ is comparable to the dynamical timescale of the turbulence on the outer scale $t_0$. Both the $v/\Exp{v^2}{\V}^{1/2}$ and $v_{\perp}/\Exp{v_{\perp}^2}{\V}^{1/2}$ panels are aesthetically similar between the simulations, but the $v_{\parallel}/\langle v_{\parallel}^2\rangle_{\V}^{1/2}$ panels are vastly different, with only low-$k$ $v_{\parallel}$ modes left in the plasma when the bulk viscosity is strong.}
         \label{fig:helmholtz_decomposition}
     \end{figure*}

    \begin{figure*}
        \centering
        \includegraphics[width=\linewidth]{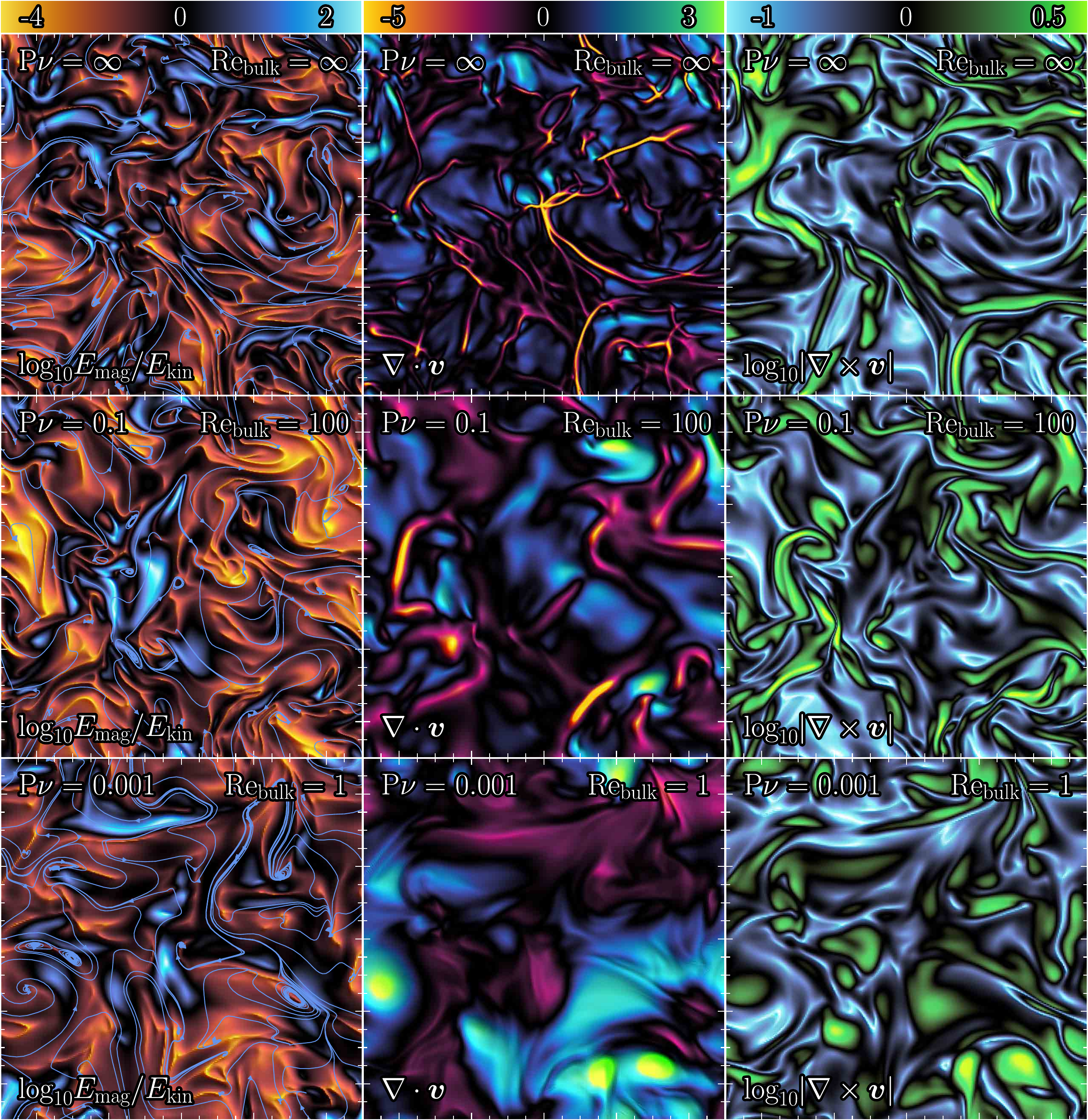}
        \caption{Two-dimensional slices in the kinematic stage of the dynamo ($t/t_0 = 30$) for $\emag/\ekin$ (left column), $\bnab\cdot\vecB{v}$ (middle column) and $|\bnab\times\vecB{v}|$. Each of the quantities are normalised by the root-mean-squared value of the field to focus on the structure. The least viscous simulations are in the top row, where $\Pnu = \infty$ and $\Reb = \infty$, and the dissipation of the compressible modes is controlled by the numerical viscosity \citep{Malvadi2023_numerical_dissipation}. The most bulk viscous simulations are shown in the bottom row where $\Pnu = 0.001$ (\autoref{eq:Pnu_defn}) and $\Reb = 1$ (\autoref{eq:Rebulk_defn}). Magnetic field lines are shown in blue in the $\emag/\ekin$ panels. The $\bnab\cdot\vecB{v}$ panel show the strongest effect as the bulk viscous increases down the page. Thin shocks (yellow structures) in the $\Pnu = \infty$ simulation (top row, middle column), become wider, diffusing away until they are completely absent in the $\Pnu = 0.001$ simulation (bottom row, middle column). There is a lesser effect on the vorticity, which looses some of the small-scale structure as $\Pnu$ gets smaller. A similar effect is shown for $\emag/\ekin$ and the magnetic field structure that can be seen in the streamlines.}
        \label{fig:9_panel}
    \end{figure*}
        
    \subsection{Initial field conditions}
        Following, e.g., \citet{Kriel2022_kinematic_dynamo_scales}, we initialise the velocity field to $|\vecB{v}(x,y,z,t=0)|/c_s=0$, with units $c_s=1$ and the density field $\rho(x,y,z,t=0)=\rho_0$, where the density has units $\rho_0=1$. Likewise, $\vecB{b}$ has units of $c_s\rho_0^{1/2} = 1$, and is initialised as a Gaussian field in $k$-space with a parabolic spectral power distribution matching the turbulent forcing function. The total energy in the turbulent fluctuations of the field is set such that $\Exp{E_{\text{mag},0}}{\V}/\rho_0c_s^2 = 10^{-10} = \beta^{-1}$, where $\Exp{E_{\text{mag},0}}{\V}$ is the initial volume integral energy and $\beta$ is the plasma beta. This corresponds to an incredibly weak seed magnetic field compared to the constant thermal energy, allowing for the simulations to spend a significant time in the kinematic stage of the dynamo. 

    \subsection{Helmholtz decomposition of velocity field}
        Any smooth vector field can be written as a vector sum of curl-free and divergence-free vectors -- the Helmholtz decomposition of a vector field. Because we are seeking to understand the role of bulk viscosity on the compressible modes in the supersonic dynamo, we specifically decompose our $\vecB{v}$ into orthogonal compressible (longitudinal) $\vpar$ ($\bnab \times \vpar = 0$) and incompressible (transverse) $\vprp$ ($\bnab \cdot \vprp=0$) modes, such that,
        \begin{align}\label{eq:H_decomp}
            \vecB{v} = \vpar + \vprp.
        \end{align}
        This is easily done in Fourier space, where
        \begin{align} \label{eq:long_operator}
            \vpar(\bm{\ell}) = \int\d{\vecB{k}}\,[\vecB{k}\cdot\Tilde{\vecB{v}}(\vecB{k})]\vecB{k} /k^2 e^{2\pi i \vecB{k}\cdot\bm{\ell}},
        \end{align}
        is the inverse Fourier transform of the Fourier transformed velocity $\Tilde{\vecB{v}}(\vecB{k})$ projected along $\vecB{k}$. Then $\vprp = \vecB{v} - \vpar$, ensuring that \autoref{eq:H_decomp} is maintained (see \aref{app:implementation} for a further example where we construct the compressible and incompressible energy equations utilising \autoref{eq:long_operator}). Throughout this study we utilise the 1D shell-integrated (isotropic) power spectra for the magnetic and kinetic energy,
        \begin{align}
            \emag(k) &= \frac{1}{8\pi} \int_{\Omega_k} \d{\Omega_k}\,  4\pi k^2 |\tilde{\vecB{b}}(\vecB{\vecB{k}})|^2, \\
            \ekin(k) &= \frac{\rho_0}{2} \int_{\Omega_k} \d{\Omega_k}\,  4\pi k^2 |\tilde{\vecB{v}}(\vecB{\vecB{k}})|^2, 
        \end{align}
        where $\d{\Omega_k}$ is the shell angle at radius $k$ and $\tilde{\vecB{b}}(\vecB{\vecB{k}})$ and $\tilde{\vecB{v}}(\vecB{\vecB{k}})$ are the Fourier transforms of the magnetic and velocity field, respectively. We likewise create the $\ekinpar(k)$ and $\ekinperp(k)$ spectrum, based on our definitions in \autoref{eq:H_decomp}, for exploring the Helmholtz decomposed kinetic energies. 
        
        We show examples of two-dimensional slices of the Helmholtz decomposed velocity fields in \autoref{fig:helmholtz_decomposition} in our $\Pnu=\infty$ (top row) and $\Pnu = 0.001$ simulations (bottom row). The first column corresponds to the total velocity field, the second column the incompressible $\vprp$ field, and the last column the compressible $\vpar$ fields, all which are coloured by the magnitude normalised by the rms. Both the $\vecB{v}$ and $\vpar$ fields qualitatively remained unchanged by the strong bulk viscosity, however the $\vpar$ field loses all of its high-$k$ structure, and becomes dominated by low-$k$ modes that are being replenished by the forcing.

        We show how this translates into changing $\bnab\cdot\vecB{v}$ (middle panel) and $|\bnab\times\vecB{v}|$ (right panel) in \autoref{fig:9_panel}, with $\Pnu = \infty$ (top row), $\Pnu = 0.1$ (middle row), and $\Pnu = 0.001$ (bottom row) simulations, along with the structure of $\emag/\ekin$ (left column). Similar to \autoref{fig:helmholtz_decomposition}, the largest effect that changing $\Pnu$ has is shown in the $\bnab\cdot\vecB{v}$ field, where strong, thin compressions (yellow) that form in the $\Pnu = \infty$ simulations are diffusive and thick in the $\Pnu = 0.1$ simulations, until they are completely dissipated away in the $\Pnu = 0.001$ simulation. From this plot, it is also clear both $\bnab\times\vecB{v}$ and $\emag/\ekin$ are also losing high-$k$ mode structure as $\Pnu$ decreases. To understand this process quantitatively, we turn now to the integral energy statistics.

    \begin{figure}
        \centering
        \includegraphics[width=\linewidth]{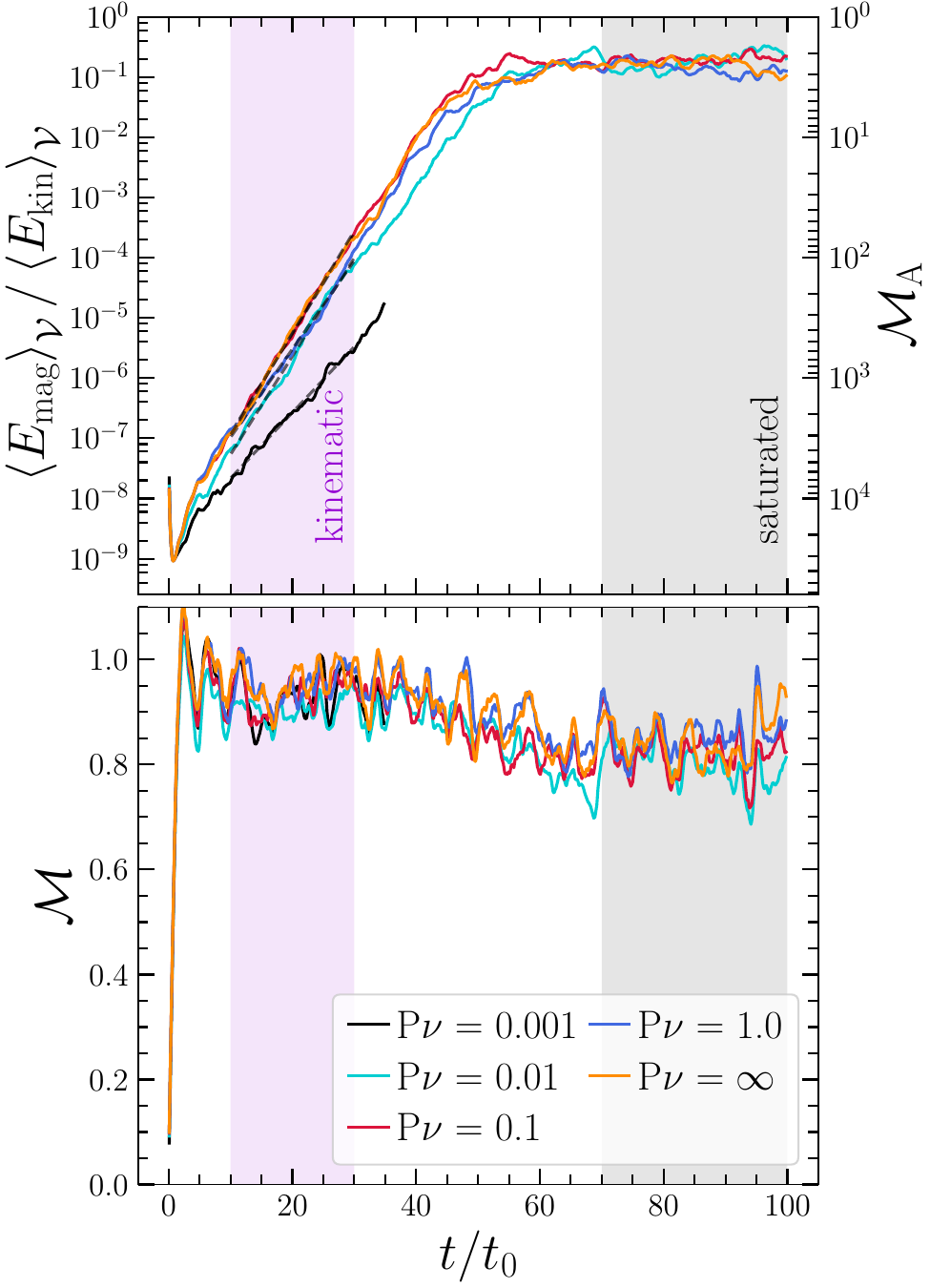}
        \caption{\textbf{Top:} the integral magnetic to kinetic energy ratio $\emagb/\ekinb$ as a function of time in units of turbulent correlation times $t/t_0$ on the driving scale $\ell_0$, coloured by $\Pnu$. The purple region indicates where we define the kinematic stage of the dynamo, $10 \leq t/t_0 \leq 30$, which we use for the entirety of our study. This is where $\emagb$ exhibits exponential growth (\autoref{eq:kinematic_growth}), which we fit an exponential model to shown with the black-dashed lines. We tabulate growth rates from these fits $2\gamma_1$ for each curve in \autoref{tab:sims}. The grey region indicates the saturated stage of the dynamo (\autoref{eq:saturated_no_growth}). Similarly, we tabulate the $\emagb/\ekinb$ ratios averaged over this region as well as the $\Ma$ in \autoref{tab:sims}. We find that for fixed $\Res$ and $\Rm$, moderate $\Pnu$ ($\nub \sim \nus$) does not have a strong influence on the growth rates or final $\emagb/\ekinb$ saturation. However, for low $\Pnu$ ($\nub > \nus$), the viscosity begins to have a strong, negative effect on both statistics. \textbf{Bottom:} The same as the top plot but for the $\M$ evolution of the simulations, which moves between stationary states for the kinematic and saturated dynamo stages as the magnetic flux modifies the kinetic energy spectrum.}
        \label{fig:integral_energies}
    \end{figure}

\section{Integral energies and growth rates}\label{sec:integral_quants}
    We start the exploration for the role of the bulk viscosity on the turbulent dynamo via the integral quantities of energy, growth rates and growth rate terms ($\vecB{b}\otimes\vecB{b}$ tensor projections) that we derived in \autoref{sec:growth}.
    
    \subsection{Volume integrated energies}
        We define the volume integral energies as,
        \begin{align}
            \emagb = \frac{\Exp{b^2}{\V}}{8\pi}, \quad \ekinb = \frac{\Exp{\rho_0 v^2}{\V}}{2},
        \end{align}
        which we will use throughout this entire study. We plot the integral energy ratio $\emagb/\ekinb$ (top) and $\M$ (bottom) in \autoref{fig:integral_energies}, colouring by $\Pnu$. These colours will be maintained throughout the entire study. In log-linear space, we immediately identify the kinematic (exponential growth; highlighted in dark violet) and saturated (stationarity; highlighted with grey) stages of the dynamo in the $\emagb/\ekinb$ plot, which all experiments undertake (i.e., all simulations are above the critical $\Rm$ for small-scale dynamo growth; \protect{\citealt{Ruzmaikin1981_rm_crit,Haugen2004_dynamo_mach_and_crit_rm,Schekochihin2004_critical_dynamo,Federrath2014_supersonic_dynamo,Seta_2015_STFdynamo_rm_crit}}). In log-linear scale, the nonlinear stage (\autoref{eq:linear_growth}) between the two regimes is quite obfuscated. However, it is clearly observed in the $\M(t/t_0)$ plot, shown in the bottom panel of \autoref{fig:integral_energies}. The transition happens between $30 \leq t/t_0 \leq 70$ between kinematic and saturated stages. This corresponds to the backreacting magnetic field modifying the kinetic energy cascade as $\bnab\cdot(\rho\vecB{v}\otimes\vecB{v}) \sim \bnab\cdot(\vecB{b}\otimes\vecB{b}) + \bnab b^2$ throughout the $k$ modes in the cascade (see Fig. B1 in~\citealt{Beattie2023_GorD_P1} and theory in \citealt{Schekochihin2002_saturation_evolution}).     
        We tabulate all of the $\emagb/\ekinb$ saturation statistics in \autoref{tab:sims}, averaged over the saturated $\emagb/\ekinb$ and $\Ma$ in the grey region, $t/t_0 \geq 75$, all of which indicate a quite small effect throughout $0.01 \leq \Pnu \leq \infty$ range, always maintaining $\emagb/\ekinb \approx 0.15$, corresponding to $\Ma \approx 2.5$, consistent within $1\sigma$ with $\Pnu = \infty$ calculations in \citet{Federrath2011_mach_dynamo}. This indicates that the bulk viscosity does not have a large effect on the low-$k$ modes, which dominate the integral energies \citep{Beattie2022_ion_alfven_fluctuations}. This is suggestive that previous $\M$ dependencies upon small-scale dynamo saturation, e.g., \citet{Federrath2011_mach_dynamo,Seta2021_saturation_supersonic_dynamo,Sur2023_dynamo}, are being controlled (more specifically the magnetic field being suppressed) by the lowest $k$ modes. Performing the same calculation with instead the Helmholtz decomposed integral kinetic energies,
        \begin{align}
            \ekinparb = \frac{\Exp{\rho_0 \vpar^2}{\V}}{2}, \quad \ekinperpb = \frac{\Exp{\rho_0 \vprp^2}{\V}}{2}, 
        \end{align}
        provides a clearer picture for how the bulk viscosity effects the integral plasma quantities. 
    
\begin{figure}
            \centering
            \includegraphics[width=\linewidth]{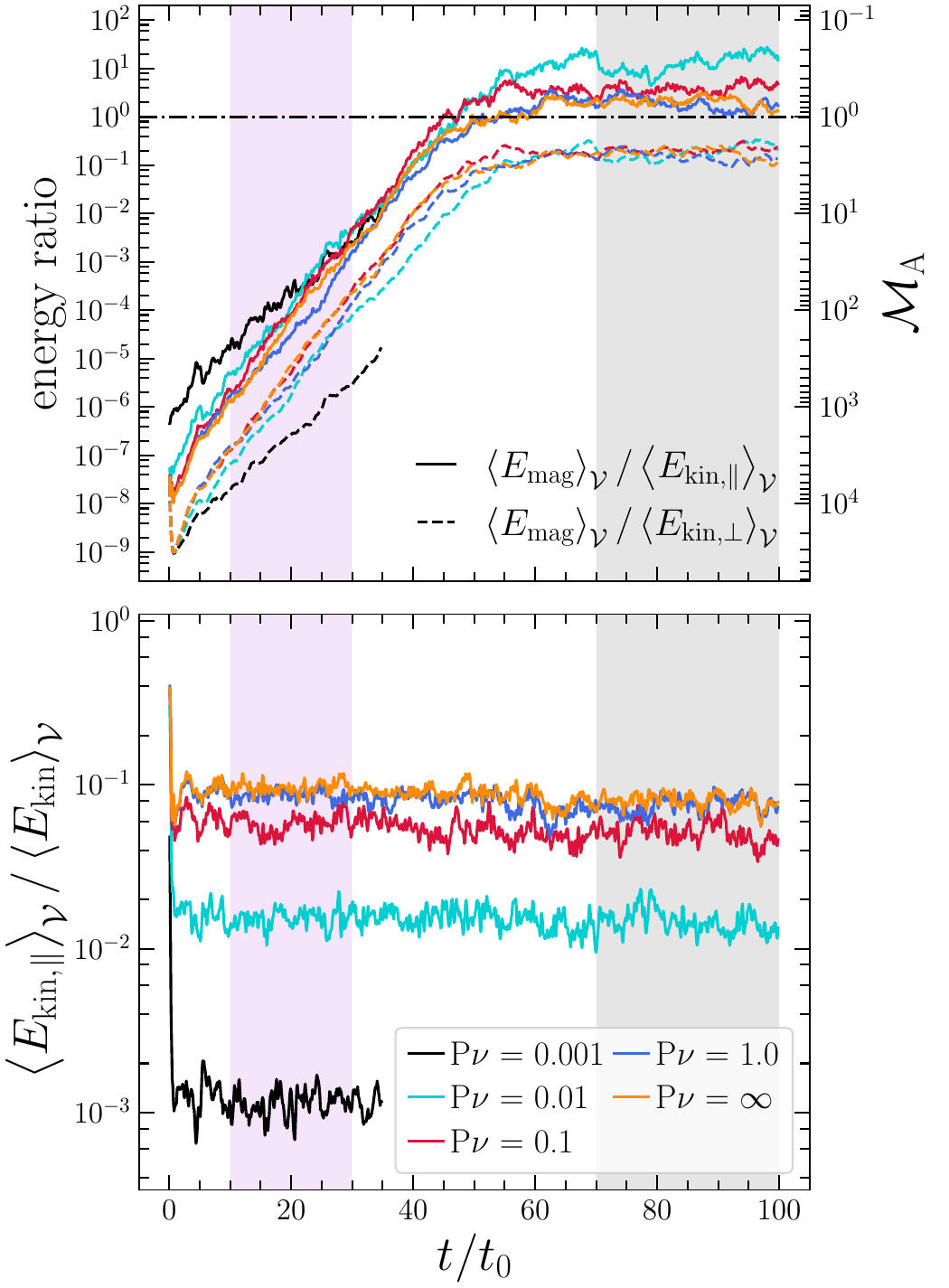}
            \caption{\textbf{Top:} the same as the top panel in \autoref{fig:integral_energies}, but separated into incompressible (solid) and compressible (dashed) components of $\ekin$. The final saturated state for the incompressible modes is below equipartition $\ekinparb/\emagb < 1$, whereas it is significantly above equipartition for the compressible modes $\ekinparb/\emagb > 1$. \textbf{Bottom:} the integral compressible kinetic energy $\ekinparb$ ratios with respect to the total kinetic energy $\ekinb$. We plot only $\ekinparb/ \ekinb$ because $\ekinperpb / \ekinb \approx 1$ for all simulations. As $\Pnu$ decreases, the total energy contained within the compressible component of the kinetic energy significantly decreases from $\ekinparb / \ekinb \approx 10^{-1}$ to $\ekinparb / \ekinb \approx 10^{-3}$, roughly following $\ekinparb / \ekinb \sim \Pnu$ for $\Pnu < 1$.}
            \label{fig:kinetic_energy_decomposed}
        \end{figure}

        In \autoref{fig:kinetic_energy_decomposed} we show the $\emagb/\ekinparb$ (solid lines) and $\emagb/\ekinperpb$ (dashed lines) in the top panel; and $\ekinparb/\ekinb$ (solid lines) and $\ekinperpb/\ekinb$ (dashed lines) ratios in the bottom panel, coloured in the same fashion as \autoref{fig:integral_energies}. We turn our attention to the bottom panel. For all $\Pnu$, we find the turbulence is dominated by the incompressible motions, even though we are replenishing the compressible modes with the same amount of energy as the incompressible modes in $\vecB{f}$. \citet{Federrath2010_driving} showed this is the case previously for hydrodynamical turbulence -- i.e., the forcing modes and the momentum modes cannot be treated analogously. For $\Pnu = \infty$, we find that $\ekinparb/\ekinb \approx 10^{-1}$ (consistent with what has been measured previously in \citealt{Federrath2011_mach_dynamo}), down to $\Pnu = 0.001$, where $\ekinparb/\ekinb \approx 10^{-3}$, decreasing somewhat linearly with $\Pnu$,
        \begin{align}
            \ekinparb/\ekinb \sim \Pnu,
        \end{align}
        for $\Pnu < 1$. This illustrates that as $\Pnu$ shrinks there is a strong suppression of $\vpar$ modes, even for moderate $\Pnu$, that we were not able to see before in \autoref{fig:integral_energies}.
    
        In the top panel of \autoref{fig:kinetic_energy_decomposed} we show how the kinetic energy in these modes balances with the magnetic energy. As we expect from the top panel of \autoref{fig:integral_energies}, the final saturation of the $\vpar$ modes (dashed lines) is insensitive to the bulk viscosity, with $\emagb/\ekinperpb \approx 0.1$ for all $\Pnu$. However, the same is not true for the $\vpar$ modes, which saturate to $1 \leq \emagb/\ekinparb \leq 10$ (equipartition and beyond). Hence, the turbulence is sub-to-trans-Alfv\'enic with respect to $\vpar$ modes, which means that, for example, shocked plasma will not be able to significantly change the morphology of the magnetic field due to the fact that $|\bnab\cdot(\vecB{b}\otimes\vecB{b}) + \bnab b^2| \gg |\bnab\cdot(\rho\vpar \otimes \vpar)|$. This has been recently shown to be the case even in the kinematic stage of the supersonic dynamo \citep{Kriel2025_supersonic_scales}, but we find it happens here in a more extreme manner with decreasing $\Pnu$, highlighting the importance of treating the $\vpar$ and $\vprp$ modes separately in the supersonic dynamo problem. 
        
    \subsection{Volume integrated growth terms}
        We compute the growth rate $2\gamma_1$ in units of $t_0^{-1}$ by fitting a log-linear model in the dark violet region of the $\emagb/\ekinb$ data, which we show in \autoref{fig:integral_energies}. We show the fits with a grey dashed line. This corresponds to the kinematic stage of the dynamo, which we conservatively define with the range $10 \leq  t/t_0 \leq 30$. We tabulate all of these values in \autoref{tab:sims}, along with the $\M$ averaged over the kinematic regime.
    
        Similar to the $t\rightarrow \infty$ saturation, for moderate $\Pnu$, $\gamma_1$ is quite insensitive to the bulk viscosity. The values range between $2\gamma_1 t_0 = 1.40\pm0.02 - 1.48\pm0.03$, i.e., every $1.4t_0$ ($e$-fold time), the field grows by $\approx e^n\Exp{E_{\text{mag},0}}{\V}$, where $n$ is the number of $e$-fold times. This suggests that the high-$k$ compressible modes that are being decayed do not have a strong influence on small-scale shear viscous velocities, which are responsible for growing the magnetic field \citep{Schekochihin2004_dynamo,Kriel2022_kinematic_dynamo_scales,Galishnikova2022_saturation_and_tearing}.  

        However, once the simulations become significantly bulk viscous, such that the decay time from our input $\nub$ becomes faster than the dynamical timescale on the turbulent outer scale $t_{\nub} \sim (\ell_0^2/\nub)^{-1} \sim t_0$, which happens at $\Reb \leq  10,\;\Pnu \leq 0.01$ at our $\M$, then $\gamma_1$ becomes significantly affected. The fact that there is an impact at all suggests that either $\nub$ plays a direct role in the exponential stage of the dynamo processes, directly by suppressing the growth term, or through an indirect role by modifying the small-scale incompressible modes \citep[as demonstrated in][]{Touber2019_2d_turbulence}, which are the modes responsible for growing the field in $\Pms\geq 1$ turbulent plasmas \citep[e.g.,][]{Brandenburg1995_magnetic_flux_tubes,Schekochihin2004_dynamo,Federrath2011_mach_dynamo,Kriel2022_kinematic_dynamo_scales,Kriel2025_supersonic_scales}. Curiously, this is the opposite of the effect that one observes when increasing $\M$, or enhancing the injection of compressibility in the forcing function \citep{Federrath2011_mach_dynamo,Chirakkara2021,Seta2021_saturation_supersonic_dynamo,Sur2023_dynamo,Kriel2025_supersonic_scales}. We will show in \autoref{sec:decomposed_spectra} that we can explain this phenomena through $k_{\nus}$ coupling, and that changing $\nub$ (modifying the high-$k$ modes) and changing $\M$ (modifying the low-$k$ modes) is not equivalent, hence there is no tension with these previous studies. Now we turn to the growth terms derived in \autoref{sec:dynamo_theory}.
        
\begin{figure}
        \centering
        \includegraphics[width=\linewidth]{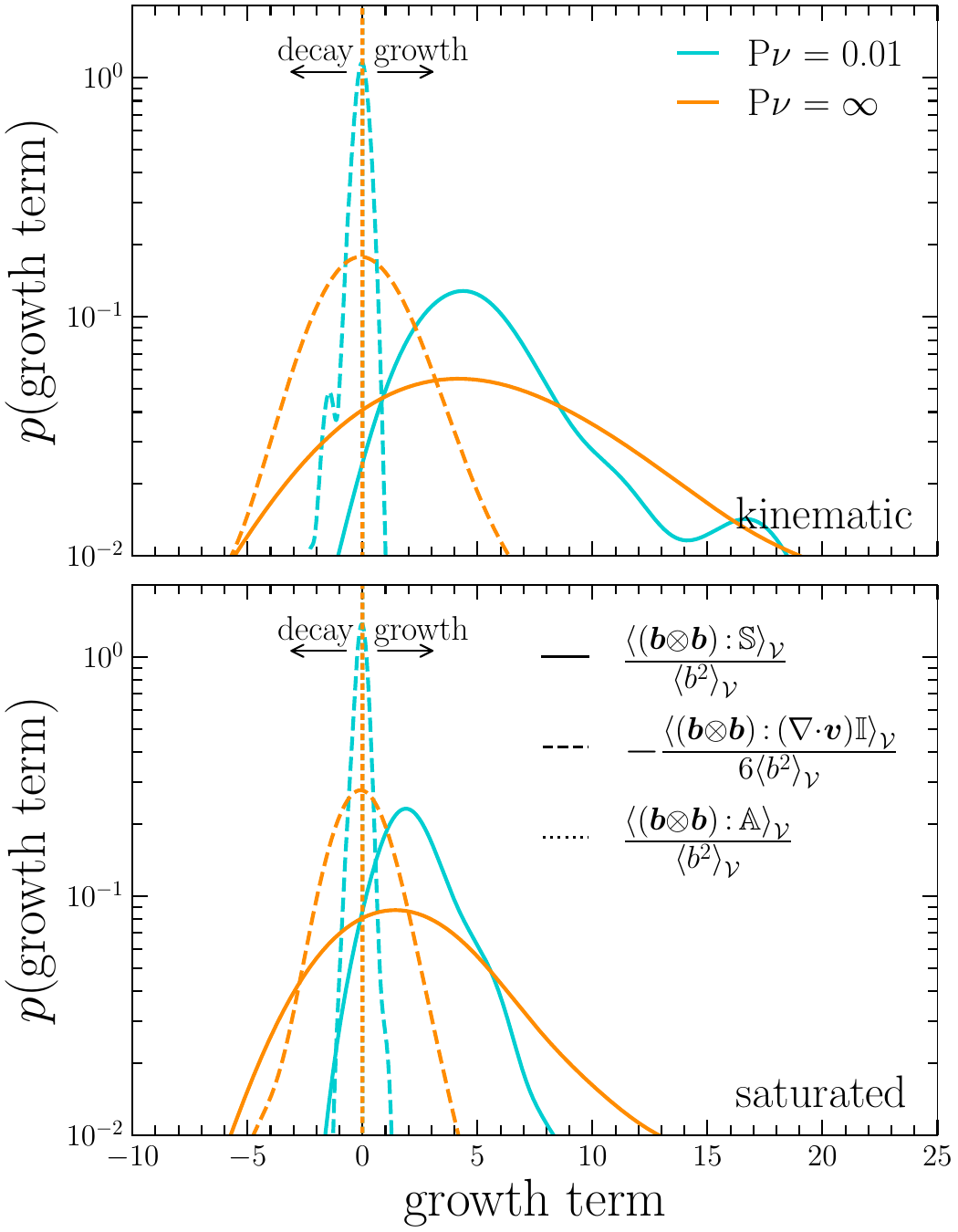}
        \caption{The temporal probability distribution functions of the stretching, winding and compression terms in the magnetic growth, \autoref{eq:growth_rate_decomp} (shown with different line styles), in the kinematic stage of the dynamo (top), and the saturated stage (bottom), coloured by $\Pnu$, showing only the two $\Pnu$ extremes ($\Pnu = 0.01$ and $\Pnu = \infty$). Each distribution function is constructed through kernel density estimation using a Gaussian function basis. For both stages, the winding term is zero (as predicted by \autoref{eq:no_winding_growth}, i.e., the rate of rotation \autoref{eq:A_vort_tensor}, an antisymmetric tensor, is everywhere orthogonal to the magnetic field, a symmetric tensor). The growth rate is therefore an interplay between the compression term ($\Exp{(\vecB{b}\otimes\vecB{b}):(\bnab\cdot\vecB{v})\It}{\V}/(6\Exp{b^2}{\V})$; dashed line style; which is suppressed for large bulk viscosities) and the stretching term ($\Exp{(\vecB{b}\otimes\vecB{b}):\St}{\V}/\Exp{b^2}{\V}$; solid line style). In both stages (kinematic and saturated) the stretching term is on average positive (net growth) and dominant over the compression. The compression approximately opposes and enhances growth in equal amounts, which is shown in more detail for all simulations in \autoref{fig:growth_cdfs}.}
        \label{fig:growth}
    \end{figure}

\begin{figure*}
        \centering
        \includegraphics[width=\linewidth]{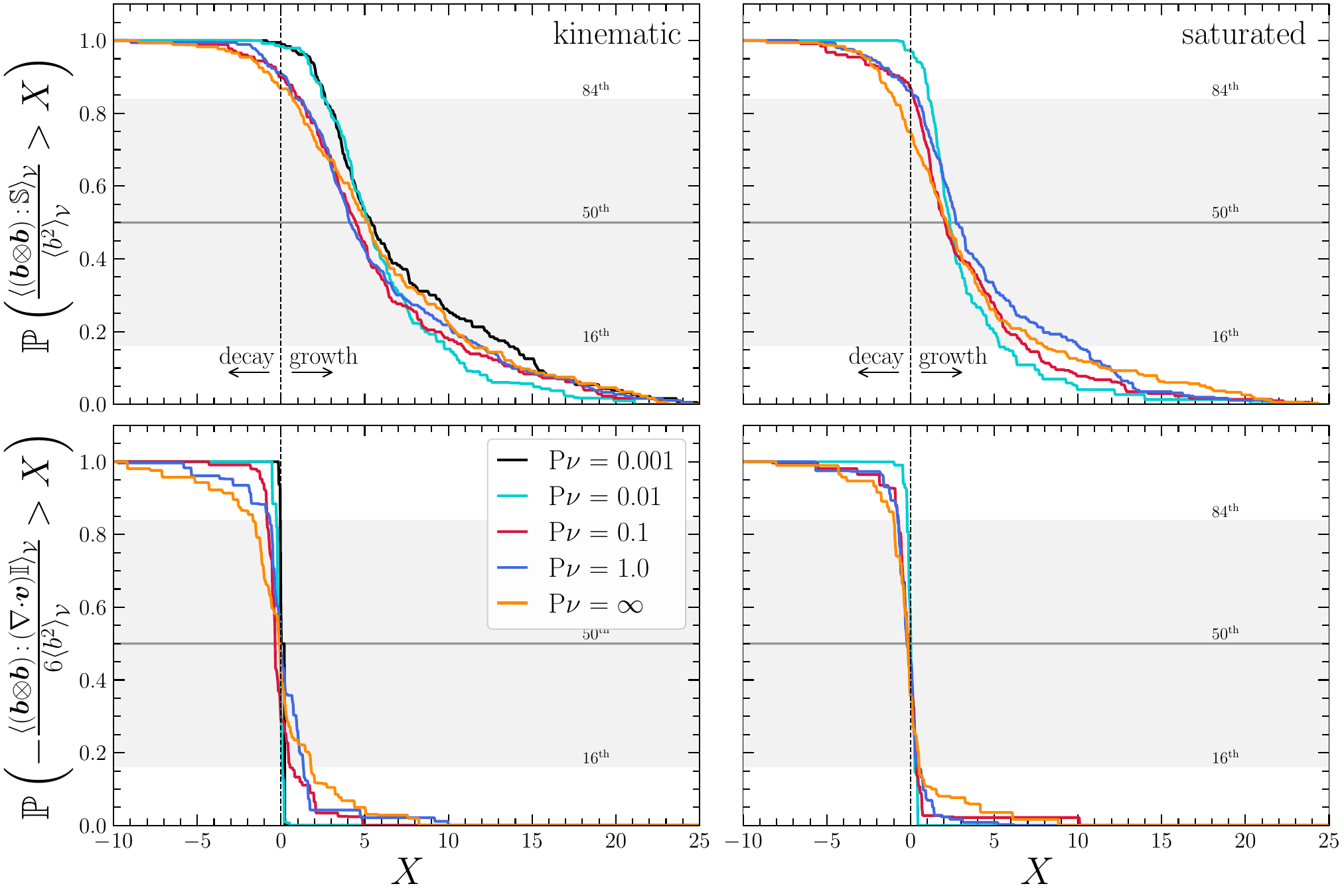}
        \caption{The complementary cumulative density functions \autoref{eq:ccdf} for the stretching growth term ($\Exp{(\vecB{b}\otimes\vecB{b}):\St}{\V}/\Exp{b^2}{\V}$; top panels) and the compression growth term ($\Exp{(\vecB{b}\otimes\vecB{b}):(\bnab\cdot\vecB{v})\It}{\V}/(6\Exp{b^2}{\V})$; bottom panels), respectively, throughout the kinematic stage (left column) and saturated state (right column) of the dynamo, binned by the $t/t_0$ ranges in \autoref{fig:integral_energies}. We use vertical, dashed, black lines to separate the values into growth and decay states, as we did in \autoref{fig:growth}, as well as adding horizontal lines to highlight the $16^{\rm th}$, $50^{\rm th}$ and $84^{\rm th}$ percentiles of a standard Gaussian distribution function. We find that the stretching terms dominate the growth states, growing the field for all quantiles above the $16^{\rm th}$ percentile, while the compression terms are slightly more predisposed to decay states. Both growth terms contract at the onset of saturation (become closer to Heaviside functions), and the stretching term shifts towards $X = 0$ by a factor of $\approx 2$, recovering the result that growth terms are suppressed during the onset of saturation \citep[e.g.,][]{Schekochihin2004_dynamo_saturation_via_anisotropisation,Seta2021_saturation_supersonic_dynamo,Sur2023_dynamo}.}
        \label{fig:growth_cdfs}
    \end{figure*}

        In \autoref{fig:growth} we plot the temporal probability distribution functions of each of the three volume-averaged growth terms in \autoref{eq:growth_rate_decomp}, indicated by changing the line style of the distribution function. In the top plot, we show the distributions in the kinematic stage, and the bottom plot, the saturated stage, maintaining the same time window as we did in the previous section and shown in \autoref{fig:integral_energies}. Furthermore, we show the cumulative density function (CDFs),
        \begin{align}\label{eq:ccdf}
            \mathbb{P}(x > X) = 1 - \int_{-\infty}^{X} \d{x}\;p(x),
        \end{align}
        for the $x = \Exp{(\vecB{b}\otimes\vecB{b}):\St}{\V}/\Exp{b^2}{\V}$ (top rows) and $x = \Exp{(\vecB{b}\otimes\vecB{b}):(\bnab\cdot\vecB{v})\It}{\V}/(6\Exp{b^2}{\V})$ (bottom rows) for each simulation in the kinematic stage (left columns) and saturated stage (right columns) in \autoref{fig:growth_cdfs}.
        
        Let us begin in the kinematic stage, focusing on the top panel of \autoref{fig:growth}. Firstly, both simulations show that the $\Exp{(\vecB{b}\otimes\vecB{b}):\At}{\V}/\Exp{b^2}{\V}$ distribution function (dotted line) is identically zero, as we predicted from \autoref{eq:no_winding_growth}. This means that the tensors are on average orthogonal to one another. Because the contribution from the winding term is therefore zero, we ignore it for the remainder of the study and focus on the other two tensor projections.  

        The next largest contribution to the growth rate is $\Exp{(\vecB{b}\otimes\vecB{b}):(\bnab\cdot\vecB{v})\It}{\V}/(6\Exp{b^2}{\V})$, which is almost symmetric about zero. This means, on average, in the kinematic stage, compressions do not have a significant contribution to the growth rates at our $\M$. However, temporal fluctuations, which are approximately Gaussian, show that the compressions make contributions to both growing the $\emagb$ and decaying $\emagb$ in approximately equal amounts, most likely due to the fact that $\Exp{\bnab\cdot\vecB{v}}{\V} = 0$, by definition. This highlights the dual role that compressible modes have in the dynamo problem -- approximately\footnote{Indeed, any net effect to explain the changes to the growth rates and $\emagb/\ekinb$ saturation in \citet{Federrath2011_mach_dynamo} and more recently \citet{Sur2023_dynamo} and \citet{Kriel2025_supersonic_scales}, at least from a mathematical perspective, must be that changing $\M$ shifts $\Exp{(\vecB{b}\otimes\vecB{b}):(\bnab\cdot\vecB{v})\It}{\V}/(6\Exp{b^2}{\V})$ towards more negative values due to the increasing skewness, $\Exp{(\bnab\cdot\vecB{v})^3}{\V}$ (first or second moments will not change $\gamma$) as the strongest, positive divergence structures become less and less space-filling, further inhibiting the growth and saturation, on average. In our study, we focus on a single $\M$ and the effects of bulk viscosity, so we leave testing this hypothesis for future studies that utilise our decomposition.} equally decaying and growing the field at $\M\approx 1$. As $\Pnu$ gets smaller (and $\Reb$ gets smaller), the fluctuations of the compressions reduce (the CDFs become closer to Heaviside functions in \autoref{fig:growth_cdfs}), but the time-average is maintained close to zero. This describes a process similar to what \citet{Chen2019_bulk_visc_homogenous_turb} found for the effects of bulk viscosity on the spatial $\bnab\cdot\vecB{v}$ distribution function -- a somewhat symmetric contraction of the $\bnab\cdot\vecB{v}$ distribution function that maintains the $\first$ moment. Hence, as the bulk viscosity increases, any fluctuations in the growth rate caused by either strong compression or rarefaction events diminish, but the average contribution of the compressions is maintained, hence the relatively weak dependence upon the growth rate for moderate $\Reb$. This is a key result from this analysis. 
    
        To understand why at very low $\Reb$ we find that $\gamma_1$ changes quite dramatically, we turn our attention to the growth rate associated with stretching $\Exp{(\vecB{b}\otimes\vecB{b}):\St}{\V}/\Exp{b^2}{\V}$. Because the distribution function (solid line in \autoref{fig:growth}) is on average positive, the interaction between the magnetic field and the $\St$ structures on average grow the field exponentially from \autoref{eq:magnetic_energy_m1}. As is evident in \autoref{fig:growth_cdfs}, its not only the average, but as low as the $16^{\rm th}$ percentile of $\Exp{(\vecB{b}\otimes\vecB{b}):\St}{\V}/\Exp{b^2}{\V}$ that is able to grow the magnetic field. This is completely consistent with the key idea behind kinematic turbulent dynamo theories -- stretching plays the dominant growth role in the kinematic stage of the dynamo \citep[e.g.,][]{Moffatt1969_fast_dynamo_growth,Haugen2004_random_stretching,Zeldovich1984_fast_dynamo_growth,Brandenburg1995_magnetic_flux_tubes}, which we show is even the case for the supersonic dynamo. This is why \citet{Kriel2025_supersonic_scales} was able to show numerically that the $k_{\eta}/k_{\nus} \sim \Pms^{1/2}$ relation holds even in $\M > 1$ fluids.
    
        Finally, we turn to the bottom panel of \autoref{fig:growth} and the right column in \autoref{fig:growth_cdfs} to explore the differences between the kinematic and saturated stages of the dynamo under these tensorial decompositions. Compared to the kinematic stage, the $\Exp{(\vecB{b}\otimes\vecB{b}):\St}{\V}/\Exp{b^2}{\V}$ terms both shift on average, with the $50^{\rm th}$ percentiles moving towards zero, reducing by $\approx 2$ across all $\Pnu$, and contract, best shown in the bottom panel of \autoref{fig:growth}. The $\Exp{(\vecB{b}\otimes\vecB{b}):(\bnab\cdot\vecB{v})\It}{\V}/(6\Exp{b^2}{\V})$ term similarly contracts in the saturated stage. \citet{Seta2021_saturation_supersonic_dynamo} found a similar trend, but their $\M$ was significantly higher than ours and the difference between the saturated and kinematic stage statistics were not as significant (we find a reduction of roughly a factor of two in the $\first$ moments for $\Exp{(\vecB{b}\otimes\vecB{b}):\St}{\V}/\Exp{b^2}{\V}$). In their decomposition, at $\M = 3$, \citet{Sur2023_dynamo} also find a factor of $\approx 2$ reduction in the $\first$ moments between the kinematic and the saturated stages. This highlights that our decomposition retains the well-known result that the growth terms are suppressed in the dynamo saturation.

    \begin{figure}
        \centering
        \includegraphics[width=\linewidth]{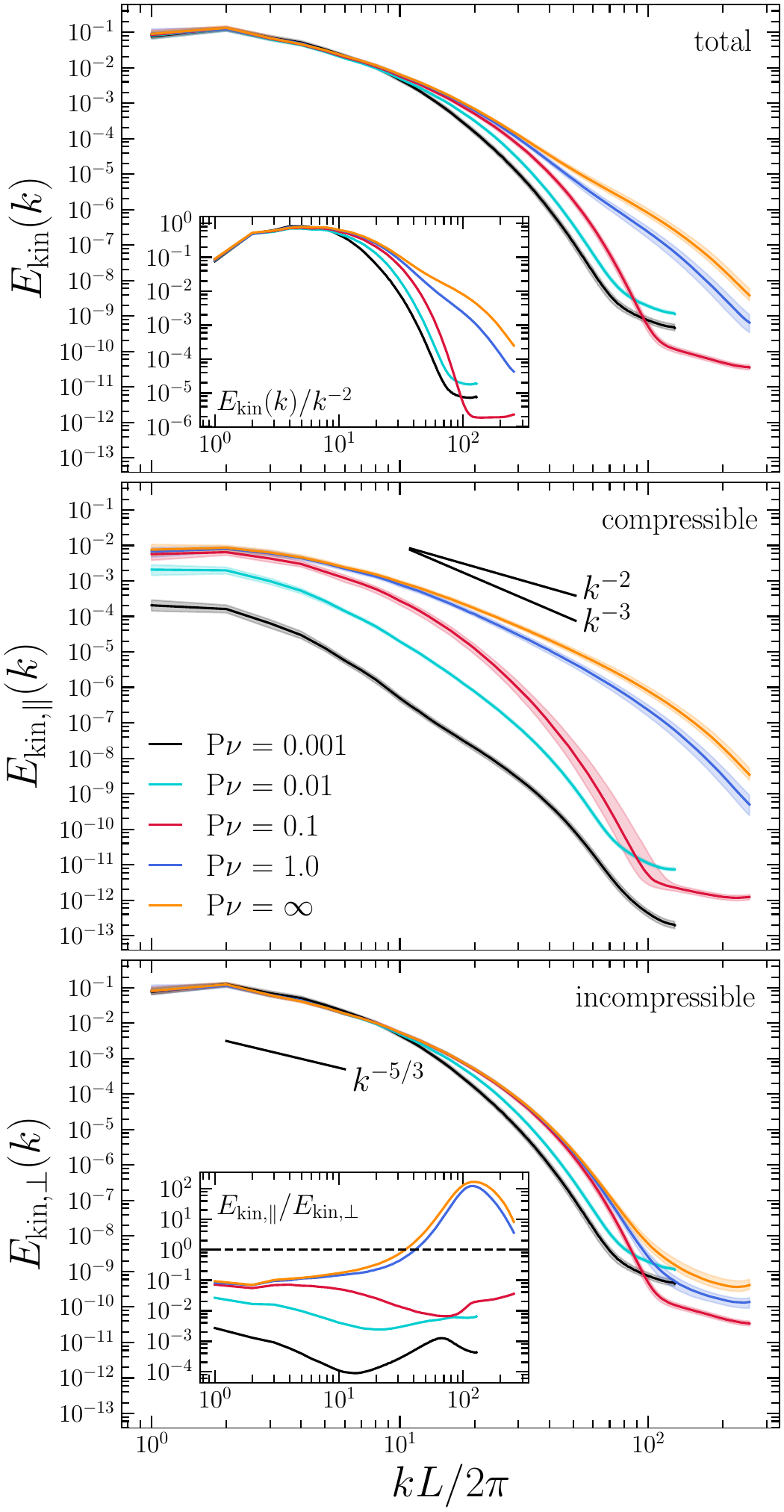}
        \caption{The shell-integrated kinetic energy spectra averaged over the kinematic stage of the dynamo. \textbf{Top:} The total kinetic energy spectrum, coloured by $\Pnu$, averaged over the kinematic regime (see \autoref{fig:integral_energies}), with inset showing flattening of the spectra under the \citet{Burgers1948} scaling, $\ekin(k) \sim k^{-2}$. The bulk viscosity acts to suppress the high-$k$ mode structure in the total energy, suggestive that compressible modes occupy significant energies on the smallest scales, contributing to a bottleneck effect. \textbf{Middle:} The compressible (longitudinal) kinetic energy spectrum, demonstrating that for $\Pnu > 1$, the dissipation in the compressible modes is not resolved or controlled until $\Pnu \approx 0.1$ (a resolved exponential cut-off), and that in general, $\Pnu$ principally controls the dissipation of the compressible modes. We annotate both the $k^{-2}$ \citet{Burgers1948} spectrum for shocks, and a $k^{-3}$ spectrum for the pseudo-acoustic mode spectrum \citep{Wang2017_psuedosound_spectrum}. \textbf{Bottom:} The incompressible (transverse) kinetic energy spectrum with the ratio of the $\ekinpar(k)$ and $\ekinperp(k)$ in the inset. For fixed $\Res$, reducing $\Pnu$ suppresses the high-$k$ compressible modes, acting more strongly on the compressible spectrum for lower $\Pnu$. The incompressible spectrum is also influenced, with much less of an effect, with coupling starting deep in the viscous scales, but slowly moving to lower $k$ modes as the decay timescale, $t_{\nub}$, becomes shorter.}
        \label{fig:decomp_vel_spect}
    \end{figure}

\section{Kinetic Energy spectra}\label{sec:kin_spectrum}

    \subsection{Total spectra}\label{sec:total_spectra}
        We plot the total kinetic energy spectra in the top panel of \autoref{fig:decomp_vel_spect}, coloured in the same fashion as the previous plots. At low-$k$, all simulations have similar spectral energy contained on those scales, which flattens under a $\ekin(k)\sim k^{-2}$ \citet{Burgers1948} scaling as shown in the inset panel, as expected for supersonic turbulence dominated by singularities \citep{Federrath2013_universality,Verma2017,Federrath2021,Kriel2025_supersonic_scales}. However, as expected for a viscosity-like process (a process that is scale-separated from the large-scales and starts by influencing the small-scale dynamics), there is a significant difference between simulations in the high-$k$ modes. Because $\second$ moments have integrals dominated by low-$k$ modes \citep{Beattie2022_ion_alfven_fluctuations}, this highlights why the $\emagb/\ekinb$ did not have a strong impact from the bulk viscosity in \autoref{sec:integral_quants}.
        
        For both the $\Pnu = 1.0$ and $\Pnu = \infty$ simulation, we observe an excess amount of power piling up at high-$k$, which could be qualitatively described as a bottleneck effect. This is what we would call the sub-shear-viscous range \autoref{eq:sub-shear-bulk-viscous_range}, which ought to appear when $\Pnu > 1$, as in the right panel of \autoref{fig:spectra}. However, as $\Pnu$ decreases to $\leq 0.1$, the sub-shear-viscous range and bottleneck completely disappear and we are left with a smooth, rather featureless decay into dissipatively-dominated, low-power Fourier energy modes, as suggested in the left panel of \autoref{fig:spectra}. That is, once the bulk viscosity is large enough ($\Reb$ and $\Pnu$ are low enough, which ought to be $\Pnu \approx 1/3$ based on our definition; see \autoref{footnote:Pnudefn}), the physical dissipation of them allows them to be removed faster than they are able to move down the cascade (which is fast $\ekin(k) \sim k^{-2}$; \citealt{Seta2021_saturation_supersonic_dynamo,Federrath2013_universality,Federrath2021,Kriel2025_supersonic_scales}), and hence the bottleneck and sub-shear-viscous range is destroyed. 

    \subsection{Helmholtz decomposed spectra}\label{sec:decomposed_spectra}
        As we showed with the integral statistics, to truly observe the $\Pnu$ effect on the plasma we plot the Helmholtz decomposed spectra in the middle (compressible) and bottom (incompressible) panels in \autoref{fig:decomp_vel_spect}, also showing the ratio between the two spectra in the inset in the bottom panel. Confirming what we discussed in the previous section, for $\Pnu > 1$, $\ekinpar(k)/\ekinperp(k)$ varies from $\ekinpar(k)/\ekinperp(k) \approx 10^{-1}$ on large-scales, to $\ekinpar(k)/\ekinperp(k) \approx 10^2$ on small-scales, highlighting how energetically dominant the compressible modes become in the sub-shear-viscous range when $\Pnu$ is large (or bulk viscosity is not included at all). Once $\Pnu < 1$, the apparent bottleneck and sub-shear-viscous range disappear, and $\ekinpar(k)/\ekinperp(k)$ becomes relatively scale-invariant, with some systematic shallowing from large to small scales, indicating that we are now resolving the viscous truncation of the compressible modes. Now we turn to the individual $\ekinpar(k)$ and $\ekinperp(k)$ spectra.
        
        As expected, the $\ekinpar(k)$ spectrum (middle panel of \autoref{fig:decomp_vel_spect}) is strongly influenced by $\Pnu$, with the dissipative effects beginning at high-$k$ and moving towards low-$k$ as the $\Pnu$ decreases. This is expected since $\nub\bnab\cdot(\bnab\cdot\vecB{v})\It = \nub\bnab\cdot(\bnab\cdot\vpar)\It 
        =  \nub \bnab^2 \vpar$ ends up having the local properties of a Laplacian operator on the $\vpar$ modes in the compressible mode momentum flux (see \aref{app:energy_equations} where we derive the viscous-scale momentum equations for each mode). As discussed in \autoref{sec:growth}, in the lowest $\Pnu$, where $t_{\nub} \sim t_0$, the low-$k$ modes are significantly suppressed by roughly $10^{2}$ in total power compared to the $\Pnu = \infty$ simulation, indicating that the viscosity is strong enough to now even act on the outer scales of the turbulence. 

        As we suggested for the $\M \approx 1$ regime in \autoref{fig:spectra}, in the top panel of \autoref{fig:decomp_vel_spect} we find $\ekinpar(k) \sim k^{-2}$ is consistent with the low-$\nub$ simulations, as opposed to $\ekinpar(k) \sim k^{-3}$ found in \citet{Wang2017_psuedosound_spectrum} and \cite{Yoshiki2023_compressible_turbulence_spectrum}. Further supporting the schematic in \autoref{fig:spectra}, we observe roughly the \citet{Kolmogorov1941} spectrum, $\ekinperp(k) \sim k^{-5/3}$, over a very truncated range of incompressible modes in the bottom panel of \autoref{fig:decomp_vel_spect}.
        
        In \autoref{fig:growth} and \autoref{fig:growth_cdfs}, we saw that the $\second$ moments of both the shear and compression growth terms, \autoref{eq:growth_rate_decomp}, contracted and the CDFs steepened under the impact of the bulk viscosity. We hypothesised that this may be due to the correlation between compressible and incompressible modes. We now see another signature of the mode correlation via $\ekinperp(k)$ (bottom panel of \autoref{fig:decomp_vel_spect}), where a Laplacian-like viscosity (starting at high-$k$) encroaches upon the spectra, and more strongly as $\Pnu$ decreases. This corresponds to a growing sub-bulk-viscous range of $k$, as we illustrated in \autoref{fig:spectra}. It is suggestive that in this $\Pnu < 1$ regime $k_{\nus} \sim k_{\nub}$, and\footnote{In that $k_{\nus}$ depends upon $k_{\nub}$, and the dependence does not need to be linear. However, we leave the determination of such relations for future works.} that the most viscous $\vpar(k)$ and $\vprp(k)$ modes are becoming increasingly coupled as the sub-bulk-viscous range increases in size. 
        
        However, we show explicitly in \aref{app:energy_equations} that the $\vpar(k)$ and $\vprp(k)$ become independent at the viscous scales, with integral energy equations in the Stokes limit
        \begin{align}
            \frac{\partial \ekinparb}{\partial t} &= (\Res^{-1} + \Reb^{-1}) \Exp{\bnab \otimes\vecB{v}_{\parallel}:\bnab \otimes\vecB{v}_{\parallel}}{\V}, \label{eq:compressible_kin} \\
            \frac{\partial \ekinperpb}{\partial t} &= \Res^{-1} \Exp{\bnab \otimes\vecB{v}_{\perp}:\bnab \otimes\vecB{v}_{\perp}}{\V}, \label{eq:incompressible_kin}
        \end{align}
        that show no explicit dependence between $\ekinpar$ and $\ekinperp$ on these scales. Hence, the correlations we find are not directly caused by the viscous modes themselves. A similar type of correlation was previously shown in two dimensions in \citet{Touber2019_2d_turbulence}, where they found that in the viscous limit bulk viscosity would modify the small-scale enstrophy production (in the opposite way for two dimensions), which indeed is likely the same effect as we are observing, and highlights that in three dimensions the dominant effect of bulk viscosity is to dissipate dilatational kinetic energy, reducing the energy available for small-scale enstrophy production through $\bm{\omega}(\bnab\cdot\vecB{u})$. We explore this correlation in more detail in the following section. 

    \begin{figure}
            \centering
            \includegraphics[width=\linewidth]{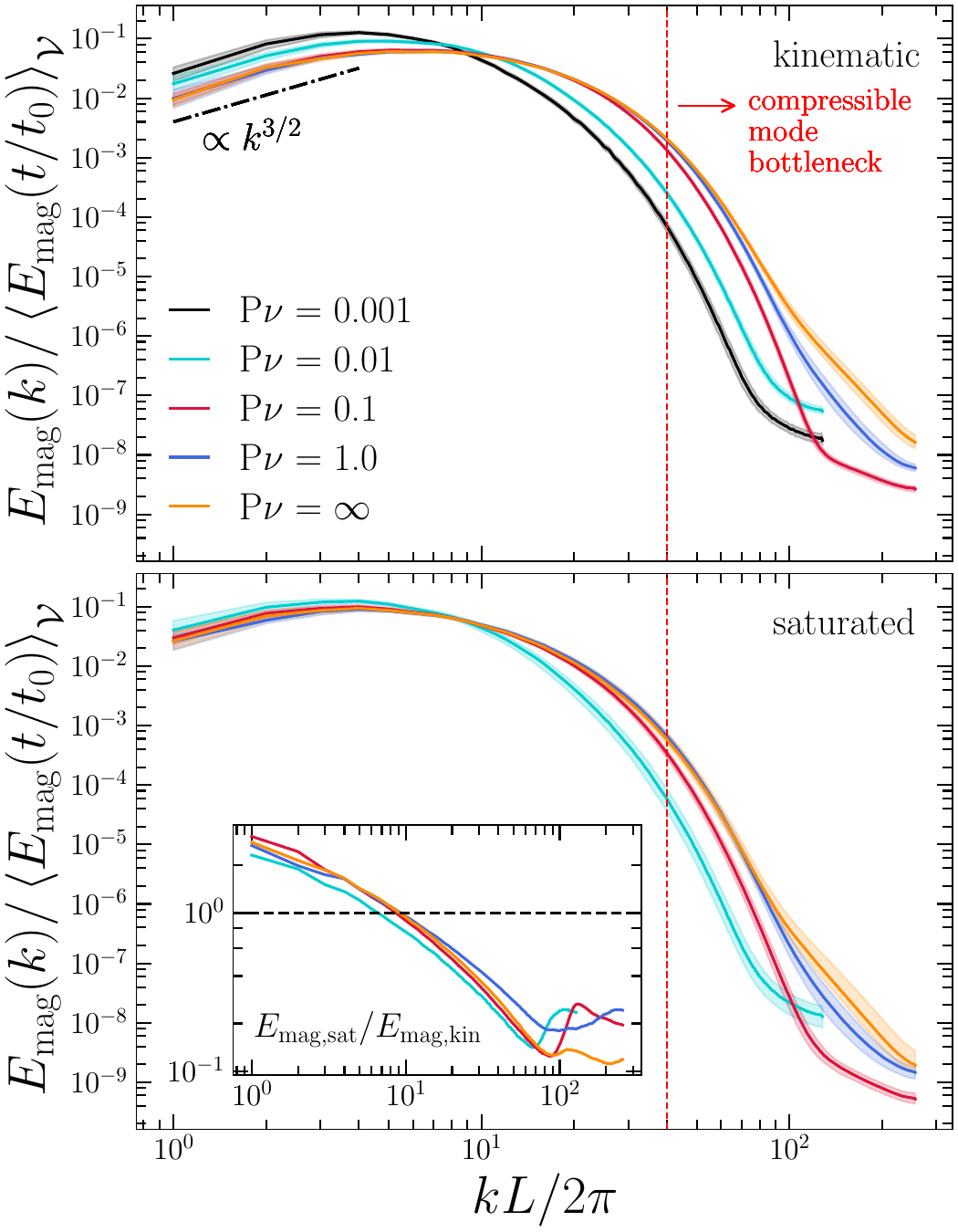}
            \caption{The same as \autoref{fig:decomp_vel_spect}, but for the magnetic energy spectrum normalised by time-dependent integral magnetic energy. \textbf{Top:} The normalised magnetic energy spectra shown in the kinematic stage of the dynamo. We show the \citet{Kazantsev1968} $\emag(k) \sim k^{3/2}$ power law in the top panel, which all simulations, regardless of $\Pnu$, exhibit up until the peak of the energy spectra. \textbf{Bottom:} The same as the top panel but for the saturated stage of the dynamo. In the inset panel we show the ratio between the normalised saturated and kinematic magnetic field spectra, highlighting the different spectral distributions of energy in each of the regimes (more large-scale in the saturated stage, and vice-versa in the kinematic stage). With a red dashed, vertical line, we show the extent of the apparent compressible mode bottleneck (the sub-shear-viscous range) from the inset in \autoref{fig:decomp_vel_spect} (only present in simulations with $\Pnu > 0.1$). This shows that the suppression of the bottleneck and the sub-shear-viscous range through the bulk viscosity results in a suppression of high-$k$ modes in $\emag(k)$. }
            \label{fig:mag_spect}
        \end{figure}
    
        \begin{figure}
            \centering
            \includegraphics[width=\linewidth]{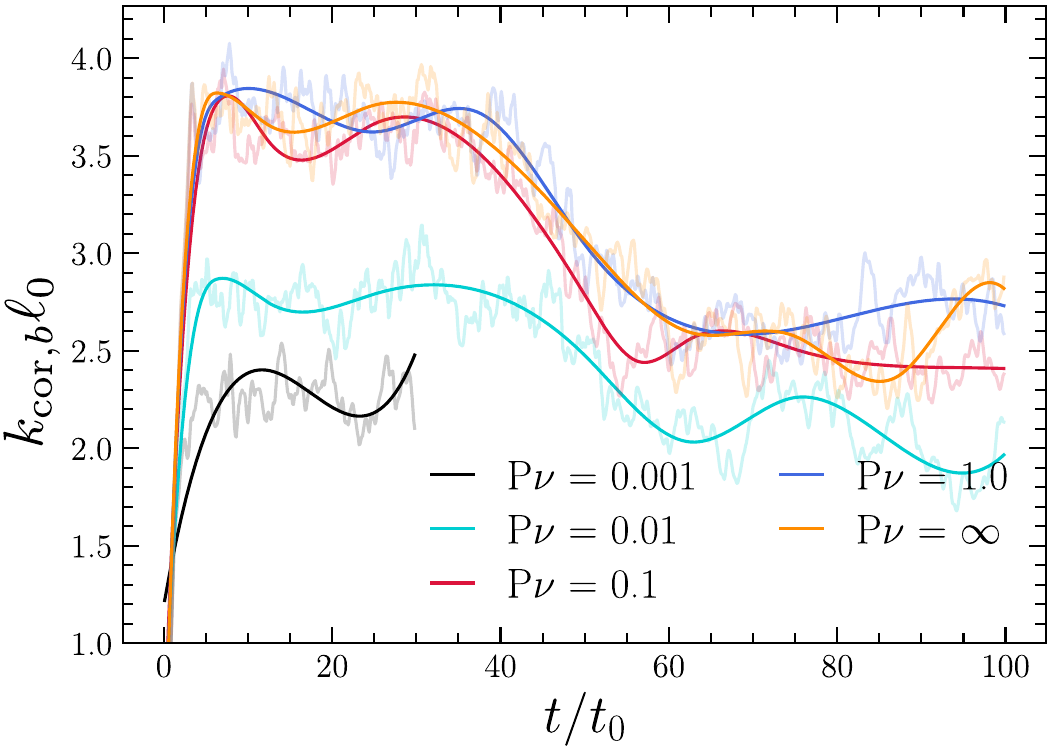}
            \caption{The integral scale of the magnetic field $k_{{\rm cor},b}$ in units of the driving scale $\lo$, coloured by $\Pnu$. As has been shown before, the kinematic stage of the dynamo quickly builds a small-scale field ($k_{{\rm cor},b}\ell_0 > 1$), likely though magnetic field-folding, which is present in both $\M < 1$ and $\M > 1$ dynamos \citep{Schekochihin2004_dynamo,Kriel2025_supersonic_scales}. However, as the nonlinear stage approaches $\sim 20t_0$ the integral scale shifts to lower modes (larger scales) before becoming stationary again at those larger scales in the saturated stage. The effect of the bulk viscosity suppresses the small-scale field in the kinematic stage of the dynamo, by correlating to the viscous solenoidal modes $k \sim k_{\nus}$ responsible for growing the magnetic field.}
            \label{fig:b_integral_scale}
        \end{figure}

\section{Magnetic energy spectra}\label{sec:mag_spectra}
\begin{figure}
        \centering
        \includegraphics{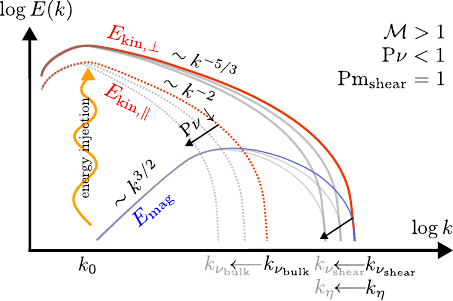}
        \caption{The same as the left panel of the spectrum schematic in \autoref{fig:spectra}, but showing that in the $\Pnu < 1$ regime, as $\Pnu$ decreases, $k_{\nub}$ shifts to lower $k$ modes, and through the viscosity correlations in the sub-bulk-viscous range, $k_{\nub}$ drags $k_{\nus}$ with it, which then shifts $k_{\eta}$, since $k_{\eta}$ depends upon the exact placement of $k_{\nus}$, $k_{\eta} \sim \Pms^{1/2}k_{\nus}$ \citep[for latest numerical confirmation, see, e.g., ][]{Kriel2022_kinematic_dynamo_scales,Kriel2025_supersonic_scales}.}
        \label{fig:spectra_diagram_with_cor}
    \end{figure}

    We plot the magnetic spectra normalised by the integral, and averaged over time, in \autoref{fig:mag_spect}, showing the kinematic stage spectra in the top panel, and the saturated spectra in the bottom panel, along with the ratio between the spectra in the inset plot. Firstly, it is apparent that the magnetic fields for the low $\Pnu$ simulations are becoming more large-scale (high-$k$ modes are being suppressed), during both the kinematic stage, where they follow the \citet{Kazantsev1968} $k^{3/2}$ power law, and in the saturated stage. We can quantify this more precisely by computing the magnetic integral (or correlation) scale. For $\emag(k)$, it is
        \begin{align}
            k_{{\rm cor},b}\lo = \frac{L}{2}\left(\frac{ \displaystyle \int \d{k} \, k^{-1} \emag(k)  }{ \displaystyle \int \d{k} \, \emag(k) }\right)^{-1},
        \end{align}
    which we compute for each time realisation, for each of our simulations, and plot across time in \autoref{fig:b_integral_scale}, fitting cubic splines to the data to extract the general low-frequency trends. We see the well-established trend that the field is small-scale ($k_{{\rm cor},b}\lo < 1$) in the kinematic stage, before becoming correlated on large-scales, reducing $k_{{\rm cor},b}\lo$ through the nonlinear and finally settling on $k_{{\rm cor},b}\lo \approx 2.5$ in the saturated stage \citep{Schekochihin2004_dynamo,Rincon2019_dynamo_theories,Seta2021_saturation_supersonic_dynamo,Galishnikova2022_saturation_and_tearing,Beattie2023_GorD_P1}. The energy redistribution can be seen on a mode-by-mode basis in the inset plot in the bottom panel of \autoref{fig:mag_spect}, where there is now a factor of $\approx 2$ more energy concentrated at low-$k$, and an order of magnitude less at high-$k$. For $\Pnu < 1$, the magnetic fields lose energy on high-$k$, potentially through the suppression of the sub-shear-viscous modes and compressible mode bottleneck, which most likely grow the field through the $\bm{\omega}(\bnab\cdot\vecB{v})$ coupling we discussed in the previous section. We annotate the rough extent of this range with red in  \autoref{fig:mag_spect}. Furthermore, as expected from \autoref{fig:mag_spect}, we find that for $\Pnu < 0.01$, where we know the compressible/incompressible mode coupling is strong, the correlation scales are pushed closer to $\lo$. This could is likely because $\ell_{\nus}$ shifts to lower $k$ modes through the coupling, which in turn reorganises the magnetic field for a fixed $\Pms$, since $k_\eta/k_{\nus} \sim \Pms^{1/2}$ \citep{Schekochihin2002_large_Pm_dynamos,Kriel2022_kinematic_dynamo_scales,Kriel2025_supersonic_scales}, and as we have discussed in \autoref{sec:decomposed_spectra}, $k_{\nus} \sim k_{\nub}$ for low $\Reb$. The key takeaway is that the bulk viscosity suppresses magnetic energy in the high-$k$ modes due to the suppression of the sub-shear-viscous range through the high-$k$ correlations, in both the stages of the dynamo. We annotate how this may work in \autoref{fig:spectra_diagram_with_cor}, where we show $k_{\nub}$ shifting to lower $k$ modes, which in turn moves $k_{\nus}$, shifting $k_\eta$, destroying the high-$k$ magnetic modes \citep{Kriel2025_supersonic_scales}. The net effect of this process is that  magnetic fields are significantly more large-scale than the pure shear-viscous dynamo counterparts.

    \begin{figure*}
        \centering
        \includegraphics[width=\linewidth]{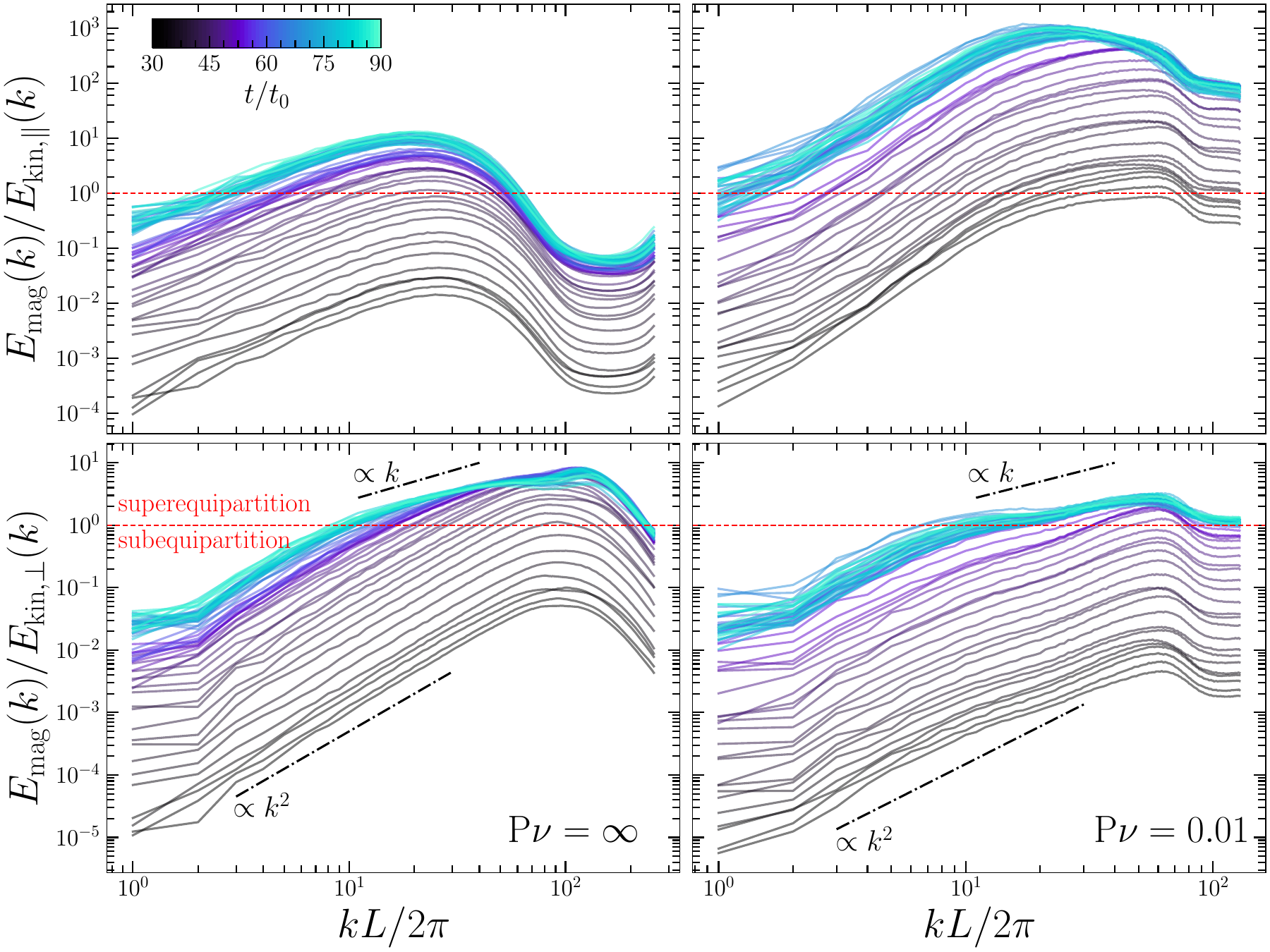}
        \caption{The magnetic $\emag(k)$, compressible ($\ekinpar(k)$; top) and incompressible ($\ekinperp(k)$; bottom) kinetic energy ratios as a function of $k$ for $\Pnu = \infty$ (no bulk viscosity; left column) and $\Pnu = 0.01$ (strong bulk viscosity; right column), coloured by $t/t_0$ (indicated in the top left panel). We start at $t/t_0 = 30$, at the end of the kinematic stage (black), and evolve to $t/t_0 = 90$, well within the saturated stage (aqua; see \autoref{fig:integral_energies} for how $t/t_0$ corresponds to the different stages). Each panel has the equipartition $\emag(k) = \ekin(k)$ drawn in red. \textbf{Compressible modes (top row):} for $\Pnu = \infty$, the saturated state for the compressible modes is in subequipartition ($\emag(k)/\ekinpar(k) < 1$) at low and high-$k$. The high-$k$ mode subequipartition energy is associated with the sub-shear-viscous range, outlined in \autoref{fig:spectra}. However, for $\Pnu = 0.01$, the sub-shear-viscous range is destroyed through the bulk viscosity, and only the low-$k$ modes are left in subequipartition, meaning that all but the lowest compressible $k$ modes ought to be dominated by the effects of the magnetic field. \textbf{Incompressible modes (bottom row):} during the kinematic stage, $\emag(k) \sim k^2\ekinperp(k)$ up to the peak scale of the spectral energy ratio, previously shown in \citet{Beattie2023_GorD_P1}. As the dynamo saturates, the energy ratio flattens to roughly $\emag(k) \sim k\ekinperp(k)$ for superequipartition $k$ modes ($\emag(k)/\ekinperp(k) < 1$), potentially due in part to $\bnab\cdot(\vecB{b}\otimes\vecB{b})$ suppressing the field-line stretching term in \autoref{eq:growth_rate_decomp}.}
        \label{fig:energy_ratio}
    \end{figure*}
    
\section{Spectral ratio functions}\label{sec:spectral_ratio}
    The final statistic we use to probe the role of bulk viscosity in the compressible turbulent dynamo are the ratios between the magnetic and compressible and incompressible kinetic energies as a function of time, \autoref{fig:energy_ratio}, revealing scale-by-scale information about the transition between the stages of the dynamo (see \autoref{sec:intro}), as previously done in \citet{Beattie2023_GorD_P1}. To further build upon \citet{Beattie2023_GorD_P1}, we split the saturation into compressible $\ekinpar(k)/\emag(k)$ (top panels) and incompressible $\ekinperp(k)/\emag(k)$ (bottom panels), with the columns organised by $\Pnu$, with $\Pnu = \infty$ (left) and $\Pnu = 0.01$ (right). We take the spectral ratios from $t/t_0 = 30$, which is within the kinematic stage, to $t/t_0 = 90$, which is well and truly saturated (see \autoref{fig:integral_energies}). The intermediate purple stage is within the nonlinear transition stage between the kinematic and saturation. We draw a dashed, red line at equipartition, $\ekin(k)/\emag(k) = 1$, which defines an energy equipartition scale, $k_{\rm eq}$ such that $\ekin(k_{\rm eq})/\emag(k_{\rm eq}) =  1$, or $\sim$ stretching\footnote{Note that this is slightly different, since the stretching scale knows about integrated magnetic energy below it \citep{Schekochihin2004_dynamo,Galishnikova2022_saturation_and_tearing}, whereas this is the instantaneous energy equipartition scale on a particular scale.} scale \citep{Schekochihin2004_dynamo}, to describe the onset of saturation (as shown in \autoref{eq:linear_growth_b_field_eq}).

    Let us first focus on the left column for the $\Pnu = \infty$ simulations. The bottom panel shows the onset of the $\emag(k)$ saturation w.r.t, the incompressible modes, which we can directly compare with the incompressible, $\Pms = 1$ simulations in \citet{Beattie2023_GorD_P1}. As \citet{Beattie2023_GorD_P1} showed previously, in the kinematic stage there is an extended power law $\emag(k) \sim k^2 \ekinperp(k)$ up to the peak of $\emag(k)/\ekinperp(k)$. Throughout the nonlinear stage, the modes $k > k_{\rm eq}$ are suppressed, in that $\emag(k) \sim k^2\ekinperp(k)$ becomes significantly shallower, close to $\emag(k) \sim k \ekinperp(k)$. Based on our analysis in \autoref{sec:growth}, this must be when the field growth is suppressed when the stretching is being suppressed, which is presumably happening on these scales. As the dynamo saturates, $k_{\rm eq}$ progressively moves from the peak of $\emag(k)/\ekinperp(k)$, $k \approx 100$, to $k\approx 10$, where it settles. $\hat{\vecB{b}}\otimes\hat{\vecB{b}}:\bnab\otimes\vecB{v}$ should still be operational on these subequipartition scales, but now the stretching is slow, since the stretching time scale $t_k \sim 1/(k v_{\perp,k})$ becomes longer towards lower $k$ modes \citep{Schekochihin2004_dynamo_saturation_via_anisotropisation}. In the top panel, we see the same spectral ratio shape and final saturated state is not retained in the compressible modes, as we expect based on our integral statistics in \autoref{sec:integral_quants}.

    Firstly, whilst $\emag(k)/\ekinpar(k)$ retains similar power-law like structure into the peak as $\emag(k)/\ekinperp(k)$, the peak of $\emag(k)/\ekinpar(k)$ is shifted to much smaller modes $k \approx 30$, and now a significant range of scales, the sub-shear-viscous scales, emerge at high-$k$ with $\emag(k)/\ekinpar(k) \approx 0.1$. Moreover, there is no clear change in the shape of $\emag(k)/\ekinpar(k)$ as the dynamo saturates, in contrast to the suppression we see in the bottom panel, i.e., compressible modes experience only a very weak magnetic backreaction. Naturally, there are two $k_{\rm eq}$, the low-$k$ $k_{\rm eq}$ corresponding to the transition between compressible modes that are in subequipartition, and hence strong enough to influence the field, and then the high-$k$ $k_{\rm eq}$, corresponding to the onset of the sub-shear-viscous range. Sandwiched in between are a range of scales that are in superequipartition, making the scale-by-scale saturation quite complex in the compressible modes of the $\Pnu \gg 1$ plasmas. 

    Finally, we turn to the $\Pnu = 0.01$ simulations in the right panels of \autoref{fig:energy_ratio}. For $\emag(k)/\ekinperp(k)$, a similar $k^2 \rightarrow k$ transition is found through the evolution from the kinematic to saturated state. From \autoref{fig:decomp_vel_spect}, we showed that $\ekinperp(k)$ modes $k\gtrsim 10$ are influenced by viscosity coupling. Compared to the $\Pnu = \infty$ simulation, these modes are suppressed, with $\emag(k)/\ekinperp(k) \approx 2$ at the peak energy scale in saturation. It is also clear that $k_{\rm eq}$ takes a slower journey towards the low-$k$ modes, albeit saturating to a similar value, which is probably the reason why that even at this low $\Pnu$, the integral statistics shown in \autoref{sec:integral_quants} do not vary significantly from the $\Pnu = \infty$ case. Because $\Pnu$ is small, the $\emag(k)/\ekinpar(k)$ ratio is notably different from the incompressible counterpart. No sub-shear-viscous range exists, and hence the high-$k$ modes are highly magnetised with respect to the compressible modes, finding values of $\emag(k)/\ekinpar(k) \approx 10^3$. Furthermore, $k_{\rm eq}$ saturates to $k_{\rm eq}\approx 1$. As we commented upon previously, the key take away is that both the journey towards and the role in the saturation of the dynamo varies significantly for the compressible and incompressible velocity modes. 

\section{Summary and Conclusions}\label{sec:conclusion}
    We perform three-dimensional, visco-resistive direct numerical simulations of trans-sonic, magnetohydrodynamic turbulence undergoing magnetic field amplification and saturation through the small-scale dynamo, including viscous effects from both the shear and bulk viscosity. We list the conclusions below.

    \subsection{Bulk viscosity in the turbulent dynamo}
    To appropriately place bulk viscosity into a small-scale dynamo framework, we derive and define two new dimensionless numbers to study the role of bulk (volume) viscosity, the ``bulk viscous Reynolds number" $\Reb$, \autoref{eq:Rebulk_defn}, which is the ratio between Reynolds stress and the bulk viscosity, as well as the ``viscous Prandtl number" $\Pnu$, \autoref{eq:Pnu_defn}, which is the ratio between the bulk and shear viscosities (or equivalently, Reynolds numbers). For $\Pnu < 1$, $\Pnu$ defines the size of a sub-bulk-viscous range, $k_{\nus} > k > k_{\nub}$, and for $\Pnu > 1$, the sub-shear-viscous range, $k_{\nub} > k > k_{\nus}$, which we illustrate in \autoref{fig:spectra}. Intuitively, the bulk viscosity acts like a Laplacian-like viscosity operator that primarily effects the compressible modes, starting at high-$k$ and shifting to low-$k$ as $\Pnu$ decreases. 
    
    In the $\Pnu > 1$ regime, which simulations without explicit bulk viscosity probe ($\Pnu = \infty$), there is a strong, high-$k$ build-up of compressible modes that are energetically dominant, and appear as a bottleneck in the total kinetic energy spectrum. For $\Pms = 1$ plasmas (\autoref{eq:Pm_shear_defn}), where $k_{\nus} = k_\eta$, this results in low-levels of magnetisation and high-levels of compressible kinetic energy building up on these scales. Without an explicit bulk viscosity operator, this leads to an unresolved dissipation scale (i.e., an exponential cutoff that falls outside of the computational domain) turning what is usually considered a direct numerical simulation (DNS) into an implicit large eddy simulation for the compressible modes. What is clear is that if one wants to do a compressible DNS where all viscous scales are resolved, one needs to use both $\nub$ and $\nus$, with $\Pnu \sim 1$, otherwise the grid controls the dissipation of the compressible modes.
    
    Beyond performing a strict DNS, where all dissipation scales are resolved, whether or not including bulk viscosity into astrophysical plasma model is of course sensitive to the microphysics of the plasma of interest. We estimate the bulk viscosity for the cold molecular ISM and show that both for the ambient CO and the shocked H$_2$ the bulk viscosity is not negligible, close to or stronger than the shear viscosity due to the internal rotational degrees of freedom of the H$_2$ and CO molecules. In general, the nature of bulk viscosity is largely unknown for the many other astrophysical plasmas, with a very limited amount of exploration in the literature both in the laboratory and in simulation. This may be especially important for dusty plasmas \citep[e.g.,][]{Hopkins2018_dust_instabilities,Israeli2023_dusty_instabilities,Soliman2024_dusty_cold_gas}, with complex internal degrees of freedom that will directly influence the nature and evolution of compressible modes in the plasmas.
    
    In the $\Pnu < 1$ regime, the compressible and incompressible modes become coupled in the sub-bulk-viscous range. The coupling shows that as $k_{\nub}$ shifts to low-$k$, $k_{\nus}$ shifts with it, which for fixed $\Pms$ also brings $k_{\eta}$ to lower $k$ modes. When $\Pnu$ is very low, such that $t_{\nub}\sim t_0$, there are significant changes in the shape of all energy spectra, dynamo integral energy statistics, and of course, an extremely strong suppression of compressible modes and $\bnab\cdot\vecB{v}$. When $\Pnu$ is closer to one, the impact the bulk viscosity has on the dynamo is only in the high-$k$ modes, and such, integral energies, growth rates, and saturations are largely unaffected. Future work is required to determine the relations for the scaling between $k_{\nub}$, $k_{\nus}$ and $k_{\eta}$ in the $\Pnu < 1$ regime, where the coupling is strong, but at outer scale $\M \approx 1$, the compressible mode spectrum appears as $\ekinpar(k) \sim k^{-2}$, which means that $\ell_0 \sim \Re^{2/3}\ell_{\nub}$, following \citet{Kolmogorov1941}-style arguments \citep{Schober2012_saturation_Re_Rm_dependence}.  

    \subsection{Compressibility in the turbulent dynamo}
    To understand the role of compressible modes in the kinematic and saturated dynamo stages, we derive the $\first$ moment magnetic energy equation that separates out the effects of compressible and incompressible modes, in the sense that they are truly orthogonal contributions to the growth rate and saturation. This includes three main terms: volume-preserving stretching, volume-preserving winding and volume-changing compressions (\autoref{eq:B_v_grad_tensor}--\ref{eq:A_v_grad_tensor}, respectively). Winding is orthogonal to the magnetic field tensor, $b_ib_j$, hence one cannot locally wind a turbulent magnetic field to grow it (see \autoref{eq:no_winding_growth}). By directly computing the growth terms (\autoref{fig:growth}), we find that stretching is always the most dominant pathway for growing the magnetic field, even in the supersonic dynamo, and irrespective of the phase of the dynamo. This explains why $\ell_{\nus}/\ell_{\eta} \sim \Pms^{1/2}$, (the stretching rate at $\ell_{\nus}$ balances with the resistive dissipation rate) is true for the compressible regime, which was recently found in \citet{Kriel2025_supersonic_scales}. This is in tension with some recent results i.e., \citet{Sur2023_dynamo}, but the tension is removed if they carefully remove the trace of $\St$ in their decomposition (see \autoref{fn:tension}). Compression plays a dual, symmetric (at least at $\M = 1$) role, in both growing and decaying the magnetic field through compressions and dilatations. The largest effect on the growth rates are through correlations between the high-$k$ modes close to the viscous dissipation scales of the compressible and incompressible kinetic energy spectra. Manifestly, this means that the bulk viscosity ends up changing the growth rate by changing the shear viscous scale. This has flow on effects on the magnetic field structure. 

    \subsection{Beyond the fluid continuum limit}
    Our study is embedded within the context of fluid plasma theory, where $\lambda_{\text{mfp}}/L \ll 1$, and $\lambda_{\text{mfp}}$ is the molecular mean-free-path\footnote{For the interstellar medium, the cold neutral medium has an electron mean-free-path $\lambda_{e} \sim 5\,\text{R}_{\oplus}$, the warm neutral medium $\lambda_{e} \sim 1.3\,\text{AU}$, the warm ionised medium $\lambda_{e} \sim 0.17\,\text{AU}$ and the hot ionised medium $\lambda_{e} \sim 0.3\,\text{pc}$, with $L$ varying between 10 to 100$\,\text{pc}$ \citep{Ferriere2020_reynolds_numbers_for_ism}, hence the interstellar medium really is very well-approximated by a fluid.}. We have studied the implications of bulk viscosity $\sim \bnab\cdot(\bnab\cdot\vecB{v})\It$ in $\M \approx 1$, turbulent dynamos, where shocks are able to permeate through the fluid medium. However, for collisionless $\lambda_{\text{mfp}}/L \gg 1$, weakly collisional $\lambda_{\text{mfp}}/L \sim 1$ and \citet{Braginskii1965_MHD}-MHD the viscous properties of the plasma are significantly different from the vanilla MHD fluid form (\autoref{eq:viscous+_stress_tensor}; \citealt{StOnge2020S_weakly_collisional_dynamos,Squire2023_pressure_aniostropy_turbulence}). For example, in \citet{Braginskii1965_MHD}-MHD, which is important for plasmas in the intracluster medium \citep{Berlok2020_branskii_viscosity}, particle collisions are rare, in turn preventing the pressure tensor from isotropising. This creates anisotropic viscous stress $\sim (\hat{\vecB{b}}\otimes\hat{\vecB{b}}):\bnab\vecB{v}$ along $\vecB{b}$, taking the same form as our our dynamo growth terms \autoref{eq:growth_rate_decomp}, making the competition for dynamo growth and viscous diffusion more fierce in this regime \citep{StOnge2020S_weakly_collisional_dynamos,Squire2023_pressure_aniostropy_turbulence}. As \citet{Squire2023_pressure_aniostropy_turbulence} points out, whether or not either of these terms, $\sim \bnab\cdot\hat{\vecB{b}}\otimes\hat{\vecB{b}}[(\hat{\vecB{b}}\otimes\hat{\vecB{b}}):\bnab\vecB{v}]$ (Braginskii viscosity) or $\sim \bnab\cdot(\bnab\cdot\vecB{v})\It$ (bulk viscosity), are viscous dissipative or non-dissipative (i.e., akin to the thermal response of $\bnab p$), depends strongly upon the aforementioned plasma regime (see footnote on page 10 in \citealt{Squire2023_pressure_aniostropy_turbulence} for a longer discussion). Interestingly, if we use our decomposition in \autoref{eq:growth_rate_decomp} for the Braginskii-MHD viscosity, we find that there is a bulk-viscous-type term $\sim (\hat{\vecB{b}}\otimes\hat{\vecB{b}}):\Bt$, which will act to preferentially damp fast and slow magnetosonic waves along $\vecB{b}$. Furthermore, it has been seen in controlled laboratory astrophysics experiments that kinetic instabilities of the \citet{Weibel1959_intabilities}-type can play a critical role in the dynamics of collisionless shocks \citep{Fox2013_filament_instability,Huntington2015_weibel_in_a_lab}. Whether such mechanisms can be parameterised at all in terms of viscosity coefficients remains an open question. We leave the detailed exploration of these bulk viscous-type operators in the context of these different plasma regimes for future studies.
    
\section*{Acknowledgements}
    We thank the anonymous reviewer who significantly helped to improve the analysis presented in the study, especially the mode correlation analysis and compressible mode spectrum discussion. We thank Bart Ripperda, Aditya Vijaykumar, Janosz Dewberry, Elias Most, Philip~K.S.~Kempski, Amit Seta, Mark Krumholz, Chris Thompson and Drummond Fielding for the helpful discussions. 
    J.~R.~B.~acknowledges financial support from the Australian National University, via the Deakin PhD and Dean's Higher Degree Research (theoretical physics) Scholarships and the Australian Government via the Australian Government Research Training Program Fee-Offset Scholarship and the Australian Capital Territory Government funded Fulbright scholarship.
    C.~F.~acknowledges funding provided by the Australian Research Council (Discovery Projects DP230102280 and and DP250101526), and the Australia-Germany Joint Research Cooperation Scheme (UA-DAAD). A.~B. and J.~R.~B. also acknowledge the support from NSF Award 2206756. 
    We further acknowledge high-performance computing resources provided by the Leibniz Rechenzentrum and the Gauss Centre for Supercomputing (grants~pr32lo, pr48pi and GCS Large-scale project~10391), the Australian National Computational Infrastructure (grant~ek9) and the Pawsey Supercomputing Centre (project~pawsey0810) in the framework of the National Computational Merit Allocation Scheme and the ANU Merit Allocation Scheme.
    
    The simulation software, \textsc{flash}, was in part developed by the \textsc{flash} Centre for Computational Science at the Department of Physics and Astronomy of the University of Rochester. We utilise the \textsc{TurbGen} code for the turbulent forcing function, $\vecB{f}$ \citep{Federrath2010_driving,Federrath2022_turbulence_driving_module}. Data analysis and visualisation software used in this study: \textsc{C++} \citep{Stroustrup2013}, \textsc{numpy} \citep{Oliphant2006,numpy2020}, \textsc{matplotlib} \citep{Hunter2007}, \textsc{cython} \citep{Behnel2011}, \textsc{visit} \citep{Childs2012}, \textsc{scipy} \citep{Virtanen2020},
    \textsc{scikit-image} \citep{vanderWalts2014}, \textsc{cmasher} \citep{Velden2020_cmasher}.

\section*{Data Availability}
    The data underlying this article will be shared on reasonable request to the corresponding author.

\bibliographystyle{mnras.bst}
\bibliography{Jan2022.bib} 
\appendix
\section{Non-dissipative property of $\At$, the rate of rotation tensor}\label{app:A_tensor}
While the rate of strain tensor \autoref{eq:S_v_grad_tensor} contributes to shearing deformation through isochoric transformations of fluid parcels, the rate of rotation tensor \autoref{eq:A_vort_tensor}, instead represents rigid body rotations about a point. The latter can be shown to neither grow nor dissipate energy\footnote{Note that even though $\At$ does not couple to $\partial_tb^2$, it does couple to $\partial_t \hat{\vecB{b}}$ through $\hat{b}_i\At_{ij}$, where $\hat{\vecB{b}}$ is the magnetic unit vector. So even though $\At$ does not directly change the amplitude, it can change the orientation of the magnetic field, which is undoubtedly a key aspect of the turbulent dynamo \citep{Seta2021_saturation_supersonic_dynamo,Sur2023_dynamo,Beattie2025_10k}.}. We show this by constructing the $E_{\text{kin}}$ equation by left-multiplying $\mathbf{v} \cdot$ to the strong form of the MHD momentum equation \autoref{eq:momentum} with zero external forcing ($\rho\vecB{f} = 0$),
\begin{align}\label{eq:mom_app}
           \frac{\partial E_\text{kin}}{\partial t} &+ \vecB{v} \cdot \bnab\cdot\Bigg[\rho \vecB{v}\otimes\vecB{v} -\frac{1}{4\pi}\vecB{b}\otimes\vecB{b} + p\It
            - \rho\bm{\sigma} \Bigg] = 0,
\end{align}
where $p$ is the total (thermal and magnetic) pressure. We direct our attention solely to the term involving the fully decomposed viscous stress tensor with $\At$ (indices explicitly shown for clarity), 
\begin{equation}
 \bm{v} \cdot \bnab \cdot (\rho\At) =  v_ i \At_{ij} \partial_j\rho + \rho v_i \partial_j \At_{ij}, 
\end{equation}
where $\rho v_i \partial_j \At_{ij} = -\rho v_i \partial_j  \At_{ji}$ and $ v_ i \At_{ij} \partial_j\rho =  -v_ i \At_{ji} \partial_j\rho$, where we have applied the skew-symmetric property of $\At$, i.e., $\At_{ij} = -\At_{ji}$. Expanding the sum of either $\rho v_i \partial_j \At_{ij}$ or $v_ i \At_{ij} \partial_j\rho$, one then finds that due to the skew-symmetric property of $\At_{ij}$, each term has a corresponding term that cancels the other out, e.g.,
\begin{align}
   \rho v_i \partial_j \At_{ij} &= \rho\big(\underbrace{v_x \partial_y \At_{xy}+ v_y \partial_x \At_{yx}}_{0} + \hdots\big) = 0.
\end{align}
Therefore, $\At$ has no contribution of the evolution (growth or dissipation) in the viscous stress term in the kinetic energy evolution, and hence need not be considered in the isotropic-deviatoric decomposition of the viscous stress tensor, \autoref{eq:viscous+_stress_tensor}.
\\

\section{Bulk viscosity in the molecular ISM}\label{app:bulk_viscosity_in_ISM}
    Let us consider the bulk viscosity of two common molecules in the cold phases of the ISM: H$_2$ and CO.

\subsection{The bulk viscosity of H$_2$ in the ambient and shock-heated cold medium}
    Retaining only the velocity divergence terms from a Chapman-Enskog expansion of the Boltzmann equation about a local Maxwellian distribution function shows that the bulk viscosity is directly related to the relaxation times of the internal degrees of freedom of the gas or plasma,
    \begin{align}\label{eq:bulk_viscosity_via_boltzmann}
        \nub = (\gamma - 1)^2\sum_i^N \frac{c_{V,i}}{R}P_{\rm eq}\tau_i,
    \end{align}
    where $R$ is the gas constant, as discussed in \autoref{sec:intro}. In the cold phase (temperatures $T \approx 10-30\,\rm{K}$), only the energetically lowest transitions can be excited. For H$_2$, this is the $J=0\!\to\!2$ rotational excitation, due to the homonuclear nature of the molecule \citep{Krumholz2015}. This transition is at roughly $\theta_r \approx 510\,\rm{K}$ above the ground state energy, which means only a small fraction $e^{-\theta_r/T} \approx 10^{-22}$ at $T = 10\,\rm{K}$ of the H$_2$ population is in this state. This prevents the bulk viscosity from ever being significant at $T = 10\,\rm{K}$ in the ambient H$_2$. 
    
    However, because $c_s \propto \sqrt{T} \approx 0.2 \,\rm{kms}^{-1}$ is so low \citep{Beattie2022_ion_alfven_fluctuations}, the cold ISM is easily able to facilitate shocked flows, which locally shock-heat the medium up to $T \approx 10^3\,\rm{K}$ and beyond \citep{Godard2019_irridated_shocks}. Focusing only on the $J=0\!\to\!2$ transition for simplicity (vibrational degrees of freedom are only excited at $T \gtrsim 2\times 10^3\,\rm{K}$), now most of the H$_2$ population is excited into the $J=0\!\to\!2$ state, i.e., $e^{-\theta_r/T} \approx 0.6$. Directly from the rotational partition function, we can estimate the heat capacity for the rotational degrees of freedom of a linear rotator,
    \begin{align*}
        \frac{c_{V,r}(T)}{R} =\left(\frac{\theta_r}{T}\right)^{\!2}\,
    \frac{\mathrm e^{\theta_r/T}}
    {(\mathrm e^{\theta_r/T}-1)^{2}}.
    \end{align*}
    where $\theta_r$ is the temperature for a rotational transition, which gives $c_{V,r}(T)/R \approx 0.98$, and $\gamma = 1 - R/c_V$, where we use $c_V = 3R/2  + 0.98 R$ for the translational and rotational degrees of freedom. This means $\gamma \approx 1.39$, very close to the ideal diatomic $\gamma$, and hence $(\gamma - 1)^2 \approx 0.15$. The final calculation we need to make is for $P_{\rm eq}\tau$ for the rotational degree of freedom. We utilize \citet{Parker1959_relaxation_of_diatomic_gases}'s collision number model, 
    \begin{align}\label{eq:parkers_formula}
        \tau_r = \frac{Z_r \mu_{\rm{dyn,shear}}(T)}{P_{\rm eq}},
    \end{align}
    where $Z_r(\text{H}_2) \approx 174$ is the collision number for H$_2$ at $T \approx 10^3\,\rm{K}$ \citep{Parker1959_relaxation_of_diatomic_gases,ONeal1962_thermal_conduct}, and $\mu_{\rm{dyn,shear}} \approx 1.8 \times 10^{-5}\,\rm{Pa\,s}$ is the dynamic shear viscosity using \href{https://www.lmnoeng.com/Flow/GasViscosity.php}{Sutherland's gas viscosity formula}. Using $P_{\rm eq} \approx n\,k_B T = 1.38 \times 10^{-11}\,\rm{Pa}$ with $n \approx 10^3\,\rm{cm}^{-3}$ and $T \approx 10^3\,\rm{K}$, $\tau_r \approx 2.3 \times 10^{8}\,\rm{s} \approx 7.4\,\rm{yrs}$. We now have everything we need to estimate $\nub$ in H$_2$ that has had its internal degrees of freedom excited via a $T \approx 10^{3}\,\rm{K}$ shock. Directly using \autoref{eq:bulk_viscosity_via_boltzmann}, we find $\nub / \nus \approx 30$, showing that $\nub$ cannot be neglected in the supersonic medium. The primary caveat being that the cooling time of the shock-heated plasma is short. However, \citep{Godard2019_irridated_shocks} showed that even when cooling times are short, temperatures and UV pumping and chemical heating add additional excitation channels while the shock is still hot, quickly shifting the H$_2$ population into high-J and vibrational levels, creating an effective $\nub$.

    \subsection{The bulk viscosity of CO in the ambient, cold molecular medium}
    CO is unlike H$_2$ because it is not homonuclear and hence the $J=0\!\to\!1$ internal degree of freedom can be excited. Furthermore, it is excited at low temperatures, $T \approx 5.5\,\rm{K}$ \citep{Krumholz2015}. Vibrational modes are not excited until a few thousand Kelvin, hence we again focus on solely the rotational mode. This means that at $T \approx 10\,\rm{K}$, $c_{V,r}(T)/R \approx 0.97$ is almost completely saturated, making $(\gamma - 1)^2 \approx 0.15$, as before. Due to a larger moment of inertia and cross-section from inelastic scattering, the collision number for CO is much smaller than for H$_2$, with $Z_r(\text{CO}) \approx 2$, noting that the laboratory measurements of $Z_r$ are confined to larger temperatures than desired for the $T \approx 10\,\rm{K}$ medium \citep{Malinauskas1970_rotational_collision_numbers}. Using \autoref{eq:parkers_formula}, we find $\tau_r \approx 7.1 \times 10^6\,\rm{s} \approx 3\,\rm{months}$, significantly shorter than shock-heated H$_2$. The short relaxation timescales translate into smaller bulk viscosities compared to H$_2$, and at $T \approx 10\,\rm{K}$, this means $\nub /\nus \approx 1$ for CO. This means that the bulk viscosity in CO spread throughout the ambient medium is just as important as the shear viscosity.

\section{Expressing the $\first$ moment magnetic energy equation in terms of tensor operations}\label{app:first_moment_mag_eq}
     We present \autoref{eq:magnetic_energy_m1} in our main study, which is a compressible version of $\first$ moment of the magnetic energy equation shown in \citet{Schekochihin2004_dynamo}. \citet{Seta2020_saturation_high_Pm} and \citet{Sur2023_dynamo} derive a similar $\first$ moment magnetic energy equation. However, so that we can apply our $\bnab\otimes\vecB{v}$ decomposition, we explicitly derive it in terms of tensor products, unlike either of these studies. \autoref{eq:growth_rate_decomp}. The induction equation with Ohmic dissipation is,
    \begin{align}
        \frac{\partial \vecB{b}}{\partial t} = \bnab \times (\vecB{v} \times \vecB{b}) + \eta \bnab^2 \vecB{b}.
    \end{align}
    Now taking the scalar product with $\vecB{b}$, 
    \begin{align}
        \frac{\partial b^2}{\partial t} = 2\vecB{b}\cdot\bnab \times (\vecB{v} \times \vecB{b}) + 2\eta \vecB{b}\cdot\bnab^2 \vecB{b},
    \end{align}
    where
    \begin{align}
        \bnab \times (\vecB{v} \times \vecB{b}) = \vecB{b}\cdot\bnab\otimes\vecB{v} -\vecB{b} (\bnab \cdot \vecB{v}) - \vecB{v}\cdot\bnab\otimes\vecB{b}, 
    \end{align}
    and,
    \begin{align}
        \vecB{b}\cdot\bnab \times (\vecB{v} \times \vecB{b}) &= \vecB{b}\otimes\vecB{b}:\bnab\otimes\vecB{v} \\
        & \quad - \vecB{b}\otimes\vecB{b}:(\bnab \cdot \vecB{v})\mathbb{I} - \vecB{b}\otimes\vecB{v}:\bnab\otimes\vecB{b}.
    \end{align}
    The last term can be simplified,
    \begin{align}
        \vecB{b}\otimes\vecB{v}:\bnab\otimes\vecB{b} &= \frac{1}{2}\bnab\cdot(\vecB{v}b^2) -\frac{1}{2}\vecB{b}\otimes\vecB{b}:(\bnab \cdot \vecB{v})\mathbb{I},
    \end{align}
    which means
    \begin{align}
        \vecB{b}\cdot\bnab \times (\vecB{v} \times \vecB{b}) &= \vecB{b}\otimes\vecB{b}:\bnab\otimes\vecB{v} \\
        & \quad - \frac{1}{2}\vecB{b}\otimes\vecB{b}:(\bnab \cdot \vecB{v})\mathbb{I} - \frac{1}{2}\bnab\cdot(\vecB{v}b^2).
    \end{align}
    The dissipation term is simply, 
    \begin{align}
        \vecB{b}\cdot \bnab^2 \vecB{b} = \vecB{b}\cdot \bnab \cdot (\bnab \otimes \vecB{b}).
    \end{align}
    Hence the dynamical equation for $b^2$ is
    \begin{align}
        \frac{\partial b^2}{\partial t} &=  2\left[ \vecB{b}\otimes\vecB{b}:\bnab\otimes\vecB{v} - \frac{1}{2}\vecB{b}\otimes\vecB{b}:(\bnab \cdot \vecB{v})\mathbb{I} - \frac{1}{2}\bnab\cdot(\vecB{v}b^2) \right]\nonumber \\
        &\quad + 2\eta\left[  \vecB{b}\cdot \bnab \cdot (\bnab \otimes \vecB{b}) \right],
    \end{align}
    Now we take the $\first$ spatial moments of the equation and divide by $8\pi$ to put the time-derivative in our energy density unit system,
    \begin{align}
        \frac{\partial \emagb}{\partial t} &=  \frac{2}{8\pi}\Exp{\vecB{b}\otimes\vecB{b}:\bnab\otimes\vecB{v} - \vecB{b}\otimes\vecB{b}:\frac{1}{2}(\bnab \cdot \vecB{v})\mathbb{I})}{\V} \nonumber \\
        &\quad + \frac{2\eta}{8\pi}\Exp{\vecB{b}\cdot \bnab \cdot (\bnab \otimes \vecB{b})}{\V},
    \end{align}
    where $\Exp{ \frac{1}{2}\bnab\cdot(\vecB{v}b^2)}{\V} = 0$ because there is no net magnetic flux through the triply periodic boundary. This is in every way equivalent to \citet{Seta2020_saturation_high_Pm}, but we retain the full $\bnab\otimes\vecB{v}$, instead of just $\St$ (see \autoref{sec:growth} for more discussions on how $\bnab\otimes\vecB{v}$ also contains tensorial compression effects), as we mentioned in the main text. We do not retain the advection term in the $\first$ moment, as in \citet{Sur2023_dynamo}. Now we simplify the dissipation term to get to the form expressed in \citet{Schekochihin2004_dynamo}. Firstly,
    \begin{align}
        \Exp{\vecB{b}\cdot \bnab \cdot (\bnab \otimes \vecB{b})}{\V} = \int_{\V} \d{\V} \, \vecB{b}\cdot \bnab \cdot (\bnab \otimes \vecB{b}),
    \end{align}
    which by parts $\int u\d{v} = uv - \int v \d{u}$, where $u = \vecB{b}$, $\d{u} = \bnab\otimes\vecB{b}$, $\d{v} = \bnab \cdot (\bnab \otimes \vecB{b})$, $v = \bnab \otimes \vecB{b}$ is
    \begin{align}
        \int_{\V} \d{\V} \, \vecB{b}\cdot \bnab \cdot (\bnab \otimes \vecB{b}) =& \cancelto{0}{\vecB{b}\cdot\bnab\otimes\vecB{b}\bigg|^{\infty}_{-\infty}} \\
        & \quad- \int_{\V}\d{\V}\, (\bnab \otimes \vecB{b}) :  (\bnab \otimes \vecB{b}),
    \end{align}
    where the first term in the integral goes to zero due to periodicity. Hence, 
    \begin{align}
         \Exp{\vecB{b}\cdot \bnab \cdot (\bnab \otimes \vecB{b})}{\V} = - \Exp{(\bnab \otimes \vecB{b}) :  (\bnab \otimes \vecB{b})}{\V}.
    \end{align}
    Finally, we put this expression back into our evolution equation, and multiply and divide by the integral energy density to get 
    \begin{align}
        \frac{\partial \emagb}{\partial t} &=  \frac{2\emagb}{\Exp{b^2}{\V}}\Bigg\langle(\vecB{b}\otimes\vecB{b}):(\bnab\otimes\vecB{v}) \nonumber\\
        &\quad\quad\quad - (\vecB{b}\otimes\vecB{b}):\left(\frac{1}{2}(\bnab \cdot \vecB{v})\mathbb{I}\right)\Bigg\rangle_{\V} \nonumber \\
        &\quad - \frac{2\eta\emagb}{\Exp{b^2}{\V}}\Big\langle(\bnab \otimes \vecB{b}) :  (\bnab \otimes \vecB{b})\Big\rangle_{\V},
    \end{align}
    the $\first$ moment equations shown in \autoref{eq:magnetic_energy_m1}.

\section{Resolution study}\label{app:convergence}
    \begin{figure}
        \centering
        \includegraphics[width=\linewidth]{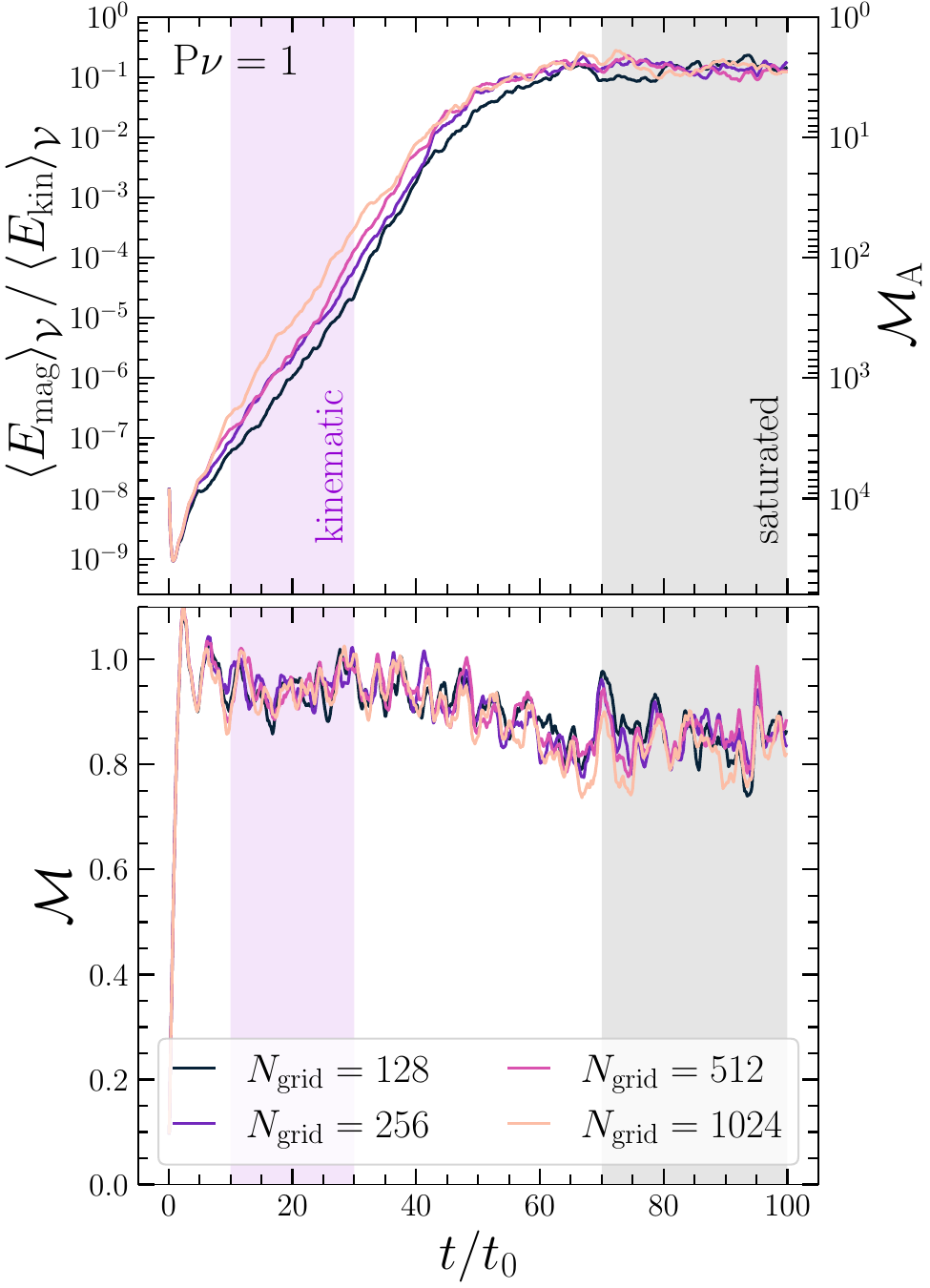}
        \caption{The same as \autoref{fig:integral_energies} but instead of colouring by $\Pnu$, we fix $\Pnu = 1$ and change grid resolution of the simulation, $N_{\text{grid}}$. We list the growth rate $\gamma_1$ and the final saturation of $\emagb/\ekinb$ in \autoref{tab:sims}, and find good convergence for both $\gamma_1$ and $\emagb/\ekinb$ saturation at $N_{\text{grid}} \approx 256$.}
        \label{fig:integral_energy_convergence_Pnu1}
    \end{figure}
    
    \begin{figure}
        \centering
        \includegraphics[width=\linewidth]{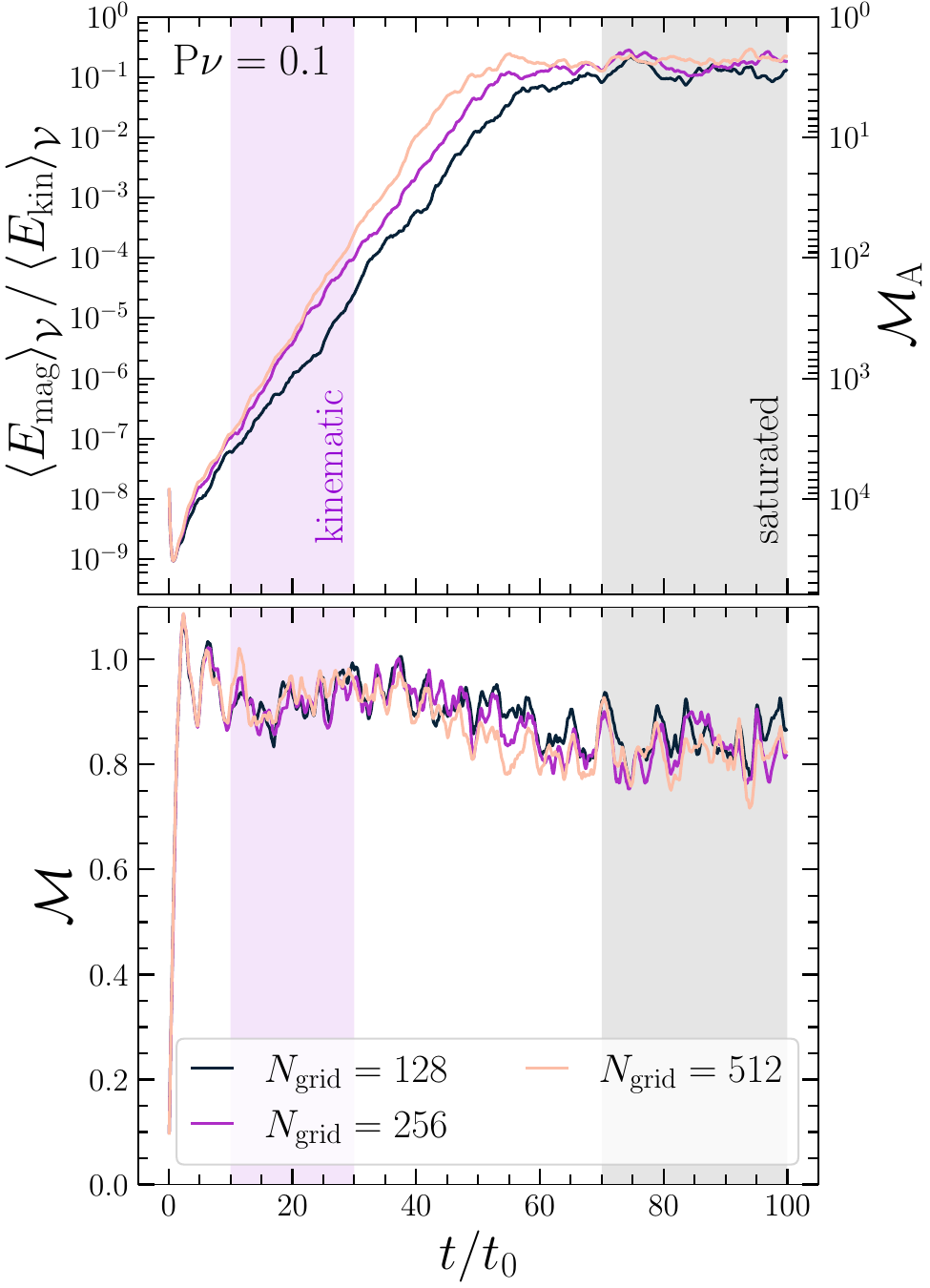}
        \caption{The same as \autoref{fig:integral_energy_convergence_Pnu1} but for $\Pnu = 0.1$. At low $\Pnu$, we similar convergence properties as the $\Pnu = 1$ simulation, for both $\gamma_1$ and the $\emagb/\ekinb$ saturation.}
        \label{fig:integral_energy_convergence_Pnu0.1}
    \end{figure}

\begin{figure}
        \centering
        \includegraphics[width=\linewidth]{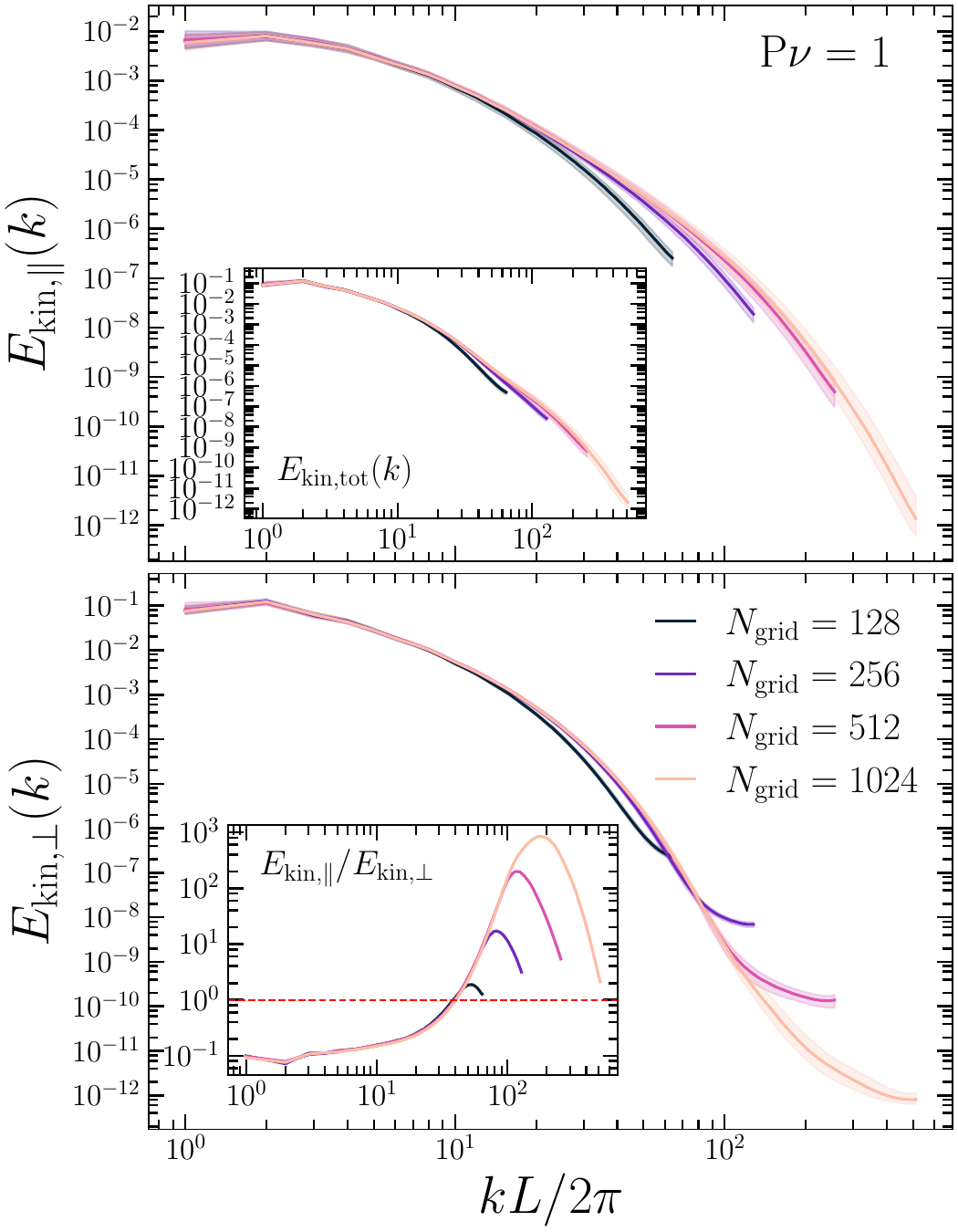}
        \caption{The compressible ($\ekinpar(k)$; top) and incompressible ($\ekinperp(k)$; bottom) kinetic energy spectra for the $\Pnu=1.0$ ensemble of simulations (the same as the middle and bottom panel in \autoref{fig:decomp_vel_spect}), tabulated in \autoref{tab:sims}. The inset in the top panel shows the total kinetic energy $\ekin(k)$, and the inset in bottom panel shows the ratio between compressible and incompressible spectral kinetic energies, $\ekinpar(k)/\ekinperp(k)$. We find good convergence in $\ekinperp(k)$, flattening out well before the Nyquist frequency, indicating that $k_{\nus}$ is resolved. However, for $\ekinpar(k)$, the dissipation scale appears to be unresolved, which results in an extended sub-shear-viscous range of scales that grows with $N_{\text{grid}}$.}
        \label{fig:power_spec_converge_P1}
    \end{figure}

\begin{figure}
        \centering
        \includegraphics[width=\linewidth]{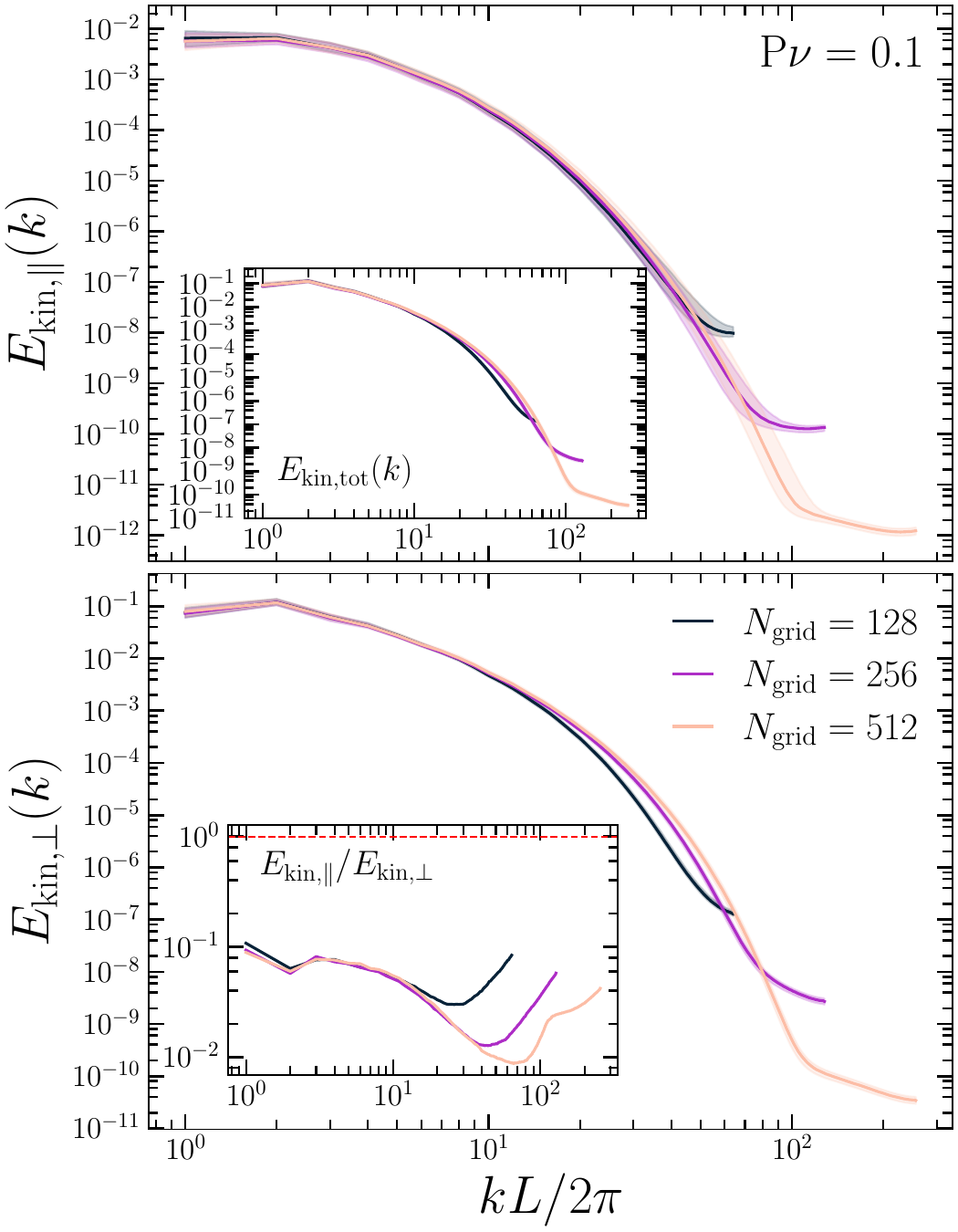}
        \caption{The same as \autoref{fig:power_spec_converge_P1} but for the $\Pnu=0.1$ ensemble of simulations, which is the $\Pnu$ where the bulk viscosity and shear viscosity are approximately equal. With the inclusion of bulk viscosity that matches the shear viscosity, $\ekinpar(k)$ now has a resolved dissipation scale that is (qualitatively) at a similar $k$ mode as the dissipation scale in $\ekinperp(k)$.}
        \label{fig:power_spec_converge_P01}
    \end{figure}

    In this section we provide a resolution study for our $\Pnu = 1$ and $\Pnu=0.1$ simulations, showing both the integral energy statistics, as in \autoref{fig:integral_energies}, in \autoref{fig:integral_energy_convergence_Pnu1} and \autoref{fig:integral_energy_convergence_Pnu0.1}, and the Helmholtz decomposed spectra in \autoref{fig:power_spec_converge_P1} and \autoref{fig:power_spec_converge_P01}, for $\Pnu = 1$ and $\Pnu = 0.1$, respectively. All statistics are coloured by the grid resolution, $N_{\rm grid}^3 = 128^3$, $256^3$, $512^3$ and (only for $\Pnu = 1$) $1024^3$. Beyond $256^3$, the integral energy statistics show strong agreement, both in saturation and in growth rates (see \autoref{tab:sims}), with very little deviation based on the grid resolution. As we have pointed out in the main text, this is because the energy integral statistics are always dominated by the low-$k$ modes \citep{Beattie2022_ion_alfven_fluctuations}, so we expect them to converge quickly, and at low grid resolutions. Similarly for the growth rates, because we have only moderate $\Res = 1000$ the growth rates, which are sensitive to $\ell_{\nus}$, also converge quickly.

    Now we compare the decomposed kinetic energy spectra in \autoref{fig:power_spec_converge_P1} and \autoref{fig:power_spec_converge_P01}. Firstly, we can confidently say that the $\Pnu = 0.1$ simulation is converged on all scales at $512^3$. That is, the physical length scales associated with the bulk and shear viscosity are properly resolved in the domain, since both $\ekinpar(k)$ and $\ekinperp(k)$ trace each other almost perfectly through all the $k$ modes, and then truncate before the Nyquist frequency. Furthermore, as we expect from our definition of $\Bt$ (bulk viscosity operator; see \autoref{eq:growth_rate_decomp}), the overall structure of the two spectra are almost identical, i.e., $k_{\nus}\approx k_{\nub}$ (which should be exact for $\Pnu = 1/3$). However, for the $\Pnu = 1$ simulation, $k_{\nus}$ looks resolved between $256^3$ and $512^3$ (as expected from our growth rates), but $k_{\nub}$ does not, i.e., at $\Pnu = 1$, we do not have sufficient resolution, even at $1024^3$, to resolve the entire sub-shear-viscous range, which we see growing in size in the $\ekinpar(k)/\ekinperp(k)$ \autoref{fig:power_spec_converge_P1} inset, reaching values up to $\ekinpar(k)/\ekinperp(k) = 10^3$ within this range of scales. This would be a problem if in our study we focused on measuring precise viscous dissipation scales, and viscous-scale relations. However, we focus on a mode-by-mode analysis, for the most part, and one can read off from \autoref{fig:power_spec_converge_P1}, that even though the physical $k_{\nub}$ is not resolved, we get relatively good convergence up to the $k\approx 100$ mode before there is significant deviation between the spectra due to numerical (compressible) viscosity. This is completely sufficient for our study. This also highlights that for \textit{no} explicit $\Bt$ viscosity ($\Pnu = \infty$), we should not expect any convergence of the high-$k$ compressible modes and $\ekinpar(k)/\ekinperp(k)$ should grow unbounded as $N_{\rm grid}^3$ increases, which exactly what one expects for an ILES.
    \newpage
\section{Compressible and incompressible $\first$ moment energy equations}\label{app:energy_equations}
    We stated the Stokes limit for the compressible and incompressible $\first$ moment kinetic energy equations in \autoref{sec:decomposed_spectra}. Here we derive the equations and take the Stokes limit to demonstrate how we got to \autoref{eq:compressible_kin} and \autoref{eq:incompressible_kin}. We start by considering the momentum equation,
    as in \autoref{eqn:momentum_eqn}, in its advective form. Because we are concerned with the viscous-scale dynamics, we may linearise the momentum equation around a stationary flow where $\Res \sim \Reb < 1$, and $\M \ll 1$,
    \begin{align}
        \frac{\partial\vecB{v}}{\partial t} =  \nus \nabla^2 \vecB{v} + \underbrace{\left(\frac{2\nus}{3} + \lambda \right)}_{\nub}\bnab(\bnab\cdot\vecB{v}).
    \end{align}
    Next, by taking Fourier transforms in space,
    \begin{align}\label{eq:fft_momentum}
        \frac{\partial\tilde{\vecB{v}}(\bm{k})}{\partial t} =  \nus k^2 \tilde{\vecB{v}}+ \nub \bm{k}(\bm{k}\cdot\tilde{\vecB{v}}),
    \end{align}
    allows us to simply apply the longitudinal Helmholtz operator $\mathbb{P}_\parallel = \hat{\bm{k}}\otimes\hat{\bm{k}} = k_i k_j / k^2$ and the transverse operator, $\mathbb{P}_\perp = \It - \mathbb{P}_\parallel$ to both sides of the equation. This is easiest in tensor notation, where $\nus k^2 \tilde{\vecB{v}} = \nus k^2 \tilde{v}_i$ and $\nub \bm{k}(\bm{k}\cdot\tilde{\vecB{v}}) = \nub k_j(k_n\tilde{v}_n)$. Starting with the longitudinal transform and applying it to both the LHS and RHS, 
    \begin{align}
        \frac{\partial\tilde{\vecB{v}}_{\parallel}(\bm{k})}{\partial t} =  \nus k_i (k_j \tilde{v}_j) + \nub k_i(k_n\tilde{v}_n),
    \end{align}
    dividing and multiplying by $k^2$ gives,
    \begin{align}
        \frac{\partial\tilde{\vecB{v}}_{\parallel}(\bm{k})}{\partial t} =  \nus k^2\frac{k_i (k_j \tilde{v}_j)}{k^2} + \nub k^2\frac{k_i(k_n\tilde{v}_n)}{k^2},
    \end{align}
    revealing that each term is just a Laplacian acting upon the longitudinal component of the velocity. Taking the inverse Fourier transform,
    \begin{align}
        \frac{\partial \vecB{v}_{\parallel}}{\partial t} = (\nus + \nub) \nabla^2 \vecB{v}_{\parallel}.
    \end{align}
    Performing the same algebra for the transverse operator acting upon,
    \begin{align}
        \frac{\partial\tilde{\vecB{v}_\perp}(\bm{k})}{\partial t} =  \nus k^2 \tilde{{v}}_{\perp,j}+ \nub \mathbb{P}_{T,ij}[k_j(k_n\tilde{v}_n)],
    \end{align}
    where
    \begin{align}
        \mathbb{P}_{T,ij}[k_j(k_n\tilde{v}_n)] = \delta_{ij}k_j(k_n\tilde{v}_n) - k_i(k_n\tilde{v}_n) = 0.
    \end{align}
    Hence, 
    \begin{align}
        \frac{\partial{\vecB{v}_\perp}}{\partial t} = \nus \nabla^2 \vecB{v}_{\perp},
    \end{align}
    only depends upon the shear viscosity. The integral energy equations are then
    \begin{align}
        \frac{\partial \ekinparb}{\partial t} &= (\Res^{-1} + \Reb^{-1}) \Exp{\bnab \otimes\vecB{v}_{\parallel}:\bnab \otimes\vecB{v}_{\parallel}}{\V}, \\
        \frac{\partial \ekinperpb}{\partial t} &= \Res^{-1} \Exp{\bnab \otimes\vecB{v}_{\perp}:\bnab \otimes\vecB{v}_{\perp}}{\V},
    \end{align}
    as we show in \autoref{sec:decomposed_spectra}. This means that, in the Stokes limit, the modes are uncoupled and evolve independently of each other, at least in the context of the viscosity operators. 

    \begin{figure}
        \centering
        \includegraphics[width=\linewidth]{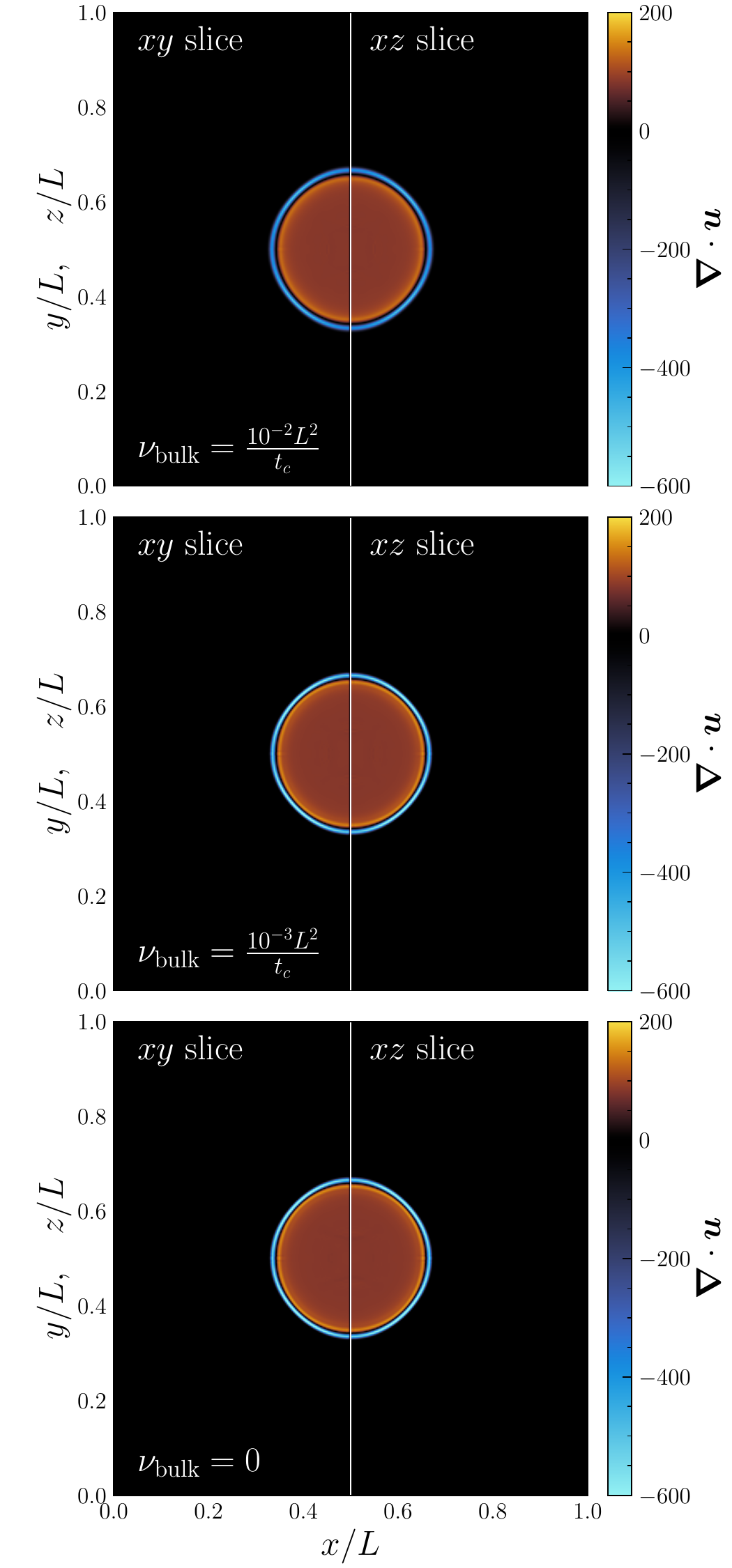}
        \caption{The velocity divergence from the three-dimensional Sedov test showing that bulk viscous fluxes are acting isotropically on the Sedov blast wave. Left of the vertical white line we show the $xy$ slice, and right of the line the $xz$ slice, with almost perfect symmetric between the two slices. Each panel as a different $\nub$ (labelled in the bottom-left corner), varying from the strongest, $\nub = 10^{-2}L^2/t_c$ (top), to the ideal limit, $\nub = 0$ (bottom). Each simulation is visualised at the same time realisation. The $\nub = 10^{-2}L^2/t_c$ simulation has a visibly thicker blast wave, with smaller $\bnab\cdot\vecB{u}$ compared the ideal case.}
        \label{fig:sedov_3d}
    \end{figure}

    \begin{figure}
        \centering
        \includegraphics[width=\linewidth]{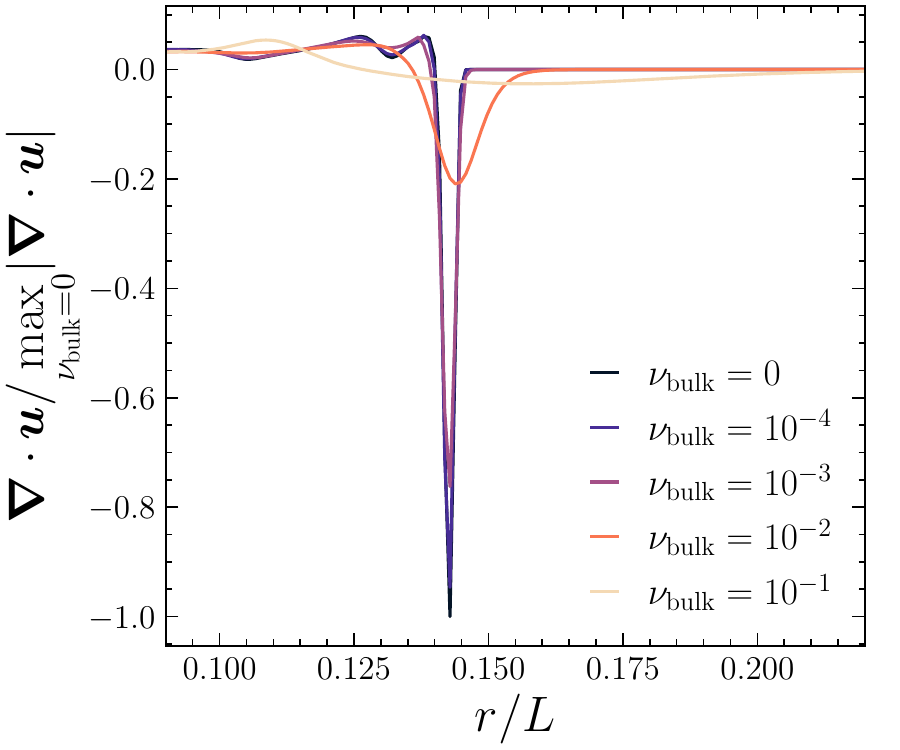}
        \caption{One-dimensional radial profiles of $\bnab\cdot\vecB{u}$ for a two-dimensional Sedov-Taylor blast wave with varying $\nub$, in units of $L^2/t_c$, all at the same $t/t_c = 10^{-2}$. The $\bnab\cdot\vecB{u}$ is normalised by the absolute maximum $\bnab\cdot\vecB{u}$ for the $\nub = 0$ simulation. As $\nub$ increases, the bulk viscosity strongly damps the compressive blast wave (strong $\bnab\cdot\vecB{u} < 0$ region) and retards the evolution of the wave.}
        \label{fig:divu_1d}
    \end{figure}

\section{Bulk viscosity implementation}\label{app:implementation}
    \subsection{Viscous fluxes and explicit time-stepping}
    We add the bulk viscous fluxes following the split, second-order explicit scheme in our \citet{Bouchut2010} solver. This adds additional fluxes to each sweep in the solver, both in the momentum equation, $\mathbb{F}(\rho v)$ \autoref{eqn:momentum_eqn} where fluxes are just directly $\bm{\sigma}$, and the energy equation, $\mathbb{F}(\rho E)$ (not discussed in the main text, since the dynamo with isothermal equation of state), where fluxes are $\vecB{v}\cdot\bm{\sigma}$. Now we systematically go through of the cell-face fluxes added in each coordinate sweep of the momentum equation. We adopt the notation that $\overline{X_{ijk}} = (X_{ijk} + X_{i-1jk})/2$ is the average of quantity $X_{ijk}$ between the cell center $X_{ijk}$ and the cell face, e..g, for coordinate $i$, $X_{i-1jk}$. In three-dimensions, the additional momentum fluxes in the $x$ sweep are,
    \begin{align}
        \mathbb{F}_{\Bt}(\rho v_x) =& - \frac{\nub}{2\Delta x} \left( v_{x,ijk} - v_{x,i-1jk}\right) \\ 
        &- \frac{\nub}{4\Delta y} \left( v_{y,ij+1k} - v_{y,ij-1k} + v_{y,i-1j+1k} \right. \nonumber\\
        &\quad\quad\quad\quad\left.- v_{y,i-1j-1k} \right) \\
        & - \frac{\nub}{4\Delta z} \left( v_{z,ijk+1} - v_{z,ijk-1} + v_{z,i-1jk+1} \right. \nonumber\\
        &\quad\quad\quad\quad\left.- v_{z,i-1jk-1} \right) \\
        \approx& \nub \left(\overline{\partial_x v_x} + \overline{\partial_y v_y} + \overline{\partial_z v_z}\right)
    \end{align}
    the $y$ sweep,
    \begin{align}
        \mathbb{F}_{\Bt}(\rho v_y) =& - \frac{\nub}{4\Delta x} \left( v_{x,i+1jk} - v_{x,i-1jk} + v_{x,i+1j-1k} \right. \nonumber\\
        &\quad\quad\quad\quad\left. - v_{x,i-1j-1k}\right) \\ 
        &- \frac{\nub}{2\Delta y} \left( v_{y,ijk} - v_{y,ij-1k} \right) \\
        & - \frac{\nub}{4\Delta z} \left( v_{z,ijk+1} - v_{z,ijk-1} + v_{z,ij-1k+1} \right. \nonumber\\
        &\quad\quad\quad\quad\left.- v_{z,ij-1k-1} \right) \\
        \approx& \nub \left(\overline{\partial_x v_x} + \overline{\partial_y v_y} + \overline{\partial_z v_z}\right)
    \end{align}
    and the $z$ sweep,
    \begin{align}
        \mathbb{F}_{\Bt}(\rho v_z) =& - \frac{\nub}{4\Delta x} \left( v_{x,i+1jk} - v_{x,i-1jk} + v_{x,i+1jk-1} \right. \nonumber\\
        &\quad\quad\quad\quad\left. - v_{x,i-1jk-1}\right) \\ 
        &- \frac{\nub}{4\Delta y} \left( v_{y,ij+1k} - v_{y,ij-1k} + v_{y,ij+1k-1} \right. \nonumber\\
        &\quad\quad\quad\quad\left.- v_{y,ij-1k-1} \right) \\
        & - \frac{\nub}{2\Delta z}  \left( v_{z,ijk} - v_{z,ijk-1} \right) \\
        \approx& \nub \left(\overline{\partial_x v_x} + \overline{\partial_y v_y} + \overline{\partial_z v_z}\right)
    \end{align}
    noting that when the sweep is in the same direction as the derivative, the average over the center to the face can be algebraically simplified to just two terms. We perform the same for the $\mathbb{F}(\rho E)$, but with the additional $\vecB{v}$ scalar product that needs to be carried through the averages. Adding diffusion to our the MHD model makes it mixture of hyperbolic and parabolic equations. Since we are using an explicit scheme, $f_j^{n+1} = f_j^{n} + \Delta t \partial_iF_{ij}^{n}$, this changes the global $\Delta t$ time-step to advance between state $f_j^{n}$ to $f_j^{n+1}$, which is necessary to ensure numerical stability. The diffusion time step becomes,
    \begin{align}
        \Delta t \leq {\rm{CFL}} \min\left\{\frac{1}{2}\frac{\Delta x^2}{\nus}, \frac{1}{2}\frac{\Delta x^2}{\nub}, \frac{1}{2}\frac{\Delta x^2}{\eta} \right\},
    \end{align}
    where $\rm{CFL}$ is the Courant–Friedrichs–Lewy number, and we have added an additional diffusion timescale for $\nub$. This becomes costly for stiff problems, such as low $\nub$, $\nus$ and $\eta$, and at high-resolution, $\Delta x^2$.

    \begin{figure}
        \centering
        \includegraphics[width=\linewidth]{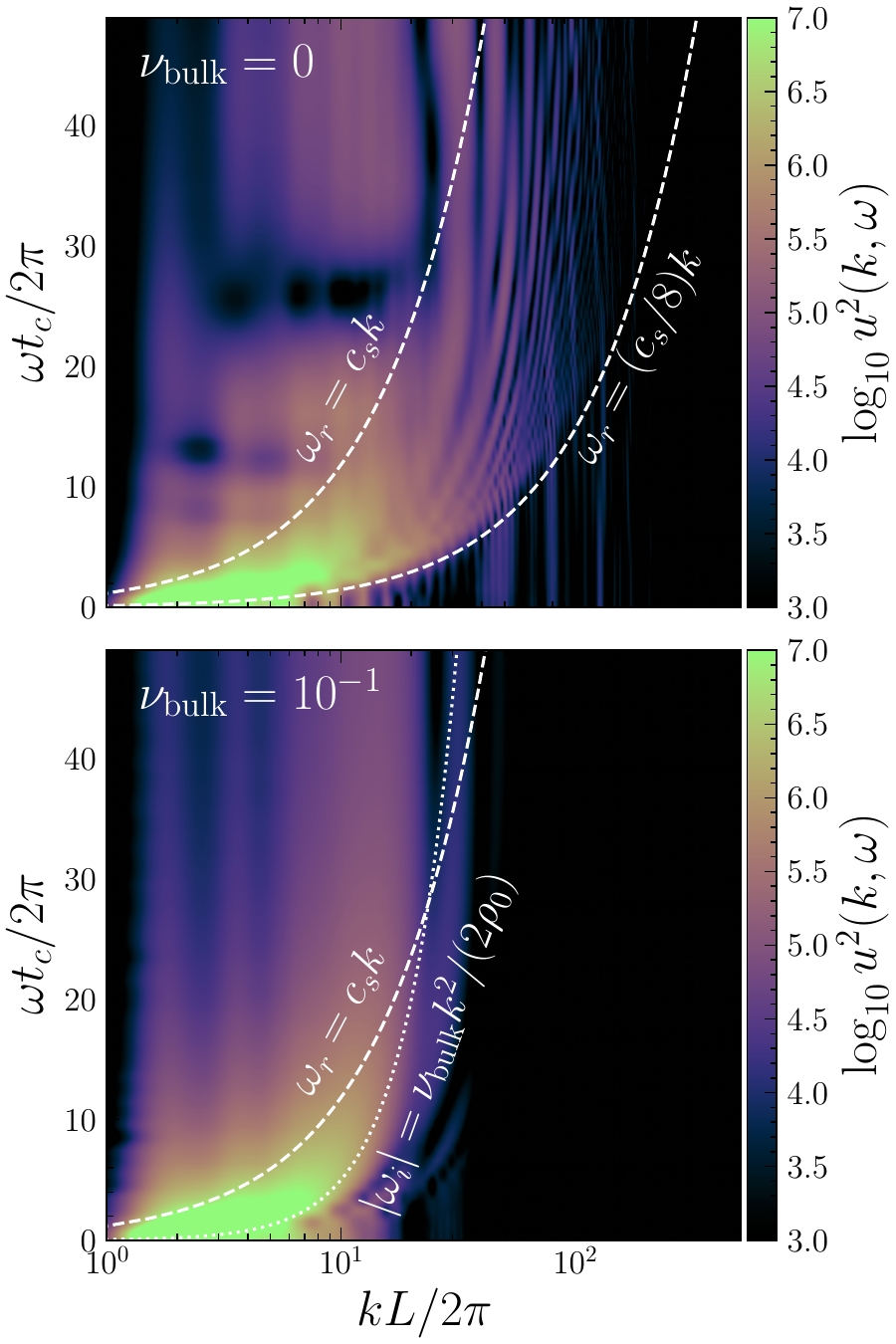}
        \caption{The velocity dispersion relation $\omega(k)$ for the two-dimensional Sedov-Taylor blast wave simulations with $\nub=0$ (top panel) and $\nub=10^{-1}L^2/t_c$ (bottom panel). We plot both real $\omega_r$ (dashed) and imaginary $|\omega_i|$ (dotted) components of the dispersion relation shown in \autoref{eq:viscous_disp}, describing a standard sound wave, $\omega_r$, and the quadratic wave damping from $\nub$, $|\omega_i|$. The top panel, where the blast wave is the strongest, shows deviations from the linear theory, due to the strong nonlinearity of the blast wave, especially at high-$k$ and high-$\omega$, which exhibit significant steepening. However, we still find with a some amount of low to moderate-$k$ and $\omega$ sound waves, with lower sound speeds than $c_s = \sqrt{\gamma p_0/\rho_0}$. These modes are completely damped in the bottom panel, where $\nub$ is the highest. The only modes left are low-$k$, low-$\omega$ sound waves, bounded by the quadratic damping.}
        \label{fig:dispersion_relation}
    \end{figure}

    \subsection{Three-dimensional viscous Sedov test}
    
    A simple numerical test that we can perform to check that the fluxes have been added correctly is a bulk viscous version of the Sedov test in three-dimensions. Because Sedov-Taylor is an intrinsically isotropic test problem, we can use it to directly explore if the additional $\nub \bnab\cdot (\rho\bnab\cdot\vecB{v}\It)$ diffusion isotropically decays the strong $\bnab\cdot\vecB{v} < 0$ shock wave from the explosion. We follow the standard Sedov setup, with a central enhancement of internal energy density, $e$, and hence pressure density $p$, in a small region $p = (\gamma - 1)e/[(4/3)\pi r^3]$, with $\gamma$, the adiabatic index and $r$ the radial coordinate away from the center of three-dimensional and two-dimensional domains $[0,L]$ discretised with $288^3$ grids in three-dimensions and $1024^2$ in two-dimensions. All of the boundaries conform to outflow conditions. We use $p = 10^5p_0$, where $p_0 = 1$ is the ambient pressure, and $\gamma = 7/5$ (diatomic $\gamma$ equation of state), such that the solution to the problem is identical to the self-similar Taylor-von Neumann Sedov blastwave solution. For ambient density $\rho_0 = 1$, the sound speed in the ambient medium is $c_s = \sqrt{\gamma p_0/\rho_0} =\sqrt{\gamma} = \sqrt{7/5}$. This provides a sound-crossing time $t_s = L/c_s = \sqrt{5/7}$, which we use to non-dimensionalise the runtime of the simulations. 

    We run three-dimensional simulations at $\nub = \left\{10^{-2}, 10^{-3}, 10^{-4}, 0 \right\} L^2/t_c$, each simulation resulting in a factor of 10 smaller $\Delta t$. We show a visualisation of the $xy$ and $xz$ slices of $\bnab\cdot\vecB{u}$ for three different simulations in \autoref{fig:sedov_3d}. The top is the most viscous simulation, $\nub = 10^{-2}L^2/t_c$, and the bottom is the ideal simulation, $\nub = 0$. All simulations are normalised to the same range of $\bnab\cdot\vecB{u}$. The low and high divergence shell around the blast wave is dampened and thickened as $\nub$ increases, as expected, but the key result of this plot is that the blast wave is isotropically symmetric, indicating that the bulk viscous fluxes have been added correctly in each sweep. 

    Since we know the fluxes have been added correctly in three-dimensions we turn to the two-dimensional problem, which is easier to run at higher resolutions and higher $\nub$. We show a one-dimensional profile in $r = \sqrt{x^2 + y^2}$ for two-dimensional simulations at $\nub = \left\{10^{-1} ,10^{-2}, 10^{-3}, 10^{-4}, 0 \right\} L^2/t_c$. The increasing $\nub$ diffuses the $\bnab\cdot\vecB{u} = 0$ away, both in the strongly $\bnab\cdot\vecB{u}<0$ blast wave itself and in the $\bnab\cdot\vecB{u}>0$ post-shock region. 

    The final code test that we do is on the wave - $\nub$ viscosity interactions. In the absence of $\nus$, we can derive the linear dispersion relation from the Navier-Stokes equation,
    \begin{align}
        \omega^2 + i\omega\frac{k^2}{\rho_0}\nub - c^2k^2 = 0.
    \end{align}
    Assuming $\omega\frac{k^2}{\rho_0}\nub  \ll c^2k^2$ (i.e., the viscosity is slow compared to the sound speed), we find the approximate solutions for bulk viscous sound waves,
    \begin{align}\label{eq:viscous_disp}
        \omega \approx \overbrace{c_s k}^{\omega_r} - \underbrace{i\frac{\nub}{2\rho} k^2}_{\omega_i},
    \end{align}
    where $\omega$ is the temporal frequency, $\omega_r$ is the real component, describing the classical dispersion relation for sound waves, and $\omega_i$ is the imaginary component, showing the wave damping from $\nub$.
    
    We plot $\omega(k)$ for the two-dimensional simulations with $\nub=0$ (top panel) and $\nub = 10^{-1}L^2/t_c$ (bottom panel) in \autoref{fig:dispersion_relation}. We plot $\omega_r$ and $|\omega_i|$ on the dispersion relation for the one-dimensional $u_i$ along $r/L$, to indicate which modes resemble linear sound waves and which modes are being quadratically damped by $\nub$. We use $c_s = \sqrt{\gamma p_0/\rho_0} = \sqrt{7/5}$ and $\rho = \rho_0$, even though both $c_s$ and $\rho$ change as a function of time and space. For the $\nub=0$ case, we see a range of signatures from linear sound waves with different temperatures (most likely due to the post and pre-shock temperatures changing $c_s \propto \sqrt{T}$) but most of the energy is in the low-$k$, low-$\omega$ sound wave. At high-$k$, we see significant steepening, which is expected for nonlinear structures like shocks. There are a large range of cooler $\sqrt{7/5}/8 \leq c_s \leq \sqrt{7/5}$ sound waves moving from low-$k$, low-$\omega$ to moderate-$k$, moderate-$\omega$. All of these modes are suppressed (indeed, truncated more or less right at $|\omega_i|$) in the strong $\nub$ case. The only feature left is the low-$k$, low-$\omega$ structure, that closely resembles a sound wave with $c_s = \sqrt{7/5}$. Hence, more or less as we show in the main text, the bulk viscosity suppresses the high-$k$ sound waves, gets rid of any nonlinear mode steepening from shocks, and leaves only the low-$k$ sounds waves. We see no significant artefacts and the process that we describe qualitatively matches the linear theory well; hence we conclude that the bulk viscosity implementation is working as it should be.

\bsp	

\label{lastpage}
\end{document}